\documentclass[%
reprint,
amsmath,amssymb,
aps,
]{revtex4-2}
\usepackage{graphicx}
\usepackage{dcolumn}
\usepackage{bm}

\usepackage{pifont}
\usepackage{xcolor}
\usepackage{soul}
\begin{document}
\preprint{APS/123-QED}

\title{Quantum Landscape of Superconducting Diodes}

\author{Muhammad Nadeem}
\email{mnadeem@uow.edu.au}
\author{Xiaolin Wang}
\email{xiaolin@uow.edu.au}
\affiliation{Institute for Superconducting and Electronic Materials (ISEM), Australian Institute for Innovative Materials (AIIM), University of Wollongong, Wollongong, New South Wales 2525, Australia.}


\begin{abstract}
This study maps the quantum landscape of superconducting diodes (SDs) \cite{nadeem23} onto the quantum technology architecture, which is currently constrained by fundamental challenges in control and scalability. In the existing non-integrated quantum technology hardware, control and scalability related issues emerge at two fronts: First, nonlinear and nonreciprocal circuit elements, which are essential building blocks for quantum processors, are often complex, bulky, and dissipative. Second, the temperature gradient between classical control electronics ($T_C\gtrsim$ K), which is also dissipative, and the quantum processor at cryogenic temperatures ($T_Q\sim$ mK) makes scalability even more challenging. The main focus is to reveal how the built-in nonlinearity, nonreciprocity, and quantum functionalities of SDs are significant for on-chip integrated circuit quantum electrodynamics (c-QED), enabling scalable integration of noise-resilient qubit and qubit-interfaces for efficient power delivery, coherent control and memory, high-fidelity readout, and quantum-limited amplification. To this end, this study will also shed light on how thermodynamic constraints and field effects can be harnessed within a quantum-enhanced SD platform, thereby enabling thermal compatibility between classical and quantum workflows, isothermal all-electrical control, and on-chip scalability. This perspective is expected to play a pivotal role in the advancement of superconducting circuit-based quantum hardware with temperature-matched classical, quantum, and hybrid workflows.
\end{abstract}
\maketitle

\tableofcontents

\section{Introduction}
Quantum technologies, their hardware performance, and critical functionalities rely on macroscopic quantum mechanisms intertwined with microscopic quantum phenomena. In quantum materials, the macroscopic properties are entangled with the microscopic behavior of the electronic wave function via bulk-boundary correspondence. However, once these quantum materials are incorporated into quantum devices and integrated circuits, c-QED \cite{xiang13,blais21} provides a framework to translate their microscopic quantum characteristics into macroscopic quantum functionalities. The overall performance of integrated circuits is determined by the efficiency with which quantum devices operate, which serve as the elementary components for qubit operations, power delivery and management, signal processing and control, data storage and memory, readout and amplification, and intervening elements for signal routing and noise isolation.

In the integrated c-QED environment, nonlinear and nonreciprocal circuit elements are the most essential building blocks for both high-end classical computation and quantum technology hardware. In current architectures, however, the nonlinearity present in qubit and qubit-interfaces is typically either dissipative, arising from Kerr effects, or it is realized as dissipationless nonlinearity only by employing complex multipole quantum circuits. At the same time, nonreciprocity is usually achieved by inserting bulky, lossy intervening elements – such as isolators and circulators – between qubit and qubit-interfaces. Consequently, despite all breathtaking progress in building scalable quantum processors, loss and noise still shorten qubit coherence times and constrain basic operations at the hardware level. Sustained progress in quantum technology hardware performance and scalability demands a device-level integration of both nonlinearity and nonreciprocity, endowed with quantum-level tunability. Addressing these challenges by culminating in a device-level quantum advantage is crucial for developing a sustainable and efficient quantum technological ecosystem.

This study seeks to address these central concepts and challenges in the quantum technology architecture by mapping the quantum landscape of SDs \cite{nadeem23,nadeem23-arXiv}. The SD is a new dipole circuit element that offers simultaneous quantum control over dissipationless nonlinearity and nonreciprocity at a device-level – capabilities that are fundamental to quantum hardware operation and highly sought after for scalable implementations. In a role analogous to that of semiconducting diodes in dissipative semiconductor electronics, an SD exhibits directional flow in non-dissipative superconducting electronics. At the same time, an SD promises dissipationless nonlinearity at the device-level, thereby helping to suppress noise and errors in qubit interfaces, which would otherwise travel backward to the qubit and disturb the quantum circuit under test. Together, this inherent nonlinearity and intrinsic nonreciprocity offer simultaneous control over information flow and suppression of noise in quantum electronics, opening a route toward scalable quantum technology architecture. Through these contributions, SDs show transformative potential to close the gap between quantum promise at the microscopic scale and quantum advantage at the macroscopic scale, bringing the quantum realm into everyday life.

\begin{figure*}
\includegraphics[scale=0.5]{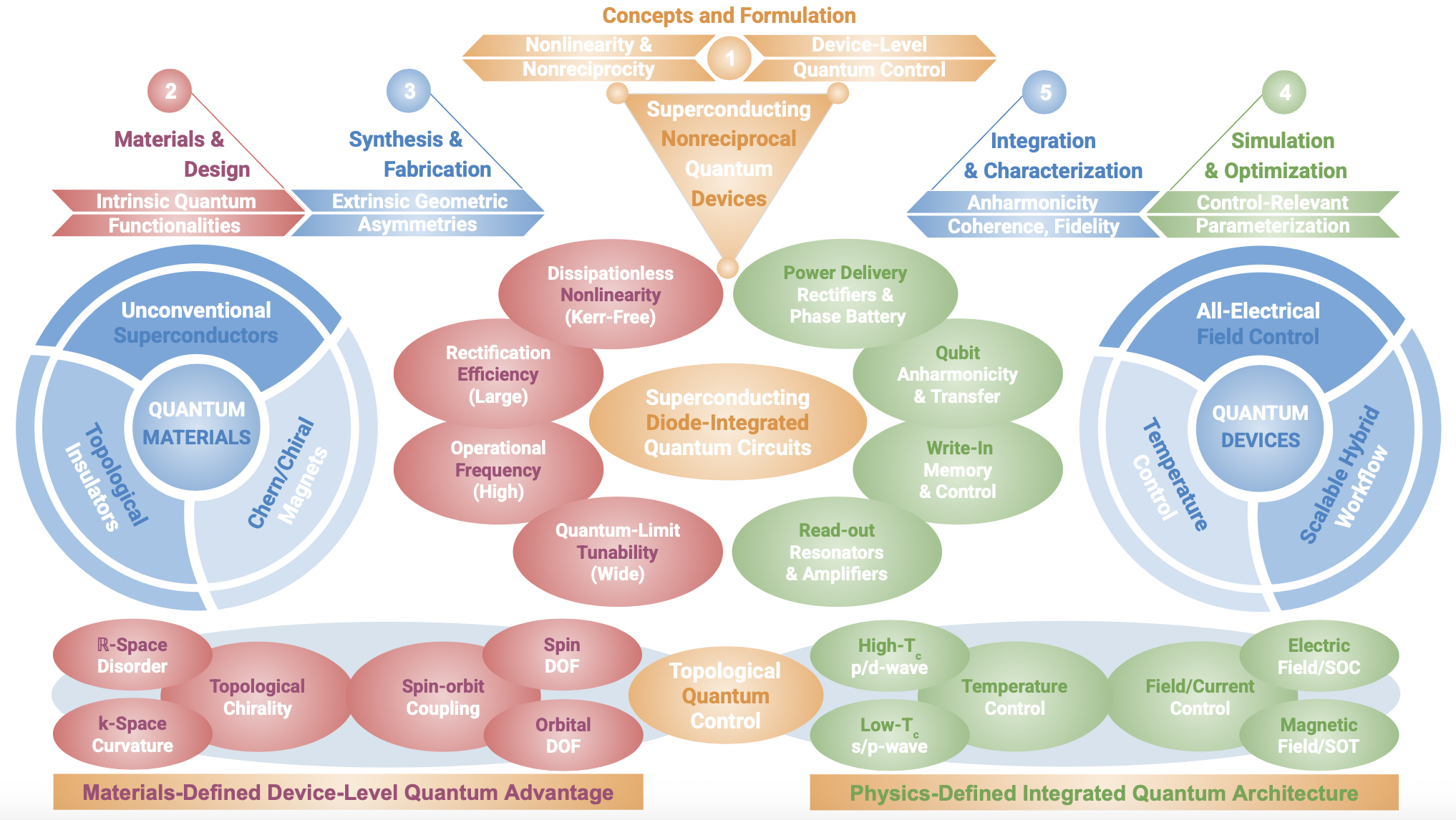}
\caption{\label{SDcQED} \textbf{Conceptual framework to achieve quantum-controlled superconducting diode platform and its system-level integration and performance characterization within c-QED.} The central objective of SD-integrated c-QED is to realize diode-induced inherent nonlinearity and intrinsic nonreciprocity directly at the device-level, avoiding the need for complex circuitry, which enriches the quantum technology hardware by providing a materials-defined device-level quantum advantage together with a physics-defined integrated quantum architecture. The fundamental functionalities of SDs, such as Kerr-free nonlinearity, rectification efficiency, and operational frequency, along with their quantum-level tunability, could become new degrees of freedom for SD-integrated c-QED for directional qubit transfer and entanglement generation, efficient power delivery and management, coherent control and memory, high-fidelity readout, and quantum-limited amplification (center panel). Quantum control across the parameter space can be achieved at different stages of material and device design and fabrication by combining intrinsic quantum functionalities with extrinsic geometric asymmetries (side and top panels).}
\end{figure*}

Following a concise discussion of the relevance and significance of nonreciprocal superconductivity and SDs, various superconducting diode-based devices are highlighted, which helps to understand how to quantify degrees of freedom and quantum-level tunability of parameter space in an SD-integrated c-QED platform. For example, rectification efficiency, operational frequency, and odd-order Kerr-free nonlinearity are the key attributes of SDs that control anharmonicity, fidelity, and coherence-relevant parameters, the core figures of merit in quantum circuits. In c-QED, as shown in Figure \ref{SDcQED}, these SD characteristics become new degrees of freedom in the quantum technology landscape, enabling on-chip integration of qubit and qubit-interfaces. The parametric framework for these prototypical examples illustrate how device-level quantum control over the inherent nonlinearity and intrinsic nonreciprocity of SDs can be harnessed to engineering SD-integrated c-QED environment that enables directional qubit transfer and entanglement generation, efficient power delivery and management via SD-based rectifiers and Josephson phase batteries, coherent control and memory operations using SD-based transistors and switching devices, high-fidelity qubit readout with nonlinear SD-based resonators, and quantum-limited amplification through nonreciprocal quantum material Josephson dipoles. In conclusion, future prospects on quantum material platforms and physics-defined device designs are presented, which could offer SD-integrated c-QED with all-electrical quantum-level tunability, achieving field-free and Kerr-free nonlinearity, which may provide a state-of-the-art all-superconducting platform for engineering on-chip quantum technology hardware with an isothermal classical and quantum hybrid workflow.

\section{Prologue: Semiconductors to superconductors}
The nonreciprocity of resistive charge transport in semiconductors allows the preferential flow of electrons in one direction over the other. This property of rectifying and controlling the flow of charge transport, known as the diode effect, is a key mechanism that underpins a wide spectrum of nonreciprocal devices, including LEDs and regulators, solar and photovoltaic cells, sensors and detectors, circulators and isolators, as well as amplifiers and transistors. The diode-based nonreciprocal devices are equally important for both high-end classical computation and quantum computing. Although semiconductor diodes and devices offer many advantages, they also face notable limitations and issues, particularly related to their dissipative transport mechanisms for asymmetric electron-hole currents, architectural design typically made of asymmetric junctions and interfaces, and operating temperature constraints. Moreover, electric field control of semiconducting diodes and devices, especially switching devices for control and memory applications, is also limited by fundamental physical laws, encapsulated by Boltzmann tyranny and thermodynamic constraints.

With these limitations, semiconducting diodes and devices are not favorable for energy-efficient control electronics and quantum technologies with ultra-low power consumption in circuits operating at ambient and cryogenic temperatures: 
At high temperatures relevant for thermionic transport, energy loss is inevitable in semiconductors because of their finite resistance. At very low (sub-Kelvin) temperatures, on the other hand, relevant for cryogenic electronics \cite{braginski19} and ultrasensitive (sub-THz frequencies) opto-electronics and detection \cite{farrah19}, semiconductors cease to work due to their large energy gap. Such temperature constraints do not favor semiconducting devices for quantum-limited operations. 
Apart from dissipation and temperature constraints on electronic transport, causing energy loss and hindering quantum-limited performance, spin-based semiconducting devices also suffer from detrimental behavior between SOI-determined control and coherence; large coherence time means less-effective control processing and vise-versa.

As a consequence, the critical challenge that poses a formidable hurdle for further advances in semiconducting technologies is our ability and readiness to handle mounting power dissipation associated with big data and its mechanization. According to the International Energy Agency (IEA) \cite{IEA24}, the total electricity consumption of data centers that process large amounts of worldwide information and in which most computation operations are concentrated has reached 460 terawatt-hours (TWh) in 2022 and could increase to more than 1,000 TWh in 2026. The surge in demand for computation and information processing is significantly outpacing the gains in power efficiency, presumed by Moore’s law, as the underlying performance of the semiconducting hardware is constrained by fundamental physical limits. Consequently, energy consumption in ICT is doubling roughly every three years and will be severely limited by global energy production within the next two decades unless new approaches are developed to store, process and compute at lower energy \cite{SRC21}. In this context, innovative ways to increase functionality and reduce power dissipation in high-end computing must be sought.

In comparison to semiconducting technologies, superconducting technologies have brought a new paradigm for energy-efficient processing of information, both with classical and quantum workflow. Due to zero-resistance supercurrent or Cooper-pair transport, low noise, high coherence, intrinsically low impedance and thereby very high supercurrent rectification, and the low characteristic energy scales associated with the superconducting gap ($\sim$meV) compared to the semiconductor energy gap ($\sim$eV), superconductors offer energy-efficient thermionic transport and cryogenic (opto)electronic devices \cite{tinkham04,braginski19}. Low-temperature superconductivity has shown improvements in cryogenic cooling efficiency and thus energy-efficient computation operations concentrated in data centers \cite{radebaugh09}. Aided by cryogenic cooling, its device applications extend to superconducting electronics \cite{tinkham04,braginski19,anders10,mukhanov19} and superconducting spintronics \cite{zutic04,linder15,mel22,cai23-AQT}, neuromorphic computing \cite{monroe24}, magnetometry and voltage standards \cite{braginski19,tafuri19}, astronomy \cite{farrah19} and dark matter \cite{dixit21} detectors, and quantum information and communication technology \cite{wendin17,liu19}, with terahertz (THz) applications \cite{yang23-NRm}. 

However, cryogenic technologies based on conventional or reciprocal superconductors also face several limitations pertaining to scalability, electrical control, and dissipation. Both high-end computation in data centers and quantum computing with superconducting quantum circuitry require the scalability and optimization of superconducting devices. 
For example, to scale superconducting energy-efficient rapid single flux quantum (RSFQ) circuits \cite{kirichenko11, mukhanov11} and to implement a quantum computing infrastructure with integrated control and measurement directly at the base temperature \cite{mcdermott18, mukhanov19}, one of the critical requirements is the mechanization for efficient dc biasing of RSFQ circuits. 
In addition, field-effect and gate tunability also remain a critical challenge in reciprocal superconductors \cite{de18-NN,golokolenov21}. That is, unlike bipolar electron-hole transport in gapped semiconductors, field control of supercurrent transport is a by far complex task in gapless superconductors due to macroscopic quantum coherence, zero electrical resistance, high density of states across the Fermi level, and unipolar nature of charge-carrying Cooper pairs. In this context, optimal diode effects for power management and regulation, along with efficient field control for switching devices, are highly required for practical superconducting technologies at cryogenic scales. 
Most importantly, even with zero resistance, reciprocal superconductors dissipate, and devices made of reciprocal superconducting materials become noisy, mainly due to switching effects and Kerr nonlinearity. Primarily, due to the difficulty in electric control in single-crystal superconductors, conventional superconductors and devices, made of superconducting-semiconducting hybrids, rely on a combination of dissipationless Cooper-pair supercurrent and dissipative quasi-particle current transport, respectively, across superconducting elements and weak links \cite{likharev79-RMP} made of normal materials. Although this hybrid transport mechanism enables various superconducting devices and functionalities based not only on current and inductance but also on voltage and resistance, it also causes dissipation and decoherence. Moreover, in Josephson junctions, which are the fundamental building blocks of superconducting technologies, like the \textit{pn} junctions in the semiconducting superconducting, dissipationless nonlinearity is only realizable through complex circuits made of multiple conventional Josephson junctions, a big hurdle for scalability in the existing superconducting quantum technology hardware. Furthermore, in conventional Josephson junctions, control processing relies on magnetic field or flux, which requires much higher energies than electrical processing.

Despite all of their advantages, compared to their semiconducting counterparts, superconducting technologies are held back because they lack a fundamental circuit element, the superconducting diode. The absence of non-reciprocity, and therefore of many non-reciprocal circuit elements that rely on the diode effect, is typically attributed to the preserved time-reversal, inversion, and thus intrinsic electron–hole symmetries that characterize a conventional BCS type-I superconducting state with a zero-momentum Cooper pairing. 

\section{Significance: Superconducting nonreciprocity}
Very recently, the SDE has reemerged as a promising mechanism which rectifies zero resistance supercurrent and allows quantum control over its tunability via various quantum mechanisms unique to unconventional quantum materials \cite{nadeem23,nadeem23-arXiv}. In the past five years, SDE have been observed in a variety of superconducting materials and structures \cite{nadeem23,nadeem23-arXiv,ma25-APR,shaffer25-arXiv}, each relying on distinct mechanisms to achieve rectification; intrinsic SDE based on asymmetry of the depairing current \cite{Ando20,Baumgartner22,Lin22, Daido22,Noah22,He22,Ilic22,Scammell22}, the vortex diode effect (VDE) based on asymmetric vortex dynamics  \cite{vodolazov05,vodolazov05-FM/SC,cerbu13, sivakov18,hou23,suri22,chahid23,ohkuma23}, and the Josephson diode effect (JDE) based on Josephson physics \cite{Baumgartner22,Davydova22,Tanaka22d-wave}. In addition, SDE can be based on spin-selective tunneling in the magnetic-superconductor tunnel junction, allowing spin filtering and spin splitting effects for superconducting spintronics \cite{Strambini22}.
The superconducting nonreciprocity offers scalable quantum circuits with quantum-enhanced functionalities and quantum-limit noise tunability: Unlike the semiconducting diode effect limited to \textit{pn} junction architecture, (i) SDE can be achieved in Josephson junctions (JJ) as well as in junction-free single crystal superconducting materials; (ii) SDE arises from the interplay between symmetry-driven transport and intrinsic quantum functionalities of the superconductors, and thus the performance of SDs is tuneable via both intrinsic quantum correlations and extrinsic stimuli; (iii) SDE is inherently defined by 3rd-order Kerr-free nonlinearity, originating from the non-sinusoidal higher-harmonic contributions in the CPR. Unlike semiconductors, which seize to work at cryogenic temperatures, and reciprocal superconductors, in which efficient gate control and scalability are hard to achieve, superconducting non-reciprocity and the diode effects offer (i) genuine quantum control to capitalize the quantum advantage at the device-level, (ii) efficient electrical control via symmetry breaking mechanisms, and (iii) cryogenic temperature control for on-chip integration of isothermal classical control and quantum processor.

For the practical applications of SDs in quantum technology hardware, dissipationless nonlinearity, rectification efficiency, and operational frequency are the three figures of merit. The rectification efficiency is defined as $\eta=(I_c^+-|I_c^-|)/(I_c^++|I_c^-|)$, where $I_c^+$ and $I_c^-$ are critical currents along positive and negative directions, respectively. When $I_c^+>I_c^-$, say, and the amplitude of the biased current is in the SD regime ($\eta\ne0$), that is, $I\in\Delta I_c=I_c^+-I_c^-$ or $I_c^-<I<I_c^+$, then the polarity of the biased current switches the SD between a superconducting state with a Cooper paired supercurrent along one direction and a conventional resistive state with a single-electron normal current along the other. In general, $\eta$ is sub-unitary due to a finite critical current along both directions, thus restricting the SD working regime to be small. Interestingly, the efficiency of the diode can be maximized to unity, reaching the limit of an ideal diode $\eta\rightarrow1$, 
in driven superconducting systems, including both bulk SDs \cite{daido25}
and Josephson dipoles \cite{souto22,valentini24,seoane24,wang25-NP,cuozzo24,shaffer25}, Josephson triode \cite{chiles23},
and supercurrent interferometers with multiple Josephson junctions \cite{souto22,bozkurt23}.
The realization of ideal SD has also been explored in bulk superconductors in thermodynamic equilibrium, such as criticality-based multiphase intrinsic superconductors \cite{hosur25-arXiv,chakraborty25,shaffer24,Noah22,banerjee24,Scammell22,Lin22} and proximity-induced superconducting metals \cite{hosur23,anh24}.

These advances indicate that SD efficiency can be significantly enhanced by employing innovative circuit designs, time-dependent biasing schemes, electric field modulation, and microwave irradiation. However, even perfectly designed ideal SDs can turn noisy in the c-QED environment because of dissipation in the resistive state through quasiparticles moving in the reverse direction. Here, we highlight how a complete quantum control, together with temperature control and electric-field control, can be realized in SDs based on unconventional superconducting structures, to enable an isothermal quantum-classical hybrid workflow.

\begin{figure*}
\includegraphics[scale=0.39]{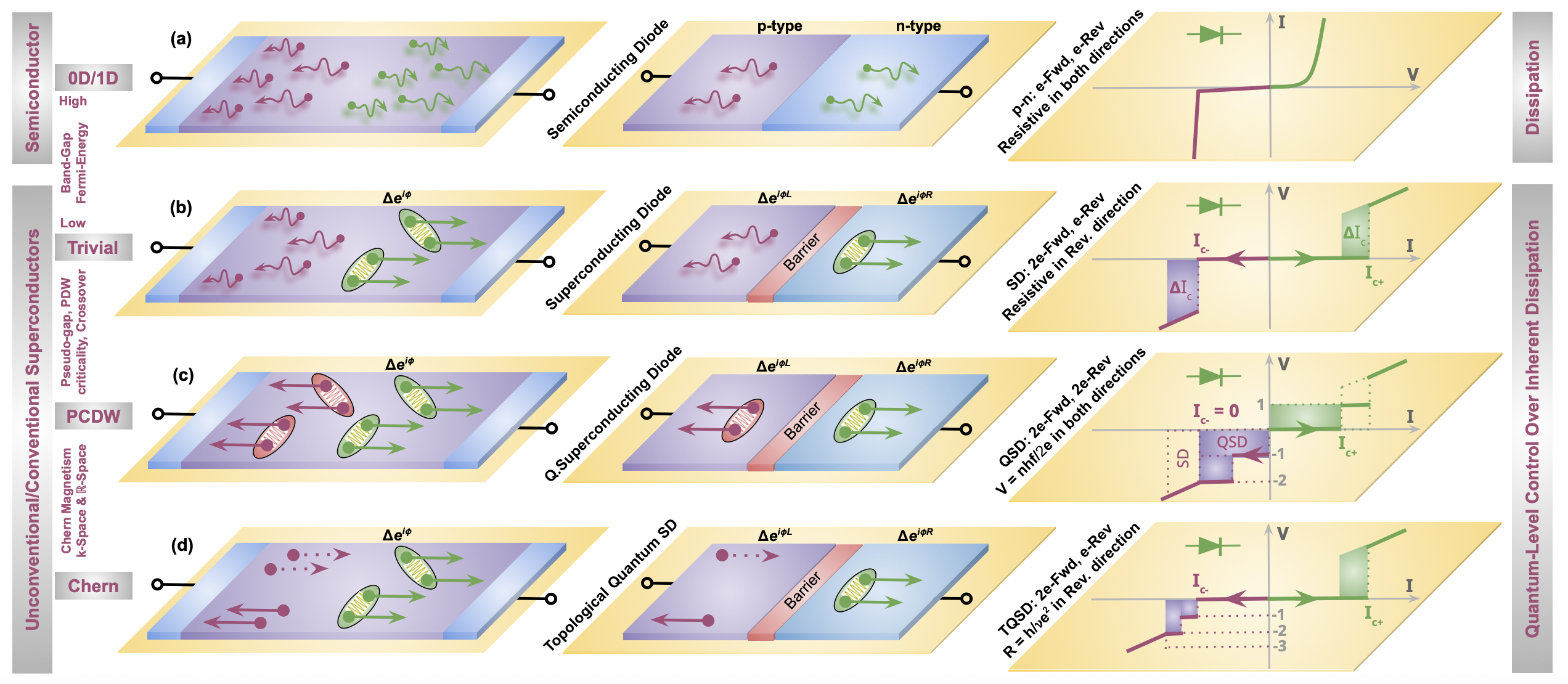}
\caption{\label{SDE-IV} \textbf{A schematic drawing of superconducting diode platforms.} \textbf{(a)} Semiconducting diode effect in bulk (left) and p-n junction (center), and the I–V characteristic (right), with a dissipative electron transport in both forward and reverse direction. \textbf{(b)} Superconducting diode effect (SDE) ($I_c^+ > I_c^-$) in junction-free bulk superconductors (left) and Josephson junctions (center), and I–V characteristic (right), with a dissipationless Cooper pair transport in the forward direction while a dissipative unpaired-electron transport in the reverse biased configuration. \textbf{(c)} Ideal quantum superconducting diode effect (QSDE) ($I_c^+ > 0, I_c^-=0$ and $\eta=100\%$) in junction-free bulk superconductors (left) and Josephson junctions (center), and I–V characteristic (right), with a dissipationless Cooper pair flow in both forward and reverse biased configuration \cite{wang25-NP}. Here plateaus in the I–V characteristic reflect the asymmetric Shapiro steps, with quantized potential values of $nhf/2e$, where $n=1,2,...$ is an integer, $e$ is electron charge, $h$ is Planck’s constant, and $f$ is the microwave frequency. In the QSD regime highlighted by purple and green, quantized voltage drops in the reversed bias only, even though Cooper pairs flow in both forward and reversed biased configurations. \textbf{(d)} Proposed topological quantum superconducting diode effect (TQSDE) ($I_c^+ > I_c^-$) in junction-free bulk superconductors (left) and Josephson junctions (center), and I–V characteristic (right), with a dissipationless Cooper pair transport in the forward direction while a dissipationless unpaired-electron chiral transport in the reverse biased configuration. Discrete steps in the I–V curve reflect the quantized Hall plateaus in the normal state. The wiggly lines represent dissipative transport while paired/unpaired arrows represent supercurrent/chiral transport in superconducting/topological phase.}
\end{figure*}

\subsection{Quantum control}
The advantages in quantum devices, surpassing the classical bounds on conventional devices, can originate either from purely quantum effects or from many-body collective effects. In the former case, the quantum advantage is referred to as the genuine quantum advantage (GQA), and it saturates the bounds imposed by underlying quantum mechanisms. While GQA guaranties miniaturization and scalability by exploiting device-level quantum effects, a quantum advantage due to collective effects in many-body systems could face challenges in achieving scalability due to many-body complex circuitry.

Superposition and entanglement are among the most celebrated quantum phenomena, recognized for providing an advantage in quantum technologies, where they coherently act together through parallel and collective processing. Alongside, quantum materials and structures that constitute the quantum technology hardware display nontrivial quantum mechanical effects on microscopic electronic states, giving rise to exotic properties at the macroscopic scale, an entanglement between bulk and boundary states generally known as bulk-boundary correspondence. Although the electron's wave function in conventional materials is also governed by quantum mechanics, it typically behaves in a trivial manner, allowing classical physics to describe most states and phases. In contrast, quantum materials exhibit quantum states and phases that are less classically describable, often distinguished by the existence of quantum phase transitions. The unique electronic, magnetic, and optical properties of quantum materials make them essential for the development of future quantum technologies, both with classical and quantum circuitry.

The quantum mechanical origin of SDE enable dissipationless quantum transport together with intrinsic quantum functionalities such as field-induced SOI and current-driven SOT, ferroelectricity and paraconductivity, chirality and chirality-induced spin-selectivity, orbitronics and altermagnetism, and finite Berry curvature and nontrivial topology, both in momentum space and in real space \cite{hess23,sinner24,sinner26}. For example, in non-centrosymmetric structures, SDE can arise from Rashba and Dresselhaus SOI \cite{Daido22,Noah22,He22,Ilic22, Baumgartner22,Baumgartner22-JPCM}, Radial and Chiral SOI \cite{kang24,costa25,costa25-jDE}, Ising and Kane-Mele SOI \cite{Wu22,Lorenz22,Pal22,dolcini15,wei23}, and a crossover between these various SOI types. When topology meets superconductivity, a topological SD (TSD) originates in topological superconducting hybrids \cite{Noah21,Legg22, Takasan22,karabassov22,Kenji19,masuko22,anh24,nagahama25} and topological JJs \cite{dolcini15,Chen18,Kopasov21,dartiailh21,Pal22,Tanaka22d-wave,legg23,lu23,chenPingbo23,cayao24,liu24,karabassov24,cuozzo24,mondal25-jde,wang25,nikodem25}.
SDE also originates from unconventional quantum effects such as chirality \cite{Hooper04,Kaneyasu10,Zinkl22, Qin17,Toshiya21,wu22-VD,he23,nunchot24,li25}, orbital Fulde-Ferrell pairing states and finite Cooper pair momentum \cite{xie23,nakamura24,zhao23,asaba24}, altermagnetism \cite{banerjee24-AM,chakraborty25,bhowmik25}, ferroelectricity \cite{Zhai22,yang25,wang25-wte2} and paraconductivity \cite{he23}. In topological quantum systems, the symmetry constraints governing SDs – simultaneous breaking of space-inversion and time-reversal symmetries – can be attributed to electronic chirality and/or magnetic chirality.

The quantum-mechanical advantage in SDs not only enables improved performance via distinctive quantum transport mechanisms but also offers noise-resilience and makes it possible to surpass the fundamental bounds in conventional systems. 
In the context of nonlinear and nonreciprocal quantum transport, topological quantum materials bridge the gap between normal electronic states (such as conventional semiconductors and ordinary metals) and superconducting states. For example, magnetochiral anisotropy (MCA) is one of the nonlinear and nonreciprocal quantum effects directly linked with SOI and Zeeman splitting. In trivial normal states, such as conventional semiconductors and metals, the MCA coefficient $\gamma_N$ is typically very tiny, usually of the order of $\sim10^{-3}$ to $10^{0}$ T$^{-1}$ A$^{-1}$ \cite{Rikken01,Rikken05,Pop14,Ideue17,rikken19}. This is because the SOI energy ($E_{soi}$) and the Zeeman energy ($\mu_B B$) in these trivial normal states is usually much smaller (by many orders of magnitude) than the semiconducting energy scale ($\sim eV$) or the Fermi energy $E_F$ that defines the kinetic energy of the electrons. On the other hand, quantum mechanical origin of MCA allows a highly controllable nonlinearity and nonreciprocity in quantum materials \cite{yasuda16,morimoto16,he18,Tokura18,yasuda20, isobe20,Legg22-nano,Zhaowei22,ye22,mei24,gao24,sakamoto24,li24-NM}, in which MCA coefficient $\gamma_Q$ ranges from $10^2$ T$^{-1}$ A$^{-1}$ in carbon nanotubes \cite{Krstic02} and topological semimetal ZrTe$_5$ \cite{wang22,wang24-PRL} to $10^5$ T$^{-1}$ A$^{-1}$ in (Bi$_{1-x}$Sb$_x$)$_2$Te$_3$ (BST) TI nanowires \cite{Legg22-nano}.
This indicates that the MCA arising from nontrivial topological quantum states can approach the magnitude of the MCA observed in superconductors, namely $\gamma_{S}\approx10^6$ T$^{-1}$ A$^{-1}$ \cite{Baumgartner22}. Overall, depending on how quantum effects manifest and what role they play in the system, MCA in quantum materials typically falls into an intermediate regime, situated between conventional semiconducting/metallic behavior and superconducting behavior, such that $\gamma_{N}\ll\gamma_{Q}<\gamma_{S}$. In other words, in contrast to semiconductors, which feature highly dissipative charge transport and therefore exhibit only weak MCA, and in a manner more analogous to superconductors, which enable dissipationless supercurrent and display large MCA, topological quantum materials allow dissipationless normal current transport with a remarkably high MCA \cite{Legg22-nano},
even without any superconducting order. 

Moreover, the dissipation induced by the quasiparticle current in the reverse direction can be eliminated, and the noise resilience of SDs can be enhanced through the quantization of transport. For example, a quantum superconducting diode (QSD) has recently been proposed in JDs \cite{valentini24,seoane24,wang25-NP}, in which the quantized Cooper pair current flows in both forward and reverse directions, and the diode efficiency can reach the maximum value of $\eta=100\%$ when microwave irradiation is applied together with a time-dependent current. A microwave controlled field-free ideal QSD with Josephson junctions made of high-$T_c$ twisted cuprates can function above the liquid-nitrogen temperature \cite{wang25-NP}.
As shown in Figure \ref{SDE-IV}(c), QSDE is characterized by the development of asymmetric Shapiro steps between the positive and negative current directions, with quantized potential values of $nhf/2e$, where $n=1,2,...$ is an integer, $e$ is the electron charge, $h$ is Planck’s constant, and $f$ is the microwave frequency. The nonreciprocity of the QSD is realized by switching the current between two distinct states, each formed by Cooper pairs: a superconducting regime in the forward biased configuration where the supercurrent flows without generating any voltage (V=0), and a dissipationless regime with quantized potential values of $V=nhf/2e$ ($V=hf/2e$ or $V=hf/e$ ) in the reverse biased configuration. Such a QSD exhibiting asymmetric Shapiro steps can be modeled using Josephson elements described by a damped Josephson potential, with their dynamics governed by the ac Josephson effect \cite{seoane24}. 
Compared to SDs and conventional Josephson junctions, the ideal QSD with quantized transport can be utilized for efficient microwave controlled logic operations and noise filtering, as demonstrated in QSD with Josephson junctions made of high-$T_c$ twisted cuprates \cite{wang25-NP}, and microwave resonators along with protected superconducting qubits, as established in QSD with Josephson junctions made of planar germanium \cite{valentini24}.

Quantization of transport, both in the forward and reverse directions, can further enhance the performance of QSDs made of topological quantum materials. Quantized topological transport, like quantized Cooper pair transport in the normal state \cite{valentini24,seoane24,wang25-NP}, can be the cornerstone for exploring a whole host of topological quantum materials where the diode effect is associated with a quantum phase transition between superconducting phase, allowing dissipationless supercurrent transport in one direction, and topological quantum phase, allowing dissipationless topological transport in the other direction. As shown in Figure \ref{SDE-IV}(d), we refer to this QSD as topological QSD (TQSD). Topological quantum materials not only provide an extra layer of protection against noise and decoherence but also allow tunability and optimization of SDE. For example, unlike conventional semiconductors, in which field effects are constrained by fundamental physical limits, topological quantum field effects in quantum materials allow the inclusion of quantum functionalities to overcome physical constraints \cite{nadeem21,fuhrer21,nadeem22,weber24}. In addition, unlike semiconductors where the energy gap ($\sim$eV) is large and SOI-determined control and coherence time are detrimental to each other, the low-energy scales associated with the superconducting gap ($\sim$meV) make SOI \cite{amundsen24-RMP,monroe24} and other quantum effects relevant. Thus, the zero-resistance and high coherence of spin-orbit coupled superconductors could be extended from charge currents to spin current, added by inherent nonlinearity and intrinsic nonreciprocity, to enable quantum-limited SOI-determined control without compromising coherence. To this end, graphene-based SDs, made of bulk superconducting structures \cite{Lin22,Scammell22,han25,chen25} and Josephson junctions \cite{Vries21,diez23,rothstein26, hu23,alvarado23, Wei22}, can offer efficient spintronic applications, where a long coherence time can be retained along with symmetry-controlled electric field switching.

\subsection{Temperature control}
Temperature constraints in the existing non-integrated architectures remain one of the most significant obstacles to scalability and control in current quantum technologies, where quantum circuits operate at cryogenic temperatures while the classical control and processing electronics reside at room temperature. The architectural separation between classical and quantum workflow introduces fundamental challenges in managing both temperature control and field control because thermal and electromagnetic noise generated by room-temperature semiconducting control electronics is inevitably transmitted to the cryogenic quantum circuits, thereby reducing coherence and limiting overall system performance. This is because the semiconductor control electronics, which is restricted by fundamental physical laws and thermodynamic constraints, produces heat dissipation and noise that exceed the quantum limit by several orders of magnitude. In addition, thermodynamic constraints are also translated into operational limitations on frequency and noise generation, even at the cryogenic temperature.

However, temperature control in SDs is more efficient, whereas thermodynamic constraints are more relaxed. For example, unlike semiconducting diodes and devices, which cease to work at cryogenic temperatures, SD efficiency typically increases as the temperature is lowered and approaches its maximum value near the absolute temperature ($T = 0$). Moreover, criticality-based thermodynamic constraints limit the efficiency of bulk SDs ($\eta\ll1$), and the ideal diode behavior ($\eta=1$) can only be achieved by fine tuning to a quantum critical point \cite{hosur25-arXiv}.
In contrast, driven nonreciprocal superconducting systems can attain ideal SD behavior even at finite temperatures.
To attain complete quantum control over the optimization and tunability of SDs, one way is to devise criticality-based quantum phase transition (QPT) models that incorporate microscopic scales characterizing the QPT at finite temperature. Critical scaling, which extends beyond the standard extraction of critical exponents, can translate the underlying principles of equilibrium criticality to non-equilibrium or dynamical quantum phase transitions \cite{heyl18-PR}.

Depending on the operating temperature, SDE can be probed in two different ways: 1) SDE arising from the nonreciprocity of the depairing current close to the critical temperature ($T \approx T_c$), which appears in the fluctuation regime of the metal–superconductor resistive transition and can be described via current–voltage characteristics and resistive measurements \cite{Ando20,Baumgartner22}; and 2) SDE originating from the nonreciprocity of the supercurrent at temperatures $T \ll T_c$, which manifests deep within the superconducting phase and can be modeled through the current–phase relation and inductive measurements \cite{Baumgartner22}. In addition to current, resistance, and inductance, nonreciprocity can also be characterized by paraconductivity in the normal state close to the critical temperature ($T \rightarrow T_c$) \cite{he23}. Moreover, resistive measurements can be used to determine the SDE by probing asymmetric vortex dynamics \cite{hou23} at temperatures significantly below the critical temperature that defines the depairing current.

Furthermore, temperature can be utilized as a control parameter to optimize the SDE. For instance, unlike conventional VDE where asymmetric vortex dynamics arises from an out-of-plane magnetic field, nonreciprocal behavior with an in-plane magnetic field dominates the nonreciprocity arising from the out-of-plane magnetic field in topological/chiral kagome superconductor CsV$_3$Sb$_5$ \cite{wu22-VD} \cite{wu22-VD}. Similarly, SDE in Rashba superconductors with an in-plane magnetic field, such as Bi/Ni bilayer \cite{cai23}, magnifies in the temperature regime associated with vortex dynamics, even though SDE vanishes under a perpendicular magnetic field. 
Interestingly, the base temperature of SD-integrated superconducting quantum circuits can be controlled through the selection of candidate materials, ranging from conventional low-$T_c$ superconductors to unconventional high-$T_c$ superconductors \cite{lee21,zhao23-Sci,ghosh24, oh24,qi25,wang25-NP}.

\subsection{All-electrical control}
All-electrical control, whether via electric fields or electric currents, is essential for achieving energy-efficient switching in devices used for information processing, data storage, and signal amplification, such as transistors, memory elements, and amplifiers. In nonreciprocal superconductors, by contrast to conventional semiconductors in which field control is largely ineffective at cryogenic scales, field control is intrinsically linked to symmetry-determined transport, which can be readily switched by reversing the gate polarity or the polarization of electric or magnetic fields. In addition, switching mechanisms can also be implemented by manipulating intrinsic magnetization and ferroelectricity, along with current-driven SOT. Moreover, unlike reciprocal superconductors in which electric-field control is difficult to realize and magnetic-field control is inefficient, nonreciprocity in superconductors effectively control the switching via symmetry-defined transport direction, instead of manipulating the unipolar electron density of states across the Fermi level.

The switching of superconducting nonreciprocity, by reversing the externally applied magnetic field or the intrinsic magnetization, applies to both intrinsic SDs, in which the MCA switches its sign, and extrinsic SDs, in which the Meissner screening current (MSC) switches its direction. In addition, it is well understood that the SDs exhibit non-monotonic behavior under a magnetic field or magnetization; the SD efficiency switches sign with increasing the strength of magnetic field or magnetization. However, magnetic field bias, and stray magnetic fields, consume a large amount of energy, which presents several challenges for superconducting qubits in the spin environment \cite{gunzler25} and magnetic/thermal burn-in and magnetic tolerance of switching devices \cite{khan18}. In this context, a magnetic field-free and field-resilient SD is highly desired that can tolerate stray fields in cryogenic superconducting circuits \cite{yang25}.

Interestingly, these losses and challenges can be evaded via all-electrical control – all without relying on magnetic fields – for switching mechanisms in SDs and corresponding superconducting control electronics. Several distinct mechanisms for all-electrical switching have been demonstrated in SDs, including gate polarity and electric field induced SOI, electric field controlled ferroelectric polarization, and current-driven SOT. For example, gate-tunable SDE has been reported in graphene-based SDs, such as gate-defined JDE realized in magic-angle twisted bilayer graphene (MATBLG) \cite{Vries21,diez23,rothstein26, hu23,alvarado23}, ge-based josephson devices with gate-controlled tunable harmonic content \cite{leblanc24}, Rashba semiconductor-based JJ (Al/InAs/Al) with gate-controlled barrier transparency \cite{mayer20gate},
topological insulator (BiSbTeSe$_2$) based JJ \cite{kayyalha20gate}, superconductor/ferromagnet multilayers with spin-SDE \cite{sun25-spin}, Dayem bridge-based SQUID \cite{paolucci23} and multi-terminal Josephson devices \cite{gupta22}. Similarly, SDE switching has also been presented by electric-field-induced control of ferroelectricity \cite{Zhai22,yang25,wang25-wte2} and current-driven SOT and the warping effect \cite{xiong24,Lu21} and dynamics of magnetic domains and skyrmions along the racetrack \cite{hess23}.

These mechanisms can be efficiently implemented in SDs exhibiting structural inversion asymmetry, where SOI and other quantum effects become important because of the low energy scales set by the superconducting gap ($\sim$meV). They are also applicable in SDs that possess electronic or magnetic chirality arising from nontrivial topological properties. Among these, all-electric field control of SDs promises efficient switching devices for superconducting and hybrid quantum technologies, since electric fields typically require much lower energies than magnetic-field- or current-based control schemes. Thus, quantum-enhanced, symmetry-driven all-electrical switching of superconducting traits motivates the incorporation of SDs into cryogenic control hardware. An efficient gate‑control responsiveness and nearly ideal SD performance at cryogenic temperatures (10mK - 4K), a regime where semiconducting control electronics cease to work, makes SD-integrated c-QED ideal for on-chip integration of all-superconducting quantum processors, with coherent classical and quantum control.

\begin{figure*}
\includegraphics[scale=0.47]{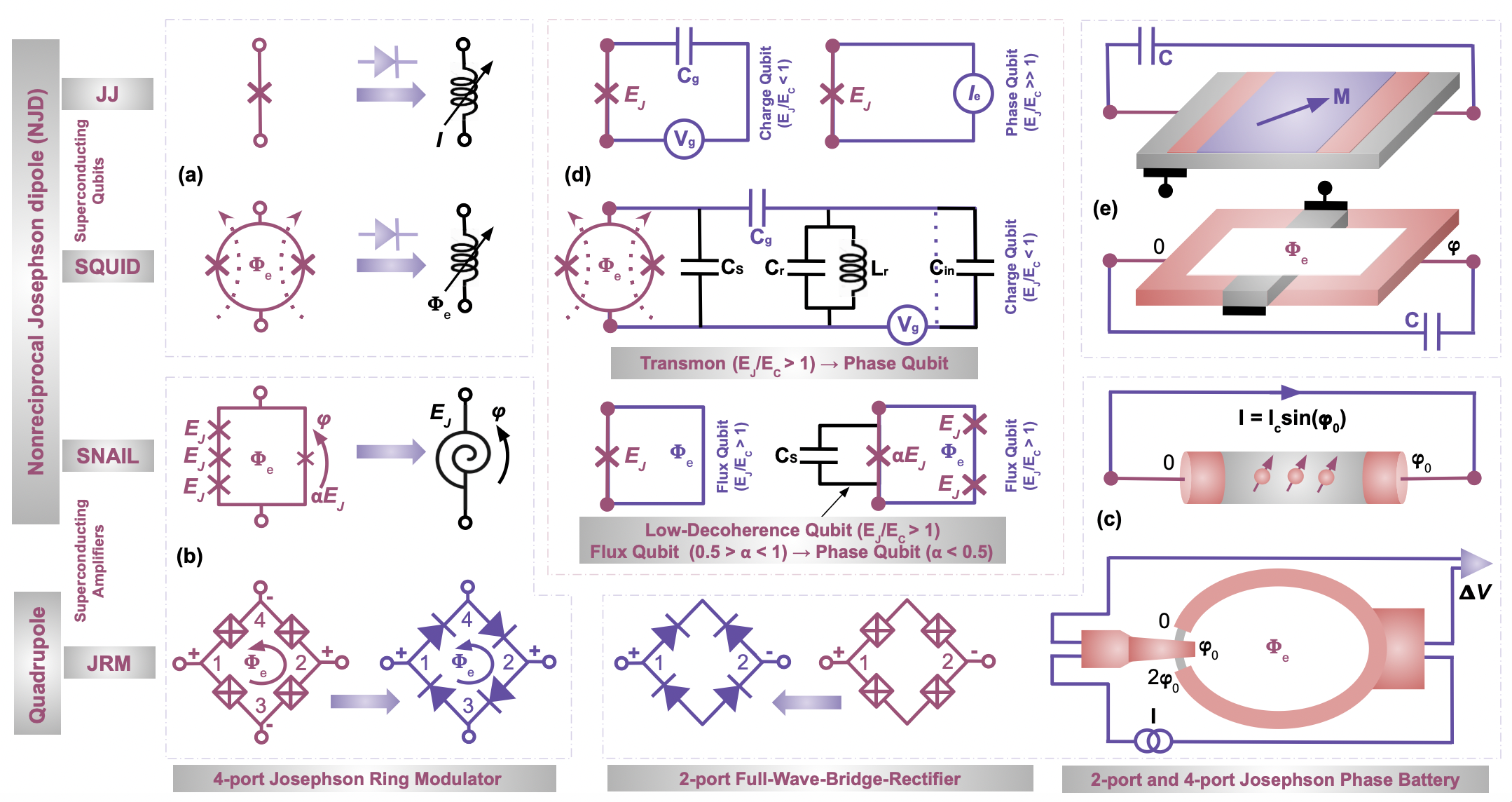}
\caption{\label{SDEjde} \textbf{Nonreciprocal Josephson dipole elements.}
\textbf{(a)} Nonreciprocal Josephson dipole (NJD), such as Josephson junction, SQUID, and SNAIL acting as current-controlled nonlinear inductor, flux-controlled nonlinear inductor, and phase-flux controlled Kerr-free nonlinear dipole, respectively. \textbf{b} Nonreciprocal Josephson quadrupole (NJQ), such as 4-port Josephson ring modulator (JRM), acting as flux-controlled Kerr-free nonlinear quadrupole.
\textbf{(c)} Josephson phase battery with Josephson junction and SQUID (right) and a schematics of 2-port full-wave bridge rectifier (left), in comparison to 4-port JRM. \textbf{(d)} Three basic types of superconducting qubit circuits (in purple color) – charge qubit, phase qubit, and flux qubit – and improved version of these different types of superconducting qubits with additional components (in black color) – transmon and low-decoherence qubit. \textbf{e} Circuit models of gate and flux-controlled NJDs shunted to capacitor, a Josephson junction made of ferromagnetic-superconductor hybrid (top) and SQUID (bottom). In all these circuits, key for the qubit and qubit interfaces, conventional Josephson junctions are replace with NJDs or Josephson diodes (JDs) to engineer SD-integrated circuit quantum electrodynamic (c-QED) environment for quantum technology hardware.}
\end{figure*}

\subsection{Isothermal hybrid workflow}
Quantum computing technologies have come to the forefront as a more powerful alternative to classical semiconducting technologies. However, even with quantum computing systems at hand, ultra-fast and energy-efficient classical computing devices are required to control the processing of quantum computing. Nevertheless, the absence of a cryogenic diode in current quantum hardware architectures poses fundamental challenges to the control and scalability of quantum-classical hybrid workflows. On the one hand, temperature constraints necessitate a non-integrated system architecture, where quantum computing at the cryogenic temperature is controlled by classical processing at the ambient temperature, while on the other hand, inefficiency of nonlinear and nonreciprocal devices causes a major challenge for scalability of a quantum processor at the cryogenic temperature.
 
In both of these classical and quantum workflows operating at different temperatures, heat dissipation is a common challenge. In the former case, heat loss is associated with dissipative transport in semiconductors, which is constrained by fundamental physical principles. In the latter case, nonlinearity induced by Kerr-effect and the nonreciprocity delivered by bulky intervening elements generate dissipation and noise. In addition, operation of quantum computer circuitry at temperatures close to absolute zero is vulnerable to the effects of unwanted thermal and electromagnetic noise, interference, or decoherence of the quantum state, mostly coming from the wires connecting low-temperature quantum circuitry to ambient-temperature electronics circuitry. 

SD has recently emerged as a promising candidate for implementing both classical and quantum workflows.  
That is, along with directional quantum state transfer and entanglement generation with SD-integrated qubit-resonator setup, SDs also promise superconducting control electronic for classical processing, rectification, and amplification. The inherent nonlinearity and intrinsic nonreciprocity of SDs provide an ideal c-QED platform to design isothermal hybrid workflow to implement integrated qubit control and readout while minimizing backaction on the qubit, thermal load, electromagnetic noise, and other external perturbations. Thus, the SD-integrated c-QED environment offers the possibility of on-chip integration of qubit and qubit-interfaces operating at the same temperature scale, by means of a fully integrated all-superconducting hybrid circuitry in which both classical and quantum workflows can be executed concurrently to fully capitalize the quantum advantage. 

\section{Superconducting diode-integrated c-QED}
SD is a superconducting counterpart of a semiconducting diode, rectifies the zero-resistance supercurrent, and thus represents an ideal building block that allows non-dissipative superconducting technologies in the way the traditional diode is for dissipative semiconducting technologies. With inherent nonlinearity and intrinsic nonreciprocity, integrated with unique superconducting characteristics such as zero-resistance and high coherence, SDs hold transformative potential for energy-efficient and high-performance computing architectures with classical, quantum, and quantum-classical hybrid workflow. Current research efforts are broadening the scope of SD-based applications for quantum circuitry, including spin-related logic and memory applications \cite{Golod22,he23-ADI,wu26-arXiv}, spin diodes \cite{Strambini22,sun25-spin} and heat engine \cite{araujo24}, Josephson devices with tunable harmonic content \cite{leblanc24} and phase battery \cite{strambini20}, superconducting antenna \cite{Zhang20}, sensors \cite{sinner24} and detectors \cite{geng23-Rev}, superconducting chiral/neuronal transistors \cite{hess23,xiong24}, superconducting full-wave bridge rectifiers for quantum circuitry \cite{castellani25,ingla25}, diode-controlled transmon qubit chain circuit \cite{zhong25,dirnegger25}, and directional logic operations \cite{xiong24,hu25-adma,dirnegger25}. Collectively, these studies indicate that establishing a quantum-enhanced parametric framework for SDs can significantly advance the performance of qubits and qubit interfaces for on-chip quantum processors, along with isothermal classical control processing.

Here, we revisit the progress of various SD-integrated quantum circuits, discuss their potential advantages over conventional superconducting counterparts, and highlight the potential ramifications that SDs and SD-based nonreciprocal circuits can bring to quantum circuits to revolutionize superconducting electronics and spintronics for integrated quantum technology hardware. To elucidate and assess the influence of SDs within SD-based quantum circuits, both qualitatively and quantitatively, we systematically examine the physical characteristics, material specifications, and quantum functionalities of SDs themselves and SD-integrated quantum circuit architectures. Given that Josephson junctions and their associated Josephson circuits constitute key building blocks of qubits and qubit interfaces, as illustrated in Figure \ref{SDEjde}, we place particular emphasis on the analysis of nonlinearity and nonreciprocity in JDEs.   

\subsection{Qubit: anharmonicity and entanglement}
In quantum processors, the qubit circuitry is interfaced with the power delivery system, control processor, memory elements, and readout resonators. In such a c-QED environment, qubit anharmonicity, one-way transfer of qubits, directional qubit-qubit coupling, and nonreciprocal entanglement generation are fundamental ingredients to the recipe of coherent qubit circuit operation, shielding the qubit from noise and operational errors, thereby enabling high-fidelity state readout. Quite recently, it has been demonstrated that the incorporation of SDs into quantum circuits can facilitate tunable anharmonicity, high-fidelity qubit transfer, and directional entanglement generation \cite{zhong25, dirnegger25}. Here, we discuss the progress and potential ramifications that SDs can bring to superconducting qubits and superconducting quantum circuits embedded with these device-level nonreciprocal elements.\\

\subsubsection{Anharmonicity and qubit transfer}
\begin{figure*}
\includegraphics[scale=0.45]{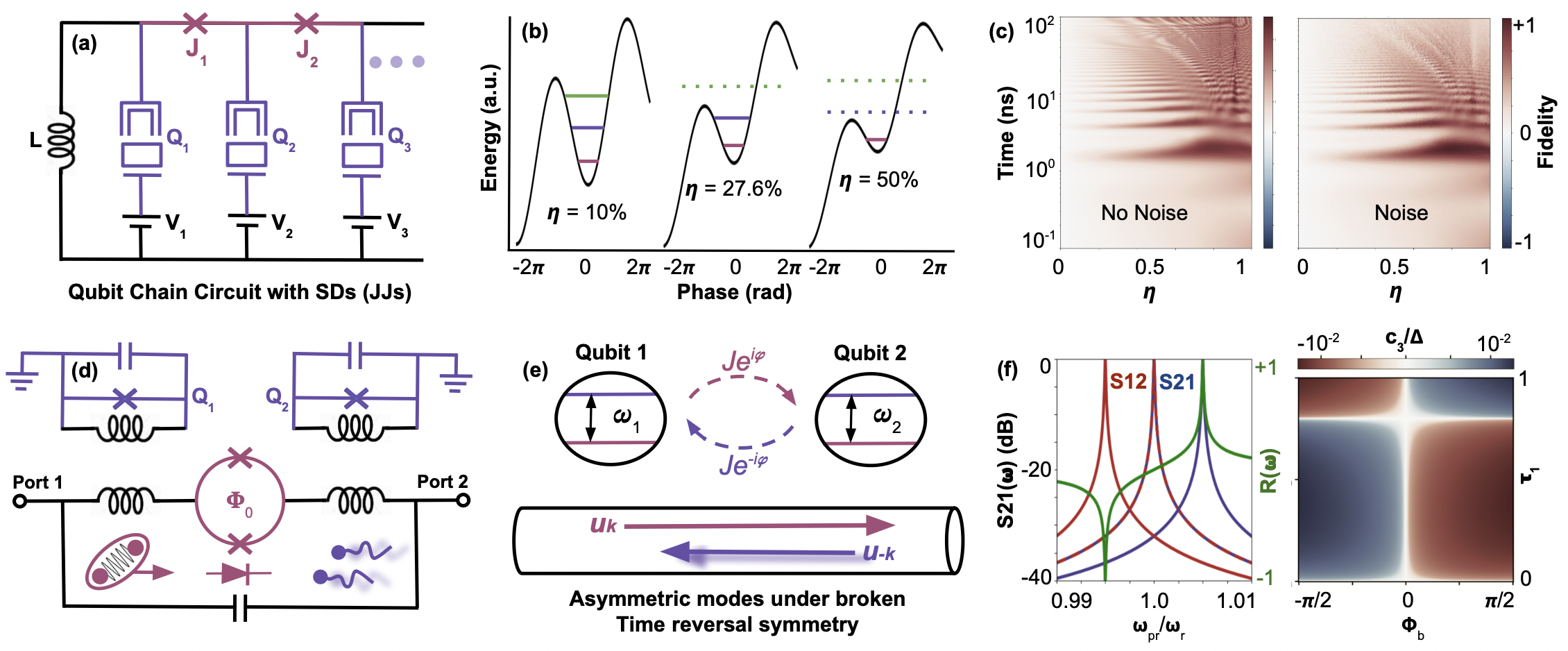}
\caption{\label{SDQ} \textbf{Superconducting diode-controlled nonreciprocal qubit transfer (a-c) and qubit-qubit coupling (d-f).} \textbf{(a)} Transmon qubit chain circuit with SDs made of twisted NbSe$_2$ bilayers in the form of Josephson junctions. \textbf{(b)} A characteristic asymmetric multi-well configuration revealed by the potential-energy landscape that arises from an interplay between the Josephson coupling energy and the phase bias introduced by the SDE. Corresponding quantized energy eigenstates, represented by the discrete horizontal lines, are overlaid on the potential energy profiles for $\eta=10\%$ allowing three bound states in the central potential well (left), $\eta=27.6\%$ maintaining exactly two bound states in the central potential well (center), and $\eta=50\%$ supporting only one bound state in the central potential well (right). This evolution demonstrates that an effectively ideal two-level system can be achieved by systematically modifying the SDE efficiency, such that the two lowest-energy bound states that constitute the qubit levels (highlighted in red and blue) stay spatially confined within the central potential valley and remain energetically separated from higher-lying continuum states by the surrounding potential barriers. \textbf{(c)} Fidelity difference between forward and reverse qubit transfer as a function of diode efficiency and evolution time, shown in the absence of noise (top panel) and in the presence of noise (bottom panel). Figure panels (a-c) are reproduced with permission from ref.\cite{zhong25}.
\textbf{(d)} Qubit-resonator circuit showing two qubits inductively coupled to a two-port SD-resonator where flux-biased asymmetric SQUID exhibits SDE, allowing Cooper pair supercurrent along the direction of forward bias while quasiparticle current along the direction of reverse bias.
\textbf{(e)} Schematic of nonreciprocal qubit-qubit coupling driven by SD-resonator. The broken time-reversal symmetry, by external flux bias ($\Phi_0$), induces SDE in SQUID, generates asymmetric modes in the SD-resonator, and makes qubit-qubit coupling complex and nonreciprocal, $J_{ij}=|J|e^{i\phi_{ij}}$, where non-local phase $\phi_{12} = -\phi_{21} = \varphi$ is determined by the nonreciprocal response of the SD-resonator, which is induced by asymmetric mode propagation.
\textbf{f} Spectroscopic characterization of SD on the two-port network. (Left) Transmission spectrum showing scattering matrix $S(\omega)$ with respect to probe frequency $\omega_{pr}/\omega_r$, normalized by resonator frequency ($\omega_r = 5$ GHz); degenerate forward $S_{21}(\omega)$ (blue) and backward $S_{12}(\omega)$ (red) transmission for $\delta\omega=0$ and $S_{21}(\omega)$ and $S_{12}(\omega)$ with asymmetric shift for $\delta\omega=50$MHz. Nonreciprocity ratio $R(\omega)$ for $\delta\omega=50$MHz is shown in green. 
(Right) Third-order Josephson nonlinearity ($c_3/\Delta$ as a function of flux $\Phi_b$ and asymmetric SQUID transmission, $\tau_1$ ($\tau_2=0.8$).
Figure panels (d-f) are reproduced with permission from ref.\cite{dirnegger25}}
\end{figure*}

The quantum diode is a key element for controlling a one-way transfer of qubits. In a quantum circuit, the one-way excitation transfer process is essential for both qubit control and high-fidelity. In general, similar to the case in rectifiers and power management systems, the quality of the quantum diode, which is characterized by its fidelity, can be improved by increasing the rectification efficiency \cite{zhao23-npj}. However, by simulating the transmon qubit chain circuit with SDs, as shown in Figure \ref{SDQ} (a-c), Zhong et al. \cite{zhong25} demonstrated that even an intermediate SD efficiency could allow optimal performance by influencing much broader functionalities through an interplay of intrinsic quantum phenomena. It unlocks the potential of SDs, which simultaneously allow qubit anharmonicity \textemdash by preserving nonlinearity, which is key to stabilizing two-level systems, and qubit control \textemdash by suppressing backward noise flow, which is key to high-fidelity. In fact, the efficiency of SDE $\eta$ acts as a new degree of freedom to optimize the anharmonicity of the superconducting qubit and the fidelity of quantum circuits.\\

The realization of qubit anharmonicity and high-fidelity quantum circuits is achieved by stabilizing two-level systems using van der Waals (vdW) twisted SDs. The SDE is observed in twisted NbSe$_2$ bilayers, where a small twist of $\theta=1^{\circ}$ induces an efficiency $\sim27\%$, nearly an order of magnitude higher than in the pristine case \cite{Zhang24-NbSe}. However, compared to other experimental work reported on 2D vdW superconducting materials \cite{Lin22,zhao23-Sci,Wu22,lee21,chenPingbo23, chenPingbo24,Lyu21}, the rectification efficiency and the average critical current in twisted NbSe$_2$ bilayers takes an intermediate values. In addition, unlike the pristine case in which the Ising spin-orbit coupled pairing is favorable \cite{Wickramaratne20}, the dependence of SDE efficiency on both out-of-plane and in-plane magnetic fields indicates an unconventional Ising pairing in the twisted NbSe$_2$ bilayers. This could be attributed to the coexistence of Ising- and Rashba-type SOI, which is derived from a twist-induced orbital magnetization that spontaneously breaks both time-reversal and inversion symmetries in twisted NbSe$_2$ bilayers.

A one-way transfer of qubits is demonstrated through quantum simulations performed in the gate-based simulation framework of QuTip \cite{johansson12}. As shown in Figure \ref{SDQ} (a), a transmon qubit chain circuit interconnects transmon qubits (Q$_1$,Q$_2$,Q$_3$) with Josephson junctions based on twisted NbSe$_2$ bilayers (J$_1$, J$_2$), acting as Josephson diodes. The qubit anharmonicity, under varying SDE efficiency, is numerically simulated through the potential-energy landscape of the transmon qubit as a function of the phase difference introduced by the SDE, Figure \ref{SDQ} (b). The two quantum states \textemdash the ground state “0” and the first excited state “1”, energetically separated from the higher-excited continuum states \textemdash can be stabilized in the tilted cosine-like potential through precise tuning of $\eta$. In this case, in order to maintain consistency between the numerical simulations of the qubit chain circuit and the experimental observations of SDE in twisted NbSe$_2$ bilayers, strong qubit anharmonicity is maintained in the optimal parameter regime, $\eta\approx27.6\%$ and $E_J/E_C=20.0$, where exactly two-level qubit states are bound in the central potential well.

The transport fidelity, across the parameter space of $\eta$ and the evolution time, quantifies the performance of the SDE and the overall transmon qubit chain circuit, as shown in Figure \ref{SDQ} (c). In this framework, the forward (reverse) fidelity provides a quantitative measure of how accurately an excitation is transferred in the forward (reverse) direction. In ideal noiseless evolution, the SDE exhibits ideal behavior \textemdash maximum forward fidelity $F_f\rightarrow1$ and minimum reverse fidelity $F_r\rightarrow0$. The difference between forward and reverse fidelity also remains visible in noisy simulations that include realistic noise sources including parameter fluctuations of $E_J$, charge noise representing Cooper pair number fluctuations, phase noise from flux variations, energy dissipation modeling environmental coupling, and measurement uncertainty. Similar robustness persists for the overall fidelity difference between forward and reverse qubit transfer, validating the potential of SDs for practical quantum circuit applications in a typical noisy experimental environment.

\subsubsection{Qubit-Qubit coupling and entanglement}
N. Dirnegger et al, \cite{dirnegger25} used a flux-controlled superconducting diode, formed by an asymmetric SQUID, with a minimal two-qubit system and demonstrated coherent nonreciprocal qubit-qubit inductive coupling and a tunable Bell-state generation through a nonreciprocal half-iSWAP gate. As shown in Figure \ref{SDQ} (d-f), a two-qubit (Q1,Q2) circuit is inductively coupled to the SQUID-based SD-resonator, which is biased by external flux $\Phi_0$. The nonlinear inductive response, driven by flux bias in the cQED setup, enables flux-controlled nonreciprocal resonance shifts in the transmission spectrum of the superconducting diode resonator. In turn, the coupling between the two qubits captures the nonreciprocity carried by the asymmetric modes in the SD transmission spectrum, which makes the coupling between qubits complex, $J_{12}= J\exp^{i\phi_{12}}$, where J is the coupling strength and $\phi_{12}=\varphi$ is a nonlocal phase. In principle, the  approach remains valid for realizing nonreciprocal coupling between nearest neighbor qubits within a n-qubit system where qubits Q1,Q2,....Qn are inductively coupled with different ports of the same SD resonator \cite{kafri17}. 

The flux-controlled non-local phase $\phi_{12}=\varphi$, which is determined by the nonreciprocal response of SD, turns out to be an SD-driven degree of freedom to passively govern the nonreciprocal coupling between qubits. The microscopic origin and the role of the SD-resonator in nonreciprocal qubit-qubit coupling can be characterized by extracting the non-local phase in terms of reciprocal ($J_{12}^{rec}$) and nonreciprocal ($J_{12}^{nrec}$) counterparts of the coupling (where $J_{12}=J_{12}^{rec}+iJ_{12}^{nrec}$), i.e., $\phi_{12}=\tan^{-1}(J_{12}^{nrec}/J_{12}^{rec})$. The sign reversal of $J_{12}^{nrec}$ under an external bias flux ($\Phi_b$) makes the qubit-qubit coupling nonreciprocal. The nonreciprocity encoded in $J_{12}$ and $\phi_{12}$ reflects the same fundamental principles that govern nonreciprocal coupling in any quantum system with broken time-reversal \cite{labarca24,ren25}. However, in the qubit circuit coupled to an SD-resonator, the flux-dependent nonlinear inductive response and nonreciprocal resonance shifts observed in the SD-resonator’s transmission spectrum make the nonreciprocity of qubit-qubit coupling explicitly SD-determined, with the direction of excitation transfer governed by the flux-controlled sign of $\phi_{12}$.

In the SD-resonator, with flux-controlled resonance shifts, nonreciprocity of SD arises from a third-order Josephson nonlinearity.
This third-order nonlinearity is also a source of three-wave mixers, including the quadrupole Josephson ring modulator (JRM) \cite{bergeal10,bergeal10-NP,abdo13} and dipole Josephson elements such as Superconducting Nonlinear Asymmetric Inductive eLement (SNAIL) \cite{frattini17,frattini18,sivak19,miano22}, SQUID \cite{castellanos08,schrade24}, and a single Josephson junction with a magnetic weak link \cite{schrade24} or a one-junction SQUID \cite{zorin16}.
The realization of minimal third-order nonlinearity allows maximal coupling between frequency crowding modes while minimizing unwanted Kerr frequency shifts. Within a circuit QED architecture, a minimal single-junction dipole Josephson element \cite{schrade24} enables Kerr-free three-wave mixing parametric amplification while simultaneously offering enhanced power-handling capability and efficient three-wave coupling between qubits and qubit-interfaces, thus supporting higher information throughput. It shows that a Josephson diode made of quantum materials, with an intrinsic capability to provide both nonreciprocity and Kerr-free nonlinearity in a single-junction dipole device, opens a new paradigm for the realization of scalable quantum technology hardware.

\subsection{Power management: rectifiers and battery}
Rectifiers and quantum phase batteries, with persistent voltage bias and current-rectification and persistent phase bias and phase-tunability, respectively, are critical for power delivery and management within quantum circuits in an integrated environment. As with any other qubit-interface, nonreciprocity and nonlinearity also play a central role in preventing dissipation and noise flow into the qubit system. To this end, superconducting nonreciprocity offers new platforms for scalable and integrated c-QED with a superconducting diode rectifier (SD-rectifier) and a superconducting diode quantum phase battery (SD-QPB).

\subsubsection{Superconducting diode rectifier}
The superconducting diode-based full-wave bridge rectifiers have been reported recently by two independent groups of researchers \cite{castellani25,ingla25}, which advances the utilization of SDE for power management systems in scalable circuits for quantum technologies \cite{kochan25}. In both of these studies, rectifiers are based on extrinsic SDE caused by asymmetric vortex dynamics, though the underlying mechanisms that lead to the asymmetric dynamics could be different. A superconducting thin film with edge asymmetry acts as a vortex diode as a result of the asymmetric vortex surface barrier under an out-of-plane magnetic field \cite{vodolazov05,cerbu13,sivakov18,hou23,suri22,chahid23, ohkuma23}. This type of magnetic field-driven vortex diode is referred to as type-A SDE. A thin-film bilayer of a superconductor and a ferromagnetic insulator also acts as a vortex diode due to the remanent magnetic field emanating from the ferromagnetic insulator \cite{vodolazov05-FM/SC,hou23}. This type of magnetization-driven vortex diode is called type-C SDE. In this sense, type-C SDE could also be termed zero-field or field-free SDE.\\  

Matteo Castellani et al. \cite{castellani25} fabricated a superconducting full-wave bridge rectifier loop by integrating four niobium nitride micro-bridges on a single microchip, by creating triangular notches on a 13-nm thick film of niobium nitride on a SiO$_2$ substrate. Each individual micro-bridge, with a triangular notche cut into one of its edges, acts as a magnetic field-driven type-A SD due to the asymmetric vortex surface barrier under an out-of-plane magnetic field, performs half-wave rectification up to 120 MHz and achieves a peak rectification efficiency of 43 $\%$. The field-driven niobium-based on-chip bridge rectifier performs full-wave rectification of a 3 MHz \textit{continuous} signal, and AC-to-DC conversion of a signal even at higher frequencies, up to 50 MHz in burst mode (short \textit{periodic} bursts) with a peak power efficiency of 50 $\%$.

\begin{figure*}
\includegraphics[scale=0.45]{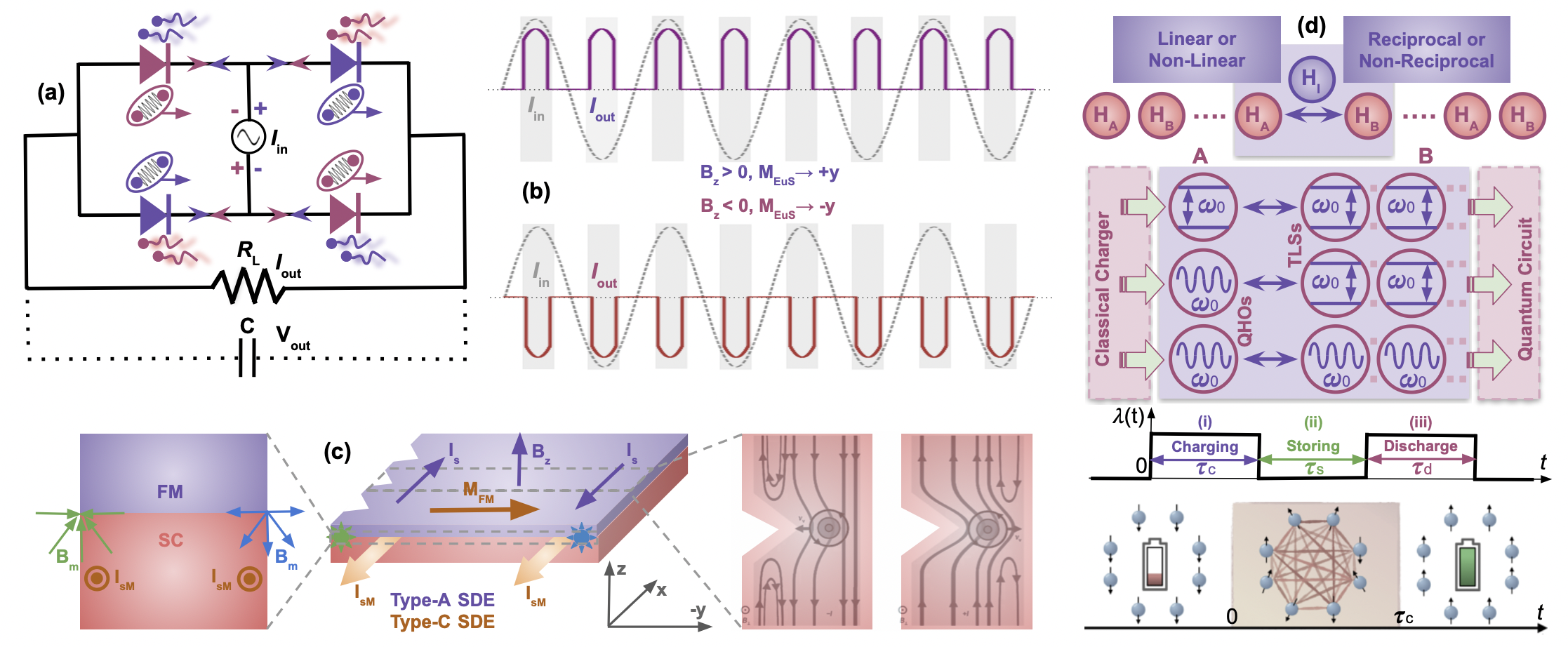}
\caption{\label{SDR} \textbf{Superconducting diode-based rectifiers (a-c) and quantum batteries (d).} \textbf{(a,b)} Schematic circuit diagram of a full-wave SD-rectifier made of SD bridge (a) and full-wave rectification where polarity of output DC signal depends upon the orientation of magnetic field and/or magnetization (b). \textbf{c} An optimized VD made of ferromagnet-superconductor bilayers (center), where left panel shows the cross-sectional view with fringing fields, while right panel depicts the asymmetric vortex flow under out-of-plane magnetic field. The system hosts both field-induced type-A SDE arising from edge-asymmetry and the externally applied out-of-plane magnetic field ($B_z$) and field-free type-C SDE arising from intrinsic in-plane magnetization ($M_{FM}$). Right panel is reproduced with permission from Ref.\cite{kochan25}. 
\textbf{(d)} A quantum battery characterized by Hamiltonian $\mathcal{H}=\mathcal{H}_0+\lambda(t)(\mathcal{H}_I-\mathcal{H}_0)$. (Top) A parallel/collective quantum charger-battery setup in which the charger ($\mathcal{H}_A$) and the battery ($\mathcal{H}_B$) are each typically modeled either as a two-level system (TLS) such as a qubit or as a quantum harmonic oscillator (QHO) in general, so that TLSs and QHOs act as the fundamental constituents of many-body quantum batteries. In a quantum charger-battery setup, where $\mathcal{H}_0= \mathcal{H}_A+\mathcal{H}_B$, the quantum charger serves as an ancilla system (or as a classical-to-quantum transducer) that links the external classical power source to the quantum battery, which subsequently discharges into a quantum circuit. Here, the classical control parameter $\lambda(t)$ is a step function that governs the cyclic unitary charging (and discharging or work extraction) processes: it switches on the interaction term $\mathcal{H}_I$ while keeping $\mathcal{H}_0$ switched off during the charging interval $\tau_c$; it then enables energy storage for a duration $\tau_s$ by turning off the interactions and switching on the battery Hamiltonian $\mathcal{H}_0$, thereby preserving the stored energy $E_B(\tau)$; finally, it initiates discharging over a time $\tau_d$ by switching the battery back off. (Bottom) Charging protocol of an SYK quantum battery (QB), where $\mathcal{H}_0$ consists of $N$ spin-1/2 cells and $\mathcal{H}_I$ denotes the SYK interaction. Bottom panel is reproduced with permission from ref. \cite{rossini20}.}
\end{figure*}

Josep Ingla-Aynés et al. \cite{ingla25} engineered a superconducting full-wave bridge rectifier by patterning four superconducting diodes on thin-film bilayers of vanadium/europium-sulfide (V/EuS), as shown in Figure \ref{SDR} (a-c). The efficiency of individual V/EuS SDs is optimized (up to 50$\%$) by combining the effect of edge asymmetry in a superconducting vanadium film (6-nm thick) and stray fields emanating from a ferromagnetic insulator EuS film (5-nm-thick), and the functionality of the fabricated SD bridge is demonstrated at cryogenic temperatures of a few Kelvin ($\approx1.7$ K). Although the SDE that originated from edge asymmetry (type-A) is induced by an out-of-plane magnetic field ($B_z\sim 4-5$ Oe), the SDE that originated from stray fields (type-C) is a consequence of an in-plane magnetization ($M_y$) of EuS. A remanent magnetic field from the EuS film, in combination with asymmetry at the edges of the vanadium film, resulted in full-wave rectification with a rectification efficiency of up to 43$\%$, and AC-to-DC signal conversion capabilities at frequencies up to 40 kHz. 

In both of these bridge rectifiers, consistent with the underlying symmetry constraints of SDE with broken time-reversal symmetry, the diode polarity can be reversed by inverting the orientation of the field or magnetization. Field-control polarity reversal is an essential capability for power management systems. However, for the integration of SDR with energy-efficient superconducting and quantum technologies, a field-free SDE is desired. In this context, SDR made of thin-film bilayers of V/EuS that leverage the remanent fields minimizes the need for an external magnetic field. On the other hand, SDR made of NbN micro-bridges is capable of operating at higher frequencies up to the megahertz range. The high rectification efficiency (up to 50$\%$) and high operating frequency (up to the megahertz range), as well as field-free full-wave rectification in these seminal reports open a window for exploring novel superconducting traits and optimized superconducting structures for efficient power delivery at cryogenic temperatures. 

\subsubsection{Superconducting diode quantum battery}
As in a classical battery, a quantum battery is a quantum mechanical system that can charge, store, and discharge energy. Typically, these three operational stages can be identified in a quantum system with a discrete and anharmonic energy spectrum of finite bandwidth, where charging and discharging takes place through a unitary transformation between ground-state energy and excited-energy state. Typically, a quantum battery can be described by a many-body interacting Hamiltonian $\mathcal{H}=\mathcal{H}_0 +\lambda(t)\mathcal{H}_I$, where $\mathcal{H}_0=\sum^N_{i=1}h_i$ represents a quantum battery made of $N$ identical cells, $\mathcal{H}_i$ is a charging Hamiltonian that characterizes interactions between battery elements, and $\lambda(t)$ is a classical control parameter that simulates external stimuli. In interacting many-body QBs, interactions among battery elements generate quantum correlations that lead to a genuine quantum advantage (GQA). The GQA, which originates from purely quantum effects rather than many-body collective effects, lies in surpassing and saturating the bounds on capacity (i.e., the maximum stored or extractable energy) and power (i.e., the minimum charging time or maximum charging speed) of QBs \cite{julia20}. 

The progress and development of quantum batteries \cite{ferraro26-NRP,campaioli24-RMP}, designed for coherent energy storage and transfer in quantum circuits \cite{kurman26}, has been expanded to include a range of solid-state material platforms \cite{camposeo25-AM,quach23-joul}. In addition, the ideas of nonlinearity \cite{andolina25,downing25} and nonreciprocity \cite{ahmadi24, khan25, guo25,sun25-NRQB,zhao25-NRQB,zafar26,lin26} in a quantum charger-battery setup \cite{andolina18,andolina19,farina19} have recently been introduced. In nonlinear bosonic quantum batteries \cite{andolina25}, a nonlinear coupling between the charger and the battery (each modeled as a harmonic oscillator) during the non-equilibrium charging process gives rise to a GQA: nonlinearity in the interacting many-body ($N$) quantum battery saturates the quantum speed limit (QSL) \cite{mandelstam45,margolus98, giovannetti03,deffner17}, so the optimal charging time matches the QSL time ($\tau_{QSL}\propto N^{-1/2}$), and the charging power scales super-linearly with $N$ ($P(\tau_{QSL})\propto N^{3/2}$), reaching the bound set by entanglement generation \cite{julia20}. In nonreciprocal quantum batteries, the one-way transfer of energy from the quantum charger to the battery leads to a significantly greater accumulation of stored energy than in conventional and reciprocal charger–battery setups \cite{ahmadi24}. The enhancement of the charging-energy dynamics due to nonreciprocity is typically achieved through reservoir engineering within the charger–battery setup, in which the quantum charger and the battery are simultaneously coupled to a common reservoir via local dissipative interactions, which establish a coherent coupling between the charger and the battery. These proposals offer a promising novel pathway for power delivery and management in quantum technologies; nonetheless, the realization of a quantum battery exhibiting dissipationless nonlinearity and nonreciprocity still remains a distant prospect.

To this end, we focus in particular on the physics and mechanization of QBs based on superconducting materials and architectures and seek how dissipationless nonlinearity and nonreciprocity can be simultaneously incorporated via diode effects at the device-level. The fundamental ideas of superconducting quantum batteries have been explored in multiple superconducting platforms, including earlier implementations based on superconducting qutrits (three-level systems) \cite{hu22}, superconducting qubits (two-level systems) \cite{gemme22-ibm,dou23,elghaayda25,hu26}, and Josephson elements with anomalous Josephson effect \cite{pal19,strambini20}. 

Interacting many-body QBs composed of two-level systems (TLSs), as shown in Figure \ref{SDR} (d), such as Dicke QBs \cite{ferraro18,andolina19} (N TLSs coupled to a single cavity mode via the Dicke Hamiltonian \cite{dicke54}) with an advantage due to many-body collective effects and Sachdev-Ye-Kitaev (SYK) QBs \cite{rossini20} (N spin-1/2 TLSs interacting via the SYK Hamiltonian \cite{rosenhaus19}) with GQA due to purely quantum effects, can be fabricated in the laboratory utilizing superconducting materials. For instance, Dicke QBs can be realized with superconducting qubits coupled to an on-chip superconducting transmission line resonator (single photon cavity) \cite{wallraff04}. On the other hand, on-chip implementation of SYK QBs can be realized in topological superconductors, hosting a collection of randomly interacting Majorana fermions (zero modes), characterized by the SYK model \cite{chew17,pikulin17}. 
Similarly, nonlinear bosonic QBs that exhibit GQA \cite{andolina25}, can be realized on superconducting c-QED platforms \cite{blais21} by employing nonlinear circuit elements, such as two superconducting $LC$ resonators coupled by a Josephson junction \cite{armour13}. Interestingly, nonlinear QBs demonstrate that GQA can likewise be implemented in a conventional charger–battery configuration \cite{andolina18,andolina19,farina19}, typically akin to SYK QBs, where charger-mediated energy transfer in quantum batteries achieves its quantum bound through a nonlinear interaction between the charger and the battery. In other words, incorporation of nonlinearity circumvents the constraints inherent to Dicke QBs \cite{ferraro18} and other analogous charger–battery architectures based on linear interactions \cite{andolina18,andolina19,farina19}. 

However, despite this progress in the charger-battery configuration enabled by nonlinear interaction, SYK QBs retain a crucial device-level advantage in terms of scalability, as both the battery Hamiltonian $H_0$ and the charging (interaction) Hamiltonian $H_I$ can be engineered within a single physical platform comprising an ensemble of interacting Majorana fermions \cite{pikulin17}. In addition, a fundamental connection between SYK interactions and correlated phases in quantum materials and unconventional superconductors \cite{camposeo25-AM} could allow the exploration of their intrinsic quantum functionalities to expand the accessible parameter space. It makes SYK QBs a purely physics-defined device, mitigating the circuit complexity. These observations motivate further exploration of schemes that enable simultaneous control of nonlinearity and nonreciprocity in both nonlinear charger–battery architectures and SYK QBs. A promising approach is to build on the symmetry-driven engineering of SDE: JDE in the Josephson junction that couples two $LC$ resonators in nonlinear bosonic QBs \cite{andolina25}, and VDE in topological superconductor hybrids \cite{pikulin17}.

Another platform for QBs with device-level nonlinearity and nonreciprocity, along with the promise of scalability, can be a Josephson phase battery. In equilibrium, a $\varphi_0$-junction characterized by a finite anomalous phase $\varphi_0\ne0,\pi$ in the ground state and a non-sinusoidal CPR $I_J(\varphi)= I_c\sin(\varphi+\varphi_0)$, acts as a Josephson phase battery that generates a constant phase bias $\varphi=-\varphi_0$ in an open circuit configuration (i.e., $I_J=0$), whereas it induces an anomalous Josephson current $I_{AJ}=I_c\sin(\varphi_0)$ in a closed circuit configuration (i.e., $\varphi=0$). Such an anomalous Josephson junction or a Josephson phase battery is only realizable when both time-reversal and inversion symmetries are broken simultaneously, i.e., $\varphi_0$ and thus $I_{AJ}$ vanishes when either one of these two symmetries is preserved.

Under these symmetry constraints, a nonreciprocal $\varphi_0$-junction can function as a nonreciprocal Josephson phase battery, provided that the finite anomalous effects ($\varphi_0\ne0,\pi$ and $I_{AJ}\ne0$) are driven by a Lifshitz-type invariant in the free energy and the corresponding CPR incorporates higher-harmonic, non-sinusoidal terms.
Such a nonreciprocal Josephson phase battery, which has so far attracted comparatively little attention, can be implemented through the broad range of mechanisms and approaches previously reported to break these symmetries, induce anomalous Josephson effects, and generate diode behavior. For example, anomalous $\varphi_0$-effects and the concept of a Josephson phase battery have recently been proposed for $\varphi_0$-junctions in which ferromagnetic polarization of the magnetic impurities is effectively converted into a $\varphi_0$ phase bias across the barrier \cite{pal19,strambini20}.

In a spin–orbit-coupled Al-InAs-Al $\varphi_0$-junction \cite{strambini20}, unpaired spins generated by surface defects in the InAs nanowire behave like ferromagnetic impurities and provide a persistent exchange interaction when their polarization is trained along a transverse direction. With a supercurrent flowing along the wire and the effective Rashba magnetic field directing perpendicular to the substrate plane, an anomalous $\varphi_0$ phase bias emerges due to the non-vanishing Lifshitz-type invariant in the free energy, which takes the form \cite{strambini20,bergeret15}: $F_L \approx f(\alpha,h)(\boldsymbol{\hat{h}}\times\boldsymbol{\hat{\epsilon}}) \cdot\boldsymbol{\hat{v}}$, where $\alpha$ denotes the SOI strength, $\boldsymbol{\hat{h}}$ is the magnetization vector breaking time-reversal symmetry, $\boldsymbol{\hat{\epsilon}}$ is the electric polarization vector responsible for inversion-symmetry breaking, and $\boldsymbol{\hat{v}}$ represents the superfluid velocity or, equivalently, the Josephson current ($\sim\boldsymbol{\hat{v}}$). A non-zero scalar triple product likewise plays a fundamental role in nonreciprocal superconductivity and the JDE, as long as it gives rise to finite-momentum Cooper pairing and odd-in-q contributions to the free energy \cite{He22}, which are linked to the higher-order non-sinusoidal terms in the CPR. This link between JDE and AJE \cite{reinhardt24} offers a more versatile platform to generate nonreciprocity in Josephson phase batteries. 

An SC-FM$_1$-spin$\_$flipper-FM$_2$-SC $\varphi_0$-junction functions as a quantized Josephson phase battery \cite{pal19}, in which quantized values of the anomalous phase $\varphi_0$ are stored in the ground state of the junction, and it also displays nonreciprocity in Josephson currents. As a result, JDE emerges, and the device simultaneously functions as a Josephson phase battery and a Josephson diode. Here, a magnetic impurity/adatom located between the two ferromagnetic layers acts as a spin-flipper. In this setup, both time-reversal and chiral symmetries are broken when the spin-flip probability is finite and the magnetizations of the two ferromagnetic layers are misaligned. The anomalous Josephson current vanishes and the Josephson current becomes reciprocal when either no spin-flipper is present and the spin-flip probability is zero, so that the exchange coupling between the electron/hole spin and the magnetic impurity is absent and the Andreev bound states remain time-reversal symmetric ($\epsilon_l(\varphi)=\epsilon_l(-\varphi)$), or magnetizations of the two ferromagnetic layers are aligned, so that both time-reversal and chiral symmetries are preserved. 
The origin of nonreciprocity and SDE here is associated with magnetic chirality, which could be further harnessed by optimizing the magnetization gradients \cite{roig24} or the inhomogeneous magnetic field profile \cite{cadorim24}, and stray fields from vortex trapping/removing \cite{Golod22}. 

These results suggest highly favorable prospects for the implementation of integrated quantum batteries based on superconducting diodes embedded within Josephson circuit architectures. Unlike dissipative reservoir engineering \cite{ahmadi24}, symmetry-breaking mechanisms alone can give rise to nonreciprocity in Josephson phase batteries, thereby realizing nonreciprocal Josephson phase batteries. However, in the operating regime, typically $I_{AJ}$ remains significantly lower than the critical current $I_c$, severely restricting the practical exploitation of the nonreciprocal response of the Josephson phase battery. The nonreciprocal Josephson phase battery can be exploited either via nearly ideal JDE enabled by unidirectional superconductivity \cite{daido25}, or via intrinsic nonreciprocal superconductivity, in which the superconducting traits remain direction-dependent well below the critical regime \cite{davydova24}. 

\subsection{Write-In: control and memory}
Switching mechanisms play a critical role in transistor functionality, memory devices, data storage systems, and signal amplification. Yet, at cryogenic temperatures, realizing efficient electrical switching in conventional semiconductors and reciprocal superconductors remains a challenging task. This challenge motivates the exploration of alternative platforms to develop more transformative algorithms and hardware, which could profoundly impact cryogenic quantum technologies. Beyond their exotic physical characteristics and quantum capabilities, SDs can also serve as highly effective superconducting switches, making them very promising for cryogenic electronic and computing architectures \cite{holmes13,soloviev17}. Here, we highlight the recently proposed SD-based designs for both transistor and memory devices.

\subsubsection{Superconducting diode transistor}
The field effect transistor (FET) architecture is a central element in controlling and processing information, both in the classical workflow and in the quantum workflow. In the classical workflow, the transistor embeds the central processing unit (CPU). In quantum-centric classical-quantum hybrid circuitry with a quantum processing unit (QPU), on the other hand, the transistor architecture is a key to both QPU and CPU, where the CPU acts as a control processor.

The electrical manipulation of symmetry-controlled nonreciprocal transport has recently been demonstrated in Josephson diode based FETs. As an illustration, in Ge-based Josephson field-effect transistors (JoFET) integrated into an asymmetric SQUID configuration \cite{leblanc24}, each JoFET fabricated from a SiGe/Ge/SiGe heterostructure functions as a Josephson diode and their CPRs display gate-tunable harmonic components. The principal harmonic corresponds to dissipationless charge-2e transport, while higher-order harmonics, second and third, respectively, reveal dissipationless charge-4e and charge-6e transport. The SQUID can be tuned from a nonreciprocal Josephson diode regime (with $\eta=27\%$) to a $\pi$-periodic reciprocal Josephson regime, where destructive interference between odd harmonics leads to a parity-protected qubit phase, effectively governed by $\sin(2\phi)$ Josephson elements with a charge-$4e$ supercurrent. It shows that gate-tunable SDE with tunable harmonic content offers Josephson devices that can act like a transistor for switching devices and control over superconducting qubits. 

In addition to the superconducting-semiconducting-superconducting (Su-Sm-Su) hybrid junctions, utilization of unconventional quantum materials could further extends the avenues for exploring quantum phenomena and quantum functionalities for SD-based transistor architectures. For example, the Josephson transistor \cite{hess23} and a Josephson diode sensor \cite{sinner24} made of chiral magnets (skyrmions and domain walls) and the nonreciprocal quantum neuronal transistor based on magnetic superconducting heterostructure \cite{xiong24} are prototypical examples that display the coexistence of the chiral phenomenon in magnetic and superconducting hybrid structures, leading to all-electrical current-driven switching.
Josephson transistors \cite{hess23} and sensors \cite{sinner24} based on chiral magnets could further motivate the exploration of SD-transistor functionalities based on exotic topological spin structures \cite{zhou25-AM}, in addition to skyrmions \cite{chen24}, for efficient write-off and read-out in unconventional nonreciprocal superconducting structures. 

\subsubsection{Superconducting diode memory}
Cryogenic memory technologies \cite{alam23}, interconnecting qubits and control processor, are critical for large-scale quantum computing architectures \cite{tannu17,hornibrook15}. With both a quantum substrate and a memory at nearly the same cryogenic temperature scale, even with non-integrated setup, the number of interconnects with a room-temperature CMOS control processor is reduced. In vortex-based SDs, a field-free switching of nonreciprocity, either by trapping and removing a vortex or by changing the bias configuration, enables in-memory functionality \cite{Golod22,golod15}. The diode-with-memory can facilitate emerging in-memory data management \cite{plattner12} for superconducting quantum technologies. The latest progress in spin-related superconducting devices for logic and memory applications, in which the underlying mechanism could be categorized as the superconducting spin diode effect, has been reviewed in Ref. \cite{he23-ADI}. More recently, an electrically-controllable superconducting memory effect has also been reported in a multiphase bulk triplet superconductor \cite{wu26-arXiv}, where electrical control drives the system into and out of a magnetic-field-induced metastable state, thereby enabling switching between high and low critical currents.

Moreover, with an intrinsic quantum control on the spin states of Cooper pairing, charge supercurrent transport, and tunable inductive CPR, different mechanisms for memories based on SDE can represent diverse potentials for scalable quantum computers. For example, electrical switching of JDE made of
chiral magnets that host magnetic domains and skyrmions along the racetrack \cite{hess23} can guide the magnetic domain-wall racetrack memory \cite{parkin08}. Moreover, quantum-level tunability of JDs, with 1D topological nanowires and 2D planar weak links, respectively, can enable miniaturization of persistent Josephson phase-slip memories \cite{ligato21} and Josephson-junction-based superconductor random access memories \cite{semenov19}. In addition, due to zero resistance and tunable rectification, SDE can also play key functions in cryotron and cryogenic flip-flop memory technologies  \cite{matisoo67,buck56}, e.g., by controlling the rectification directions of SDs placed on different arms of flip-flop memory, allowing readout with an infinite on/off ratio due to zero resistance.

\subsection{Read-Out: resonators and amplification}
Accurate quantum state measurement, supported by high-fidelity amplification processes, lies at the heart of quantum computing for translating quantum operations and converting fragile quantum information into accessible classical information processing. In this context, resonators and amplifiers provide the essential interface between quantum and classical workflow, thereby realizing the practical outcomes of quantum computation. This section highlights how the inherent nonlinearity of SDs could improve the performance of superconducting resonators and superconducting parametric amplifiers, and how the intrinsic nonreciprocity of SDs could avoid the requirement of intervening isolators and circulators between qubit-resonator and amplifier setup in the c-QED architecture.      

\subsubsection{Superconducting diode resonator}
Superconducting resonators \cite{zmuidzinas12,mcrae20,gurevich23} are fundamental elements in c-QED, providing a coherent electromagnetic environment for qubit readout \cite{wallraff04}, qubit-qubit coupling and entanglement generation, microwave detection \cite{day03}, quantum-limited parametric amplification \cite{castellanos07,bergeal10}, and coupling between qubit and qubit interfaces in hybrid quantum circuits \cite{xiang13}. In their simplest form, either a compact lumped-element circuit or a distributed transmission-line form \cite{goppl08}, superconducting resonators are fabricated by applying lithographic patterning techniques to a superconducting thin film deposited on an insulating substrate. These superconducting resonator structures, LC resonator and coplanar waveguide (CPW) resonator or transmission line resonator, play the role of cavities in superconducting circuits, and can be described by the same Hamiltonian that characterizes the dynamics of a cavity \cite{xiang13}: $H_{\text{cavity}}=\sum_k \hbar\omega_k(a^{\dagger}_k a_k+1/2)$ where $\omega_k$ is the frequency of the $k^{th}$ cavity mode while $a^{\dagger}_k$ and $a_k$ are the corresponding creation and annihilation operators, respectively.

The resonance frequency and quality factor, which quantifies energy losses, are the two key figures of merit that characterize the performance of superconducting resonators. 
The superconducting resonators, and the associated applications in c-QED setup, can capitalize on a wide variety of superconducting materials and broad range of phenomena, including magnetoelectric and kinetic inductance effect, nonlinear and nonreciprocal response, and dissipationless nonequilibrium dynamics. In addition to the exotic physical characteristics of superconducting thin films and nanowires in CPW and LC structures, superconducting resonators can be embedded with Josephson inductive elements (Josephson junctions and SQUIDs) to incorporate Josephson dynamics for tuning the resonance frequency and optimizing the quality factor \cite{palacios08,kennedy19}.

These superconducting resonators behave as quantum harmonic oscillators, characterized by well-defined and inductively controlled resonance frequencies, with a wide tunability through a rich landscape of superconducting phenomena. In the linear regime, superconducting resonators are particularly well suited for quantum nondemolition/dispersive readout and quantum memory/storage applications.
On the other hand, nonlinear superconducting resonators with a substantially improving measurement sensitivity enable quantum-enhanced signal detection, bifurcation-based readout, quantum-limited parametric amplification, and nonreciprocal coupling between qubits and qubit-interfaces. The nonlinear effects in microwave superconducting resonators, and their impact on the resonance frequency and quality factor, have been implemented via nonlinear kinetic inductance \cite{healey08,vissers15,thomas20}. 
For instance, the quadratic dependence of the resonance frequency on the magnetic field applied perpendicular to the plane of the superconducting thin film \cite{healey08}, and consequently on its fundamental and higher harmonics, is a manifestation of the nonlinear kinetic inductance of a superconducting device. The nonlinear current dependence of the kinetic inductance of superconducting thin films and nanowires is expected to be of even-order \cite{zmuidzinas12,ho12,zhao20}, just as in the case of JJs, $L_k(I)=L_k(0)[1+(I/I_{*,2})^2+(I/I_{*,4})^4+....]$ for $I\ll I_*$ or $T\ll T_c$, where $I_{*,2,4}$ are related to the critical current $I_c$ and set the scale of quadratic and quartic order of nonlinearity.
The superconducting resonators, embedded with flux-controlled SQUIDs \cite{khabipov22} or galvanically connected with a gate-controlled Josephson Junction \cite{strickland23}, also exhibit dissipationless nonlinearity with suppressed Kerr nonlinearity. In the former system, which is also non-centrosymmetric, Kerr-free nonlinearity emerges due to a second harmonic generation \cite{khabipov22}.

With simultaneously broken time-reversal and inversion symmetries, higher harmonic terms make the superconducting structure a superconducting diode and the superconducting resonator becomes a superconducting nonreciprocal resonator (SN-resonator) or a superconducting diode resonator (SD-resonator) \cite{dirnegger25}. Although superconductors are intrinsically dissipationless, however, several types of losses persist with the integration of superconducting resonators at the hardware level, including insertion loss, quasiparticle dynamics, microwave losses, and noise from two-level systems \cite{zmuidzinas12,mcrae20, gurevich23}. In this context, the SD-resonator plays a fundamental role to control the direction of noise flow. While the primary role of SD-resonator could be device-level nonreciprocity and thus a directional coupling of the resonator with qubit and qubit-interfaces, SD-resonator can also display several advantages in achieving a better temperature control, efficient gate and flux controlled cooperativity, and quantum-limited parametric amplification. For instance, superconducting resonators made of conventional low-T$_c$ superconducting materials, such as Al, Nb, and NbN, offer high quality factors. On the other hand, unconventional high-T$_c$ superconducting materials open many new possibilities for superconducting resonators that can operate in extreme conditions \cite{velluire23,ghirri16,ghirri15}; at high working temperature and under strong magnetic fields. The intrinsic microwave dissipation and the quality factor of high-T$_c$ superconducting resonators can be optimized via diode-controlled cooperativity.
The resonators made of conventional superconductors face critical challenges to attain efficient gate-control and dissipationless nonlinearity. However, SD-resonators made of superconducting diode materials allow quantum-enhanced tunability via symmetry-determined nonreciprocity and nonlinearity in the supercurrent transport.   

Both CPW and LC resonators can be engineered with SDs, so that the corresponding SD-resonators can be realized with both bulk superconducting structures (i.e., superconducting thin films and nanowires) and Josephson junction elements, respectively. This flexibility in the design allows SD-resonators to capitalize on whole host of superconducting traits in a diverse range of both conventional reciprocal and unconventional nonreciprocal superconducting materials. The symmetry breaking quantum effects, other than the fundamental physical characteristics of superconductivity, allow for a better quantum control over widely-tunable SD-resonators. In SD-resonators, with a diode-controlled interplay between linear and nonlinear regimes, SDE is a powerful resource that critically shapes resonator functionality for diverse applications in the c-QED architecture. For example, in a qubit-resonator system designed for readout and storage, SD-resonator can be tuned such that both Kerr-free and Kerr coefficients are simultaneously zero. On the other hand, for directional qubit transfer, qubit-qubit coupling and entanglement generation, and parametric amplification, the SD-resonator can be tuned such that only Kerr-free nonlinearity emerges in the parametric phase space.

\subsubsection{Superconducting diode amplifiers}
An amplification setup with low-noise, broad-bandwidth, large-saturation power and high-gain is crucial for next generation technologies requiring the accurate detection of ultra-low-amplitude signals in the microwave regime, such as qubit readout \cite{krantz19}, electron spin resonance detection \cite{xiao04,bienfait16}, electro-mechanical detection \cite{kippenberg08,teufel11}, radio astronomy \cite{bryerton13,smith13,pospieszalski18} and dark matter axion detection \cite{rybka14,caldwell17,jeong20,backes21}.
In quantum technology hardware, a state-of-the-art amplifier is characterized by the figures of merit implying quantum-limited noise, high fidelity, and single-shot quantum non-demolition readout. Achieving this requires a sophisticated engineering of readout hardware and the amplifier setup, co-locating with cryogenic qubit circuitry, such that insertion loss is removed to near quantum-limit and a signal-to-noise ratio sufficiently large for high-fidelity single-shot readout is achieved. These strict constraints on figures of merit and their implementation exclude standard solid-state amplifiers, which typically dissipate several milliwatts of power while amplifying the output signal from the readout resonator at the millikelvin stage.\\

In superconducting quantum computing, where the qubit and resonator readout are implemented through superconducting circuits, superconducting parametric amplifiers made with Josephson junctions have assumed a fundamental role in the superconducting qubit measurement. The most popular classes, typically based on negative resistance reflection or traveling-wave architectures, are known as Josephson parametric amplifiers (JPAs) \cite{aumentado20} and Josephson traveling-wave parametric amplifiers (JTWPAs) \cite{esposito21}, respectively. Both JPAs \cite{bergeal10,hatridge11,roch12} and JTWPAs \cite{macklin15} can be quantum limited, where only a minimum amount of noise, allowed by the principles of quantum mechanics \cite{caves82,clerk10}, is added by the amplification process. 
However, despite offering high gain and ultra-low noise, both JPAs and JTWAs still face critical challenges in terms of multiplexed qubit readout, scalability, and isolation. Although standing-wave JPAs built from nonlinear resonators are compact and leave a small footprint on a quantum processor, narrow-bandwidth ($\sim$MHz) and low saturation power, which are undesirable for quantum applications with multiplexed qubit readout and large power-handling capability, make JPAs suitable only for small-scale quantum setups.
In contrast, although JTWPAs provide high-saturation power and broad-bandwidth ($\sim$GHz), offering multi-qubit multiplexed readout setup \cite{macklin15,heinsoo18}, they incorporate a long chain of Josephson junctions \textemdash typically exceeding $10^3$ elements \textemdash which results in a large on-chip footprint and poses significant challenges for large-scale high-yield fabrication.

In addition, a high-fidelity qubit transfer requires a strong isolation between the qubit–resonator readout system and the amplification setup. The existing JPAs are intrinsically non-directional, they produce a gain in reflection mode, and therefore require circulators at the input to enforce directionality; separate outgoing amplified signals from their incoming counterparts, route the amplified signal toward output detection, and isolate the qubit–resonator circuit from the amplified signal and back-propagating noise. In contrast, JTWPAs are inherently directional because of their traveling-wave architecture, allowing forward-mode amplification in the direction of the pump propagation, without the need for intervening circulators to separate incoming and outgoing signals. However, a standard setup for JTWPAs still requires isolators to ensure reverse isolation, and thus does not completely eliminate the need for non-reciprocal components.
In addition to imposing nonreciprocity through the intervening elements, resonant-JPAs and JTWPAs also face different challenges in terms of controlling nonlinearity in the Josephson junctions \cite{aumentado20, esposito21}, which sets practical limits on gain, bandwidth, and power handling. 

The integration of superconducting diodes can sidestep many of these challenges faced by JPAs and TWPAs, including inherent control over directionality, coherent control over operating frequency range and bandwidth, and their intertwining with the operating temperature. These characteristics are intertwined with the quantum-controlled nonreciprocity and nonlinearity in the superconducting diodes, which can be widely tuned through both intrinsic quantum geometry and the extrinsic geometric structure of the device. The SDE can be incorporated at the device level, either by turning Josephson junctions into Josephson junction diodes so that Josephson elements become intrinsically directional or by integrating superconducting diodes as intervening nonlinear nonreciprocal elements in the Josephson junction based qubit readout circuits. The superconducting diode-controlled realization of parametric amplifiers that simultaneously achieve near–quantum-limited noise performance, high gain, broad bandwidth, and large saturation power – while exhibiting a truly directional control combined with backward-signal attenuation – would represent a significant breakthrough in qubit readout c-QED and overall quantum technology hardware.

Unlike power management, qubit control, and memory applications, the role of SDE in parametric amplification has not yet been explored, and a comprehensive study is still awaited. However, several features of the superconducting diode make it attractive for integration with parametric amplifiers, where the diode-controlled traits could solve several challenges through diode-controlled parametrization. For example, nonreciprocal superconductivity can be incorporated into Josephson junctions, where nonlinearity can be controlled well below the critical temperature $I_0$, rather than reaching the depairing limit, and thus a tradeoff between Josephson inductance, bandwidth, gain, and signal power can be acquired. The superconducting diode-controlled qubit anharmonicity via rectification efficiency \cite{zhong25} and a coherent nonreciprocal qubit-qubit inductive coupling via flux bias \cite{dirnegger25} shows that integration of SDE offers a promising route to realize a scalable circuit scheme of resonant-JPAs by optimizing the cavity architecture where superconducting diode-controlled anharmonicity, introduced by the nonlinear Josephson inductance, can be modulated through a wide-tunability of superconducting diode response. However, it is still open to address how controlling current-dependent and magnetic-flux-dependent nonlinear inductance of a Josephson junction, combined with gain-bandwidth constraint and the saturation power, could help in chasing a multi-qubit setup with compact superconducting-diode controlled JPAs.
On the other hand, a recently proposed setup for the traveling-wave parametric amplifier isolator (TWPAI), though with diplexers, \cite{ranadive25} suggests that the inherent isolation capabilities of TWPAI can eliminate the need for a commercial isolator. The SDE-driven inherent isolation could simplify the amplification setup even further. Interestingly, the second-harmonic contributions to the current-phase relation (Josephson current) affect the gain profile of a JTWPA, and the non-sinusoidal contribution enhances amplification without dispersion engineering \cite{guarcello25}. It further strengthens the potential role of intrinsic SDE in the amplification setup, which is solely based on the non-sinusoidal contribution to the CPR \cite{reinhardt24}.
More importantly, SDE may be employed for enhancing the first-stage amplification within traveling-wave architecture. Although TWPAs rely on the designing and manufacturing of a long amplification chain, the overall gain and noise performance is predominantly determined by the characteristics of the first amplifier in the chain \cite{pozar24}, which establishes the fundamental performance floor that constrains all subsequent amplifier stages. In this context, nonreciprocal and nonlinear superconducting traits at this first stage are of paramount importance to achieve quantum-enhanced gain and quantum-limited noise. 

In this direction, the superconducting diode-controlled Fourier engineering of Josephson energy-phase relations (EPRs)  \cite{bozkurt23} and dissipationless nonlinearity in Josephson diodes made of quantum materials open a new window for scalable amplification setup in the c-QED architecture \cite{schrade24}.

\begin{figure*}
\includegraphics[scale=0.4]{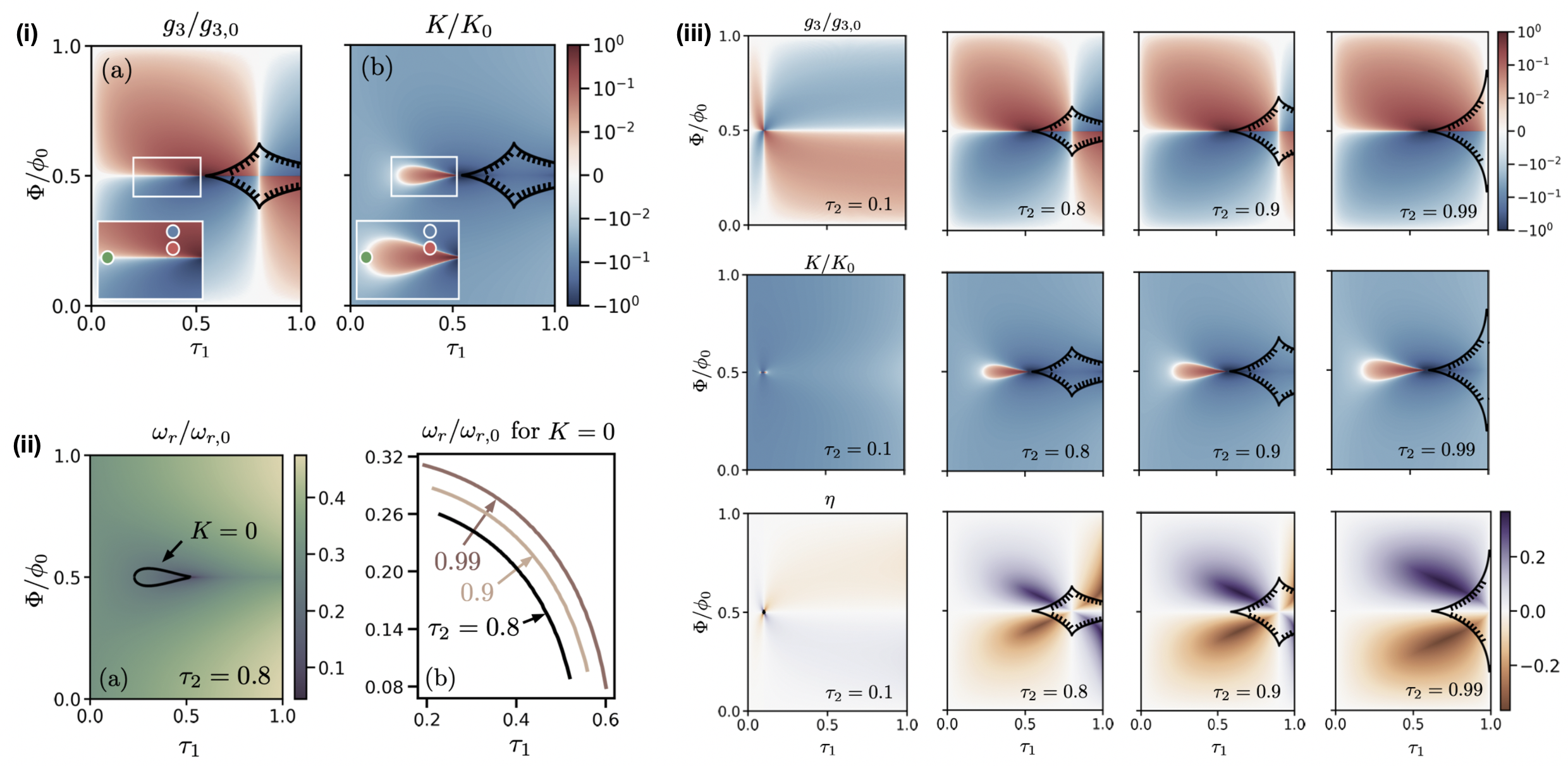}
\caption{\label{SDA} \textbf{Dissipationless nonlinearity, Kerr effect, and resonator frequency in nonreciprocal Josephson dipoles (SQUID).} \textbf{(i)} Third-order nonlinearity $g_3/g_{3,0}$ normalized by $g_{3,0}\equiv c_2\varphi_{ZPF}^3/6\hbar$ (a) and Kerr coefficient $K/K_0$ normalized by $K_0\equiv c_2\varphi_{ZPF}^4/2\hbar$ (b) as function of magnetic flux ($\Phi$) and transmission ($\tau_1$). At the green dot, with $g_3=0$ and $K=0$,  Josephson potential $U(\varphi)$ remains mirror symmetric across $\varphi=0$, whereas $U(\varphi)$ become asymmetric at the red dot $(g_3\ne0,K=0)$ and at the blue dot $(g_3\ne0,K\ne0)$. \textbf{(ii)} Resonator frequency $\omega_r/\omega_{r,0}$, normalized with $\omega_{r,0}=\sqrt{\Delta/C}/\phi_0$, as a function of $\Phi$ and $\tau_1$ with $\tau_2=0.8$ (a) and with different values of $\tau_2$ along the $K (\Phi(\tau_1),\tau_1)=0$ arc (b). \textbf{c} Third-order nonlinearity (top row), Kerr coefficient (center row), and diode efficiency (bottom row)  as a function of $\Phi$ and $\tau_1$, with different values of transmission $\tau_2$. Figure is reproduced with permission from ref. \cite{schrade24}}
\end{figure*}

\subsubsection{Field-free and Kerr-free nonlinearity}
Kerr-free third-order nonlinearity, and its interplay with fourth-order Kerr nonlinearity, has been reported in gate- and flux-tunable asymmetric SQUID-based JDs and magnetization-determined single-junction JDs \cite{schrade24}. The diode behavior in these nonreciprocal Josephson dipole (NJD) elements is characterized by the higher harmonics in CPR, which originate because of simultaneous breaking of time-reversal and inversion symmetries. In the interferometer setup, gate-control arises from the gate-tunability of the Andreev bound state (ABS) transmission in the junction. Compared to JRM \cite{bergeal10,bergeal10-NP,abdo13} and SNAIL \cite{frattini17,frattini18,sivak19, miano22}, as well as superconducting diode-based Fourier engineering of Josephson EPRs in an array of Josephson junctions \cite{bozkurt23,haenel22}, gate- and flux-controlled Kerr-free nonlinearity in the SQUID setup promise crucial benefits in terms of scalability and control. Apart from its simplicity in terms of circuit engineering, gate-voltage tunability permits efficient electrical control instead of magnetic fluxes. In addition, the resonance frequency $\omega_r$ is also gate-tunable in the dissipationless nonlinear regime, where third-order nonlinearity is realizable with vanishing Kerr coefficient ($K=0$). Compared to circuits where multiple flux-bias lines are required for frequency tunability \cite{miano22}, frequency-tunable Kerr-free nonlinearity in the quantum material SQUID setup can be realized through a single flux-bias. The gate-tunability of dissipationless third-order nonlinearity ($g_3$), the Kerr coefficient ($K$) characterizing fourth-order nonlinearity, the resonance frequency ($\omega_r$), and the superconducting diode efficiency ($\eta$) of SQUID-based NJD are shown in Figure \ref{SDA}. The Kerr-free regime ($K=0$) can be tuned by triplet ($\tau_1,\tau_2,\phi$) where $\tau_{1,2}$ is the gate-controlled asymmetric transmission of ABSs and $\phi=\Phi/\phi_0$ is the flux-controlled phase difference.

Similarly, a magnetization-determined quadruplet ($g_3,K,\omega_r,\eta$) is also realizable in single-junction JDs with a magnetic weak-link \cite{schrade24}. In the single-junction NJD, where two transversal conduction channels ensure the presence of second-harmonic terms in the CPR, field-free third-order nonlinearity arises from the interplay of Rashba SOI and intrinsic magnetization, following the exact symmetry-constraints required for the intrinsic SDE. As a result, the quadruplet ($g_3,K,\omega_r,\eta$) should be tunable via both magnetization and Rashba SOI, as well as their interplay, allowing for a whole host of intrinsic quantum functionalities, broadly characterized by Chern magnetism \cite{haldane88,weng15,liu16,shabbir18,nadeem20,bernevig22,chang23,nadeem24}, for wide-tunability of dissipationless nonlinearity. In accordance with the symmetry-constraint JDE, where sign reversal of nonreciprocity is a key characteristic, Kerr-free nonlinearity also changes its sign with gate-voltage, magnetic flux and magnetization.

\section{Parametric Framework: Quantum leap}
In hybrid quantum technology architectures, which integrate the processing of both bits and qubits, superconducting technologies offer a fully superconducting environment to be integrated into future quantum computing, communication, and sensing systems.
Although conventional superconductors promise dissipationless transport with a zero-resistance, the reciprocity in conventional superconducting components causes energy loss and decoherence. However, nonreciprocal superconductivity offers a potential platform to design superconducting circuitry that can perform scalable and high-fidelity operations by mitigating backscattering and noise. In addition, intrinsic nonreciprocity of superconductors simplifies quantum circuitry or circuit topology by avoiding the complicated architectures otherwise required to control directionality. For example, Josephson inductance $L(I)=L_0+L^{\prime}I+ L^{\prime\prime}I^2/2$ is a convenient probe for quantifying asymmetry in the CPR. A non-zero $L^{\prime}\equiv\partial_IL|_{I=0}$ is crucial in breaking CPR symmetry and leading to the nonreciprocity of supercurrent by inducing higher harmonics \cite{Baumgartner22}.
The non-zero $L^{\prime}$ is also a key element for modeling non-reciprocal superconducting devices \cite{leroux22-arXiv}. However, with conventional reciprocal superconducting components, an effective $L^{\prime}\ne0$ could only be engineered through complex superconducting circuitry \cite{frattini17,fadavi23}. On the other hand, due to the symmetry-constrained nonreciprocal supercurrent, a non-zero $L^{\prime}$ is an intrinsic feature of a single Josephson junction \cite{costa23-NN}.

With this device-level advantage, the SD-based quantum devices show that the degrees of freedom of SDs and their quantum-level tunability, while incorporating intrinsic quantum geometric effects along with extrinsic structural optimization, promise a novel parametric framework for scalable superconducting quantum technology hardware. These SD parameters act as new degrees of freedom to simultaneously control qubit and qubit interfaces, along with scalable interface coupling through inherent nonlinearity and intrinsic nonreciprocity at the device level.  

In the c-QED architecture, the physics and applications of Josephson junctions \cite{barone82,martinis04, golubov04,tafuri19} play a fundamental role in quantum engineering of superconducting qubit circuits and interfaces \cite{krantz19}, spanning from the Josephson phase battery, the transmon qubit and its control, noise mitigation and readout resonators, and amplification. For a scalable quantum technology architecture, Kerr-free nonlinearity and intrinsic nonreciprocity are the two of the most important features desired by Josephson junctions. In what follows, we shall discuss the role of JDE that not only brings Kerr-free nonlinearity and intrinsic nonreciprocity in a single Josephson junction by itself, and thus inherently in superconducting qubit and qubit-interfaces, but also JDE-based devices that could be integrated into nonreciprocal quantum circuitry as an intervening element.

First, unlike linear inductors and the corresponding passive LC resonator, the nonlinearity of the Josephson inductance allows an active anharmonic quantum oscillator, where low-energy non-degenerate qubit states are energetically spaced from the higher energy levels.  
The Josephson equations \cite{JOSEPHSON62}, $I=I_0\sin\varphi$ and $V=\frac{\Phi_0}{2\pi}\frac{d\varphi}{dt}$ where $\Phi_0=h/2e$ is the magnetic flux quantum, relate the current and voltage to a phase difference $\varphi$ and its time derivative, respectively, across the junction and to the critical current ($I_0$). These Josephson equations imply that the Josephson junctions act like a non-linear inductor with current-dependent inductance, $L_J(I)=\frac{L_{J0}}{\sqrt{1-(I/I_0)^2}}$, that diverges as the current reaches the critical current, which depends on the microscopic superconducting gap energy along with the macroscopic barrier thickness. The nonlinear Josephson inductance can be made magnetic-flux-dependent in a dc-biased superconducting quantum interference device (SQUID), a superconducting circuit made of two Josephson two junctions in a ring configuration. The magnetic flux quantization constraint on the superconducting ring, $\Phi=n\Phi_0$, makes the net inductance of SQUID a periodic function of flux, $L_{SQ}(I,\Phi)=\frac{L_J(I)}{2\cos(\Phi/\Phi_0)}$. This nonlinearity and its directional tunability (nonreciprocity) are at the core of practical superconducting c-QED, making Josephson junctions the basic building blocks in cryogenic superconducting quantum technology hardware, playing a diverse role for superconducting qubits, phase battery, switching and memory devices, and parametric amplifiers.

However, conventional JJ with sinusoidal CPR is Kerr-type nonlinear (fourth-order) and the transmission remains reciprocal. On the other hand, when both time-reversal symmetry and inversion symmetry are broken, an anomalous Josephson current with non-sinusoidal CPR could give rise to third-order nonlinearity while minimizing the Kerr effect, but the transmission still remains reciprocal in the single-channel Josephson junction \cite{yokoyama14}. Interestingly, when both time-reversal symmetry and inversion symmetry are broken and higher order harmonic terms are present in the CPR, a nonreciprocal Josephson diode (NJD) \cite{schrade24} can simultaneously exhibit both Kerr-free (pure $\phi^3$) nonlinearity and nonreciprocity.

It shows the critical importance of quantum material Josephson diodes \cite{schrade24}.
Apart from several unique characteristics and physics-imposed functionalities, one of the main features of Josephson junctions is to bring semiconducting-superconducting hybrids for efficient electric-field control through the semiconducting weak link while maintaining the coherence of superconducting states. However, the observation of SDE in Josephson junctions as well as in junction-free superconducting materials provides an alternative route for designing superconducting quantum devices based on bulk superconductors and Josephson constrictions, without requiring semiconducting hybrids. In addition, Josephson junctions made with quantum materials that exhibit nontrivial electronic and magnetic chirality could replace conventional semiconductor based Josephson junctions and offer all-electrical control over Josephson circuitry.

The integration of SD not only controls the nonlinearity and directionality at the device level, but also broadens the parameter space of c-QED environment integrating qubit and qubit-interfaces, which allows for an intrinsic quantum control and wider tunability. This is attributed to the current-dependent and flux-dependent nonlinearity in the Josephson kinetic inductance, which allows rectification efficiency and non-local phase as SD-determined degrees of freedom to control and tune the SD-integrated C-QED setup. As a proof of concept, the microscopic origin of the SD-controlled parametric space can be elucidated using the effective Hamiltonian for an SD-shunted transmon qubit chain and for a two-qubit quantum circuit inductively coupled to an SD-resonator.

The SD-determined control over parametric space characterizing the c-QED for qubit and qubit-interfaces can be understood from the nonlinear and nonreciprocal dynamics of Josephson elements. The canonical c-QED Hamiltonian for SD (JD here) shunted to the capacitor reads $H=H_C+H_U+H_J$, where $H_Q=H_C+H_U= 4E_C(\hat{n}-n_g)^2+E_JF(\hat{\varphi})$ is the transmon Hamiltonian \cite{tafuri19,blais21} and $H_J$ is the interaction between them. Here $E_C=e^2/(2C)$ is the charging (capacitative) energy with C the relevant circuit capacitance and $n_g$ a gate-induced charge offset, and $E_J=\Phi_0I_J/2\pi$ is the Josephson (inductive) energy with $\Phi_0=h/2e$ the flux quantum and $I_J$ the Josephson current. The phase function $F(\varphi)=\int f(\varphi)d\varphi$ connects the Josephson energy term with the current-phase relation (CPR) \cite{amundsen24-RMP,zhong25}, $I(\phi)=I_Jf(\varphi)=I_J\frac{d}{d\varphi}F(\varphi)$. The higher harmonic terms, other than $\sin\varphi$ in $f(\varphi)$ and $\cos\varphi$ in $F(\varphi)$, are the source of inherent odd-order nonlinearity and intrinsic nonreciprocity in the Josephson elements, a single-junction \cite{schrade24,zhong25} or SQUID \cite{schrade24,dirnegger25}, integrated in the c-QED architecture.  

The parametric role of SD can be obtained by transforming the canonical c-QED Hamiltonian to bosonic operators or qubit operators, that is, by permitting the quantization of circuit variables $(\varphi,n)$ and expanding the Hamiltonian around the global minimum of Josephson potential energy landscape. The canonically conjugated phase and number operators, with $[\hat{\varphi},\hat{n}]=i$ and $\Delta \phi \Delta n\ge1/2$, can be expressed in terms of bosonic ladder operators for creation and annihilation of the Cooper pair as $\hat{\varphi}=\varphi_{zpf}(\hat{a}+\hat{a}^{\dagger})$ and $\hat{n}=-in_{zpf}(\hat{a}-\hat{a}^{\dagger})=-i(1/2\varphi_{zpf})(\hat{a}-\hat{a}^{\dagger})$, where $\phi_{zpf}=(2E_C/E_J)^{1/4}$ is zero-point fluctuations (zpf) of the superconducting phase and $\phi_{zpf}n_{zpf}=1/2$. This uncertainty between phase and number operators and the quantization of energy landscape ($E_J-E_C$) set the quantum mechanical constraints on the transmon qubit as a weakly anharmonic quantum oscillator \cite{tafuri19,blais21}; in the transmon regime $E_J/E_C\gg 1$, where the Josephson tunneling dominates ($I_J\propto E_J$)
while the zpf in the superconducting phase are minimized ($\varphi_{zpf}\ll1$), the anharmonicity is constrained by the insensitivity of energy levels to the zpf in charge $n_{zpf}$. That is, the minimal uncertainty in the well-defined phase value $\langle\hat{\varphi}\rangle$, with a desirable protection from the suppressed charge noise, comes at the expense of reduced anharmonicity. In this context, a major challenge is a dissipative Kerr effect originating from the even-order nonlinearity.

In the presence of non-sinusoidal higher harmonics that allow odd-order nonlinearity, a small ZPF $\phi_{zpf}$ permits the Taylor expansion of the Josephson potential around its global minimum $\varphi_{min}$
and the Taylor expand transmon Hamiltonian can be expressed as
\begin{equation}
H = 4E_C(\hat{n}-n_g)^2+\sum_{m}\frac{c_m(\eta,\Phi,\tau,\omega)}{m!}(\varphi-\varphi_{\text{min}})^m
\end{equation}
where $c_m=\partial^m_{\varphi}U|_{\varphi_{min}}$ are Taylor coefficients,
that can be tuned with flux-bias parameters, gate-driven transmission, frequency or effective inductance and capacitance, and quantum-determined diode efficiency to control an interplay between odd-order and even-order nonlinear effects.
For instance, the second-order Taylor coefficient of $U(\varphi)$, $c_2\rightarrow E_J$, characterizes the harmonic transmon with $\phi_{zpf}=1/2n_{zpf}=(2E_C/c_2)^{1/4}$, Josephson inductance $L^{-1}=\phi_0^{-2}c_2$, and resonance frequency $\omega_r=1/\sqrt{LC}=\sqrt{8E_cc_2/\hbar^2}$. The third-order Taylor coefficient $c_3\rightarrow g_3$ characterizes the diode-determined third-order nonlinearity, that emerges due to the broken time-reversal symmetry, $U(-\Phi)\ne U(\Phi)$, while the Taylor coefficient of fourth-order term $\varphi^4$ determines the Kerr coefficient $K$.

In this direction, SDE with higher harmonics is required to tune and manipulate both weak anharmonicity and directionality via odd-order dissipationless nonlinearity. In SD-integrated c-QED setup, SDE can be parametrized by 
the figures of merit triplets $(\eta,\phi,\omega)$ and ($(g_3,K,\omega_r)$. The superconducting diode efficiency ($\eta$), operational frequency (of diode $(\omega$) as well as diode-controlled resonance frequency ($\omega_r$)), and dissipationless nonlinearity, including third-order Kerr-free nonlinearity $(g_3)$ and its interplay with the fourth-order Kerr coefficient $(K)$, can be controlled through external gate and flux knobs ($V_G,\Phi_b$) as well as intrinsic quantum functionalities associated with spin, orbital and valley degrees of freedom. Such an SD-integrated c-QED setup demonstrates the  enormous potential of superconducting diodes, making the spectrum of the effective potential-energy landscape rich and widely tunable with respect to the efficiency $\eta$ of SDE and the phase difference $\varphi=\phi_L-\phi_R$ introduced by SDE.

In a recent study on controlling the qubit anharmonicity and directional transport fidelity of transmon qubit chain circuits integrating SD \cite{zhong25}, nonlinearity and nonreciprocity are introduced by $F(\varphi)=-\cos(\varphi-\arcsin\eta)+\eta\varphi$, corresponding to a skew-symmetric CPR with $f(\varphi)=\sin(\varphi-\arcsin\eta)+\eta$, where non-zero anomalous phase is determined by the diode efficiency $\eta$. For the transmon qubit chain circuit with integrated superconducting diodes, as shown in Figure \ref{SDQ} (a), the effective qubit Hamiltonian in the truncated space reads \cite{zhong25}
\begin{equation}
\begin{split}
\hat{H} &\simeq 4E_C\left(\hat{n}-n_g\right)^2 \\
&+E_J\left[\frac{1}{2!}\sqrt{1-\eta^2}\varphi^2
+\frac{1}{3!}\eta\varphi^3
-\frac{1}{4!}\sqrt{1-\eta^2}\varphi^4\right] \;\label{eta}
\end{split}
\end{equation}
With any given $E_J/E_C$ ratio that characterizes the qubit circuit, an operational regime can be identified through a precise tuning of $\eta$ that supports exactly two-level qubit states inside the central potential well. Although a larger $\eta$ would be required for a larger $E_J/E_C$ ratio, an intermediate value of $\eta$, even well below the ideal case ($\eta\rightarrow100\%$), could be practically optimal to stabilize the two-level system configuration and preserve the qubit anharmonicity, which guaranties high transport fidelity by effectively suppressing the backward flow of noise.

The anharmonicity in superconducting diodes, demonstrated to improve the operation of transmon qubits through rectification-controlled phase difference \cite{zhong25}, may also be useful for optimizing other superconducting qubits through the precise tuning of inductive parameters ($L\simeq L_{J0}$ for the flux qubits) or bias currents ($I\rightarrow I_0$ for the phase qubits). The rectifying properties of SDE, for both the current-voltage characteristics in resistive measurements and the current-phase relations in inductive measurements, broaden the parametric space and allow a large tunability of the key parameters that determine the anharmonicity in the potential-energy landscape of various superconducting qubits. 
This makes superconducting diodes a key element in qubit circuits, not only because they overcome the usual reduction in qubit anharmonicity by transforming the qubit’s properties, but also because they can improve both coherence time and gate fidelity by introducing controlled anharmonicity within superconducting diodes, which are essential for scalable and noise-resilient quantum operations.

Such an interplay between odd-order and even-order nonlinear contributions, as depicted in qubit Hamiltonian (\ref{eta}), is an inherent characteristic of SD \cite{schrade24}. SDE promises third-order Kerr-free nonlinearity both in a single-junction JD heterostructure and in an interferometer setup of SQUID-based JDs \cite{schrade24}. In SQUID-based SD, JDE can be controlled by external magnetic bias flux ($\Phi_b$) \cite{souto22} and local gate-voltages \cite{schrade24}.
The asymmetric SQUID is characterized by the potential $U(\varphi)=U_1(\varphi)+U_2(\varphi-\Phi/\phi_0)$, corresponding to a flux-determined CPR $I(\varphi)=I_1(\varphi)+I_2(\varphi-\Phi/\phi_0)$, where $\phi_0=\hbar/2e=\Phi_0/2\pi$ is the reduced flux quantum and $\Phi$ is an external magnetic flux pierced through the area between the two JJs. Here $U_{1,2}$ and $I_{1,2}$, respectively, are the Josephson potentials and corresponding CPRs in the $l^{\text{th}}$ junction $(l=1,2)$. The origin of the JDE in such an interferometer setup is associated with non-vanishing higher harmonic contributions in the CPRs, $I_l(\varphi)= I^{(1)}_l\sin\varphi+I^{(2)}_l\sin2\varphi+....$ and correspondingly $U_l(\varphi)=-U^{(1)}_l\cos\varphi-U^{(2)}_l\cos2\varphi+....$, where amplitudes $I^{(m)}_l$ and $U^{(m)}_l$ represent contributions from the $m^{\text{th}}$ harmonic in the $l^{\text{th}}$ JJ. The JDE with a skew-symmetric CPR, originated from a superposition of sinusoidal and non-sinusoidal harmonics, requires (i) $\Phi/\phi_0\ne n\pi$  (n $\in$ integer) which breaks time-reversal symmetry and (ii) $I^{(m)}_1\ne I^{(m)}_2$ which break inversion symmetry. The second condition can be met by inducing asymmetry in the junctions transmission, such that the JDE arises from phase-dependent destructive (for $-\pi<\varphi<0$) and constructive (for $0<\varphi<\pi$) quantum interference between low and high transmission of the ABSs in the two junctions.

In the short-junction limit, CPR and Andreev levels emerging within the superconducting gap $\Delta$, which mediate the supercurrent, take the form \cite{beenakker91,souto22} $I(\varphi)=(e\Delta^2\tau_l/2\hbar) f(\varphi)$ where $f(\varphi)= \sin(\varphi)\text{Tanh}[\varepsilon_l(\varphi)/2k_BT]/\varepsilon_l(\varphi)$ and $\varepsilon_l(\varphi)= \Delta\sqrt{1-\tau_l\sin^2(\varphi/2)}$.
Here $\tau_l$ is the gate-tunable ABSs transmission in the $l^{\text{th}}$ junction, and the ground-state EPRs of the ABSs are given by $U_l(\varphi)=-\varepsilon_l(\varphi)$. A different transmission in the junctions, $\tau_1\ne\tau_2$, ensures that $I(\varphi)\ne -I(-\varphi+\Phi/\phi_0)$ implying $I^+_c=\text{max}_\varphi I(\varphi)\ne\text{min}_\varphi I(\varphi)=I^-_c$. However, if both junctions transmission are small, $\tau_l\ll1$, JDE vanishes because EPR $U_l(\varphi)$ is dominated by the first sinusoidal harmonic in the CPR, $I_l(\varphi)\propto\sin\varphi$ and $U_l(\varphi)\propto-\cos\varphi$. Together with $\tau_1\ne\tau_2$, if the junction transmission is large, $\tau_l\lesssim 1$, in one of the two JJs at least, then the total CPR acquires higher harmonic content and the JDE is guaranteed. 

With a high junction transmission $\tau_l\lesssim 1$ that allows odd-order contributions from higher harmonics and a small ZPF $\phi_{zpf}$ that permits the Taylor expansion of the Josephson potential around its global minimum $\varphi_{min}$, the canonical Hamiltonian for SQUID shunted to capacitor can be expressed in terms of bosonic lowing and raising operators as \cite{schrade24, frattini18} $\mathcal{\hat{H}} = \hbar\omega_ra^{\dagger}a +\sum_{m\ge3}\hbar g_m(\hat{a}+\hat{a}^{\dagger})^m$ where Kerr-free nonlinearity is expressed in terms of the third-order Taylor coefficient of $U(\varphi)$ as $\hbar g_3=(c_3/6)\varphi_{zpf}^3$ and the fourth-order Kerr term is computed as $\hbar K=1/2(c_4-5c_3^2/3c_2)\varphi_{zpf}^4$. The Taylor expanded transmon Hamiltonian up to the fourth order can be expressed as \cite{schrade24,frattini18}
\begin{equation}
\mathcal{\hat{H}}=\hbar\omega_ra^{\dagger}a
+\frac{c_3}{6}\varphi_{zpf}^3 (\hat{a}+\hat{a}^{\dagger})^3
+\frac{1}{2}\left(c_4-\frac{5c_3^2}{3c_2}\right)\varphi_{zpf}^4 (\hat{a}+\hat{a}^{\dagger})^4
\end{equation}
As shown in Figure \ref{SDA}, non-vanishing $g_3\ne0$ is realized when both time-reversal and inversion symmetries are broken ($\Phi/\phi_0\ne n\pi$ and $\tau_1\ne \tau_2$) but $g_3$ vanishes when either time-reversal symmetry or inversion symmetry is restored ($\Phi/\phi_0= n\pi$ or $\tau_1=\tau_2$).
The symmetry-driven gate- and flux-tunable Kerr-free nonlinearity $g_3\ne0$ is consistent with the underlying symmetry constraints in JDE. Moreover, within a finite regime of $(\Phi,\tau_1)$ phase diagram where the third-order nonlinearity remains finite $g_3\ne0$, the Kerr coefficient vanishes along a closed contour in the $(\Phi,\tau_1)$ phase diagram, with a gate-tunable resonator frequency along the $K=0$ arc. Interestingly, this Kerr-free trajectory, and the associated tunability of the resonator frequency $\omega_r=1\sqrt{LC}=\omega_{r,0}\sqrt{c_2}$ where $\omega_{r,0}\equiv(1/\phi_0)\sqrt{\Delta/C}$, can be further optimized by the second local electrostatic gate that controls transmission $\tau_2$, thereby enabling in situ control of the Kerr-free operating regime. Similar analysis persists for Kerr-free and field-free third-order nonlinearity in the single-junction setup \cite{schrade24}.

Sharing common origin and symmetry constraints, the strength of $g_3$ and JDE efficiency $\eta$ show congruent behavior in the phase diagram of $(\Phi,\tau_1)$, their strength increases with increasing transmission of both junctions simultaneously, while keeping the constraint $\tau_1\ne\tau_2$. However, Kerr-free nonlinearity ($K=0$) is realized only at low diode efficiency ($\eta<10\%$) around the time-reversal symmetry-restoring line $\Phi/\phi_0=0.5$ in the interferometer setup and $\eta<20\%$ in the single-junction setup. This analysis is also qualitatively persistent with the SD-integrated qubit chain circuit, where $g_3\propto\eta$ and $K\propto\sqrt{1-\eta^2}$ such that $(g_3,K)\rightarrow(0,1)$ when $\eta\rightarrow0$ and $(g_3,K)\rightarrow(1,0)$ when $\eta\rightarrow1$. In the SD-integrated c-QED setup, where a NJD shunted to a capacitor remains within the two-state working regime $E_J/E_C\gg1$, an optimized parameterization space ($c_2,c_3,c_4$) is highly required such that even a large shunting capacitance makes zero point fluctuations (zpf) of the superconducting phase small, $\phi_{zpf}=(2E_C/E_J)^{1/4} \ll1$. In this aspect. an interplay between dissipationless nonlinearity and nonreciprocity reiterates the need for quantum-controlled SDs with a widely tuneable efficiency and frequency.

The SQUID-based SD nonreciprocity, arising from an intrinsic third-order nonlinearity, has recently been introduced to demonstrate the nonreciprocal qubit dynamics; directional qubit-qubit coupling and entanglement transfer \cite{dirnegger25}. In close analogy to efficiency-controlled qubit dynamics enabled by single-junction JDs integrated in the qubit chain circuit \cite{zhong25}, SDE permits flux-controlled phase difference in the SQUID-based SD embedded in the resonator structure. Such SQUID-based SD-resonator allows SD-determined third-order nonlinearity and simultaneously supports the directional qubit dynamics, the directionality being passively mediated through nonreciprocal qubit-resonator coupling. A multi-qubit system with an inductive coupling to a resonator structure \cite{kafri17,solgun19,labarca24}, $H_{i,j}\propto\hat{\varphi}_i\hat{\varphi}_j \rightarrow\hat{a}_i\hat{a}_j^{\dagger}+\hat{a}_i^{\dagger}\hat{a}_j$, can be expressed in leading order qubit operators by projecting on the two lowest transmon states as \cite{dirnegger25}
\begin{equation}
H=\sum_i\frac{\omega_i}{2}\sigma_z^{i}+\sum_{i,j}J e^{i\tan^{-1}\left[\frac{J_{ij}^{nr}}{J_{ij}^{r}}\right]}
(\sigma_-^{i}\sigma_+^{j}+\sigma_+^{i}\sigma_-^{j})
\end{equation}
where $\sigma^i_{x,y,z}$ and $\sigma_{\pm}^{i}=\sigma^i_x\pm i\sigma^i_y$ are Pauli operators and $\omega^i$ is the excitation energy of the qubit at port $i\in \{1,2\}$. The interaction $J_{i,j}=J_{ij}^{r}+iJ_{ij}^{nr}=Je^{i\phi_{ij}}$ couples the nearest-neighbor qubits $i$ and $j$. The SD-resonator passively induces a nonreciprocal exchange coupling between the nearest-neighbor qubit modes through a direction-dependent non-local phase $\phi_{ij}=-\phi_{ji}=\varphi=\tan^{-1}(J_{ij}^{nr}/J_{ij}^{r})$. The sign of the non-local phase $\phi_{ij}$ can be controlled by external flux bias ($\Phi_b$), where $J_{ij}^{nr}$ changes sign with $\Phi_b$. With this analysis at hand, where non-vanishing $J_{ij}^{nr}$ originates from third-order nonlinearity, it is natural to ask how $J_{ij}^{nr}$ correlates with the diode efficiency in a Kerr-free regime of parameter space.

In the SQUID-based NJD, the dissipationless Kerr-free third-order nonlinearity can be parametrized through the tunability of even-order nonlinearities as
\begin{equation}
c_3=\pm\sqrt{\frac{3}{5}c_2c_4}
\end{equation}
This Kerr-free limit can be reached by varying inductive parameters $c_2=\phi_0/L$, or the resonance frequency and charging energy as $c_2=\hbar^2\omega_r^2/8E_c$, through a single flux-bias parameter. The flux-dependence, and in situ flux-tunability, of both odd order and even order nonlinearities is inherited from the EPR: while even-order nonlinearities are present even for symmetric EPR, odd-order nonlinearities emerge from the simultaneous breaking of time-reversal and inversion symmetries. As a result, quantum materials provide quantum-limited parametric framework where parameter space is comprised of both external stimuli such as gate voltage and intrinsic quantum functionalities such as SOI, orbital effects, and nontrivial Berry effects. In other words, instead of engineering the nonlinearities by varying the physical design of the device \cite{frattini18}, intrinsic quantum functionalities in the NJDs promise tunability of inherent nonlinearity. This aspect opens a new avenue to explore explicit quantum mechanization to engineer inherent nonlinearity over multiple orders of magnitude in NJD made of quantum materials.

These prototypical examples depicting the control over qubit anharmonicity, directional transfer of qubits, nonreciprocal qubit-qubit coupling and entanglement generation, which originates from the SD nonreciprocity emerging as a third order nonlinearity, promise immense potential of SDs in the c-QED setup. In the c-QED architecture with a gate-tunable magnetic NJDs \cite{schrade24} where both time-reversal and inversion symmetries are simultaneously broken, nonreciprocity and third-order nonlinearity can be attained in the absence of an external magnetic field and a Kerr-effect. This automatically extends the potential role of superconducting diodes for the realization of field-free and Kerr-free nonreciprocal qubit dynamics, resonator coupling, and three-wave mixing Josephson parametric amplification. In addition, a single-junction Josephson element does not necessitate the use of complicated circuitry. The inherent presence of field-free and Kerr-free dissipationless nonlinearity and intrinsic nonreciprocity in NJDs made of quantum materials inspires future research to explore the integration of superconducting diodes in scalable quantum technology hardware.

\begin{figure*}
\includegraphics[scale=0.45]{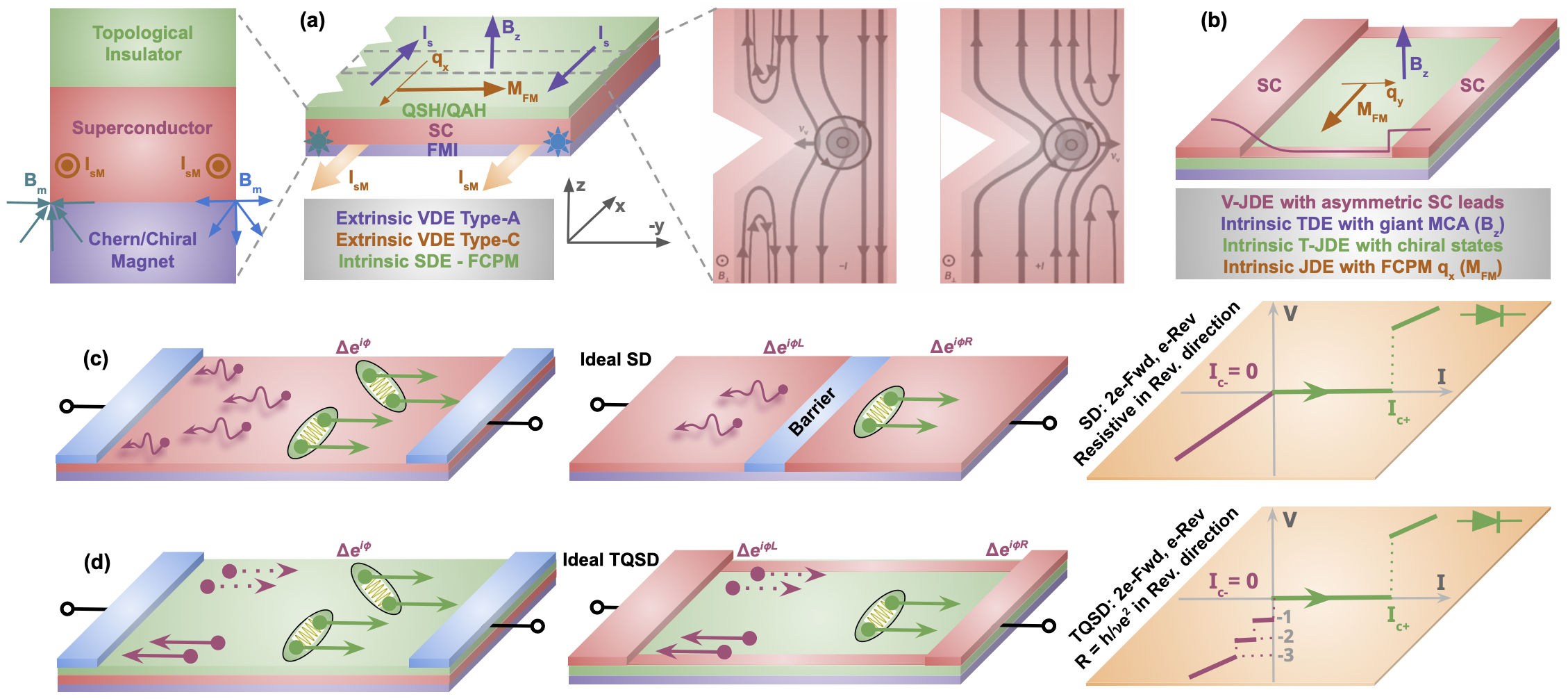}
\caption{\label{SDD} \textbf{Superconducting diode device design for optimized SDE and quantum control for technology readiness.} (a) A heterostructure of superconductor with topological insulator and Chern/chiral magnet (left) where screening currents and stray fields are drawn for the FMI-SC-QSH/QAH heterostructure (middle) originating from the asymmetric vortex dynamics (right). (b) Josephson junction made of FMI-SC-QSH/QAH heterostructure with asymmetric superconducting leads. (c) Transport mechanism in conventional SD based on FMI-SC hybrid structure, junction-free (left) and Josephson junction (middle), showing ideal diode behavior with Cooper pair supercurrent in one direction whereas single-electron normal current along the other (right). (d) Transport mechanism in TQSD based on FMI-SC-QSH/QAH hybrid structure, junction-free (left) and Josephson junction (middle), showing ideal diode behavior with Cooper pair supercurrent in one direction whereas quantized single-electron chiral current along the other (right).}
\end{figure*}

\section{Epilogue: Future perspective} 
The quantum landscape of SDs directly addresses the pressing needs at cryogenic temperature to enable on-chip integration of hybrid quantum technology hardware with temperature-matched classical and quantum workflows. Nevertheless, superconducting electronics still lack SD that is readily available for cryogenic technologies. The existing SDs face critical challenges in terms of performance and translation, whereas proposals for quantum circuits with SD-enabled quantum information processing remain in an early development stage. The current SD proposals (i) do not enable an optimized SDE where both intrinsic and extrinsic SDE mechanisms can coexist, (ii) do not allow noise-resilient signal processing in the circuit environment and remain dissipative due to normal state single-electron current in the revere direction, and (iii) do not exhibit a simultaneous quantum-level tunability of all three critical attributes required for control and scalability in cryogenic electronic and quantum technologies, that is, a wide Kerr-free dissipationless nonlinear regime in parameter space, large rectification efficiency, and broad operational frequency spectrum.

Several superconducting diode materials and structures have been reported, with different underlying mechanisms leading to a different origin of diode effects, categorized as intrinsic and extrinsic SDE. While the intrinsic mechanisms allow for wide parameter space for quantum control and tunability, extrinsic mechanisms allow for high rectification efficiency due to geometric optimization of asymmetric vortex dynamics. An optimization scheme is highly desired where high rectification efficiency driven by extrinsic mechanisms can be controlled via intrinsic quantum mechanisms. Additionally, the operational frequency of most existing superconducting diodes is very small ($\sim$ MHz). Geometric optimization is desired to enhance the frequency range, up to the micro range ($\sim$ GHz), such that superconducting diodes in this frequency range are readily available for integration with quantum hardware. Furthermore, SDE is characterized by inherent nonlinearity, which is a critical figure of merit for quantum technologies. However, dissipationless Kerr-free nonlinearity, with the absence of Kerr-effect, has not been fully explored in the existing models for SDE.

Henceforth, advancing SDs calls for an innovative quantum-enhanced SD platform whose degrees of freedom can be uniquely enhanced for optimizing SD-integrated quantum circuits, in which device-level nonlinearity and nonreciprocity will enable on-chip integration of qubit and qubit-interfaces, with perfect anharmonicity, high-fidelity, and large coherence.
Interestingly, flexibility in device design and versatility in materials selection give SDs an advantage in addressing these significant challenges faced by current SD proposals and the existing quantum technology hardware. As a potential route, we propose an innovative SD-device design and noise-resilient SD-mechanism, referred to as the TQSD. To realize the tantalizing prospects of this design and mechanism for SDs, from power management and controlled quantum processing to its storage and readout, further theoretical and experimental research is required to test the feasibility of various superconducting, magnetic, and topological materials, and fabricate geometric asymmetries. Such a quantum-enhanced SDE will make SDs a truly physics defined materials science and engineering platform.

\textbf{Proposed SD-device design:} Unlike a semiconducting diode, nonreciprocity in an SD emerges from symmetry-controlled nonreciprocal supercurrent transport. However, there is no unique device design or unique mechanism that could make SD readily available for potential integration in quantum technology hardware. A technology ready device design for quantum-enhanced SD can be fabricated via ferromagnetic/superconductor/topological-insulator heterostructure, either in the bulk structure (\ref{SDD}(a)) or in the Josephson junction (\ref{SDD}(b)). Due to the coexistence of magnetism, topology, and superconductivity, the quantum control over wide-tunability of its parameter space arises from a delicate interplay between intrinsic quantum functionalities, combined with the extrinsic geometric asymmetries in the architecture. First, this structure can exhibit an optimized SDE that incorporates different SDE mechanisms at the single platform, i.e., coexistence of intrinsic SDE based on finite Cooper pairing momentum (FCPM), extrinsic VDE due to asymmetric vortex dynamics, and JDE based on intrinsic or extrinsic mechanisms. Second, this topological structure not only offers the coexistence of intrinsic and extrinsic SDEs, but also incorporates topological quantum functionalities, a wide parameter space, and quantum-enhanced tunability. Third, this innovative design will allow coexistence of field-free SDE and field-driven SDE, so that field-training enhances SDE while leaving field-free SDE unaffected. In addition, the presence of a magnetic field and intrinsic magnetization in orthogonal directions, directed along both in-plane and out-of-plane orientations, allows inclusion of Rashba and Ising SOI, and thus a width control in topological superconducting structures \cite{yi22crossover}. 

For example, as shown in Figure \ref{SDD}, this device design with a bulk superconducting hybrid structure can host extrinsic VDE of type-A (with $B_z$) and type-C (with $M_{FM}$) as well as intrinsic SDE with FCPM $q_x$ (through $M_{FM}$). In addition, Josephson junctions made of this hybrid structure can exhibit (i) intrinsic SDE \cite{Legg22,legg23} together with the topological diode effect (TDE) with giant MCA ($B_z$) \cite{Legg22-nano}, (ii) Intrinsic JDE with FCPM $q_x$ (by $M_{FM}$) \cite{Pal22} along with the intrinsic topological diode effect (TDE) with chiral states,
(iii) extrinsic vortex-driven JDE (V-JDE) with asymmetric superconducting leads or current crowding and quantum confinement \cite{antola25}.

\textbf{Optimization of SDs:} The quantum mechanization of SDE can be engineered by incorporating intrinsic SDE combined with geometric asymmetry that produces extrinsic SDE, enabling comprehensive control for exploring the quantum advantage by harnessing intrinsic quantum functionalities and leveraging quantum confinement.
For example, field-free type‑C VDE arising from stray magnetic fields and field‑driven type‑A VDE resulting from edge asymmetry can be combined to optimize the extrinsic VDE in ferromagnetic-superconducting bilayers. This strategy has recently been demonstrated in V/EuS bilayers, in which the extrinsic VDE can be tuned by simultaneously exploiting the effect of edge asymmetry in the superconducting vanadium (V) layer and the stray fields generated by the ferromagnetic insulator EuS. In this system, the SD effect associated with stray fields (type‑C) originates from the in‑plane magnetization component ($M_{FM}$) of EuS, whereas the SDE associated with edge asymmetry (type‑A) is induced by an out‑of‑plane applied magnetic field ($B_z$). In the presence of nontrivial topology arising from the topological insulator layer, intrinsic SDE also irises, either from Rashba SOI with in-plane magnetization ($M_{FM}$) or Ising SOI with out-of‑plane applied magnetic field ($B_z$).

This coexistence of different intrinsic and extrinsic SDE mechanisms is allowed due to the same symmetry constraints leading to the different mechanisms of the SDE. nonreciprocity. For example, together with the interplay of SOI and Zeeman spin-splitting leading to MCA, FCPM can also arise due to Meissner screening currents \cite{Pal22}, thus allowing the coexistence of a wide variety of intrinsic mechanisms with MCA and extrinsic SDE mechanisms MSC. In addition, thickness and width-dependent structural optimization also enables intrinsic and extrinsic SD mechanisms to operate together in a complementary way, thus enhancing the performance of microwave devices with a broad frequency spectrum. For example, unlike conventional VDE where asymmetric vortex dynamics arises from an out-of-plane magnetic field, nonreciprocal behavior with an in-plane magnetic field dominates the nonreciprocity arising from the out-of-plane magnetic field in quasi-two-dimensional topological/chiral kagome superconductor CsV$_3$Sb$_5$ \cite{wu22-VD}. Such an anisotropic superconducting properties in a quasi-two-dimensional Rashba structure offer coexistence of both intrinsic SDE and extrinsic VDE with an in-plane magnetic field. 

\textbf{Topological quantum SD (TQSD):} Although superconductors allow zero-resistance supercurrent transport, SDs can cause dissipation due to normal state single-electron current in the revere direction, even with ideal SDs exhibiting 100$\%$ rectification efficiency. To make SD completely dissipation-free, normal state transport must also be quantized. The proposed device design can enable TQSD where the Cooper-pair current flows in the forward direction, while quantized chiral edge states enable transport in the reverse direction. First, the quantum-enabled nonlinearity and nonreciprocity in TQSD will lay foundation for a powerful platform for uncovering fundamental principles of quantum condensed matter, providing deep insight into superconductivity, magnetism, topology, and the rich physics arising from their coexistence. Second, TQSD will also allow efficient switching due to topological quantum field effects \cite{nadeem21,fuhrer21,nadeem22,weber24}. Third, the incorporation of TQSD, in which topology offers an additional layer of protection so that quantized state transfers become resilient to disorder and noise, and in which quantum functionalities arising from topological mechanisms enable a widely tuneable parameter space, offers significant advantages for conceptualizing quantum-controlled c-QED architecture.

\textbf{Device-level scalability:} The inherent nonlinearity and intrinsic nonreciprocity of unidirectional superconductors can be introduced in quantum technology hardware either by integrating SDs in the existing c-QED environments or by directly incorporating SDE in the quantum material Josephson elements that constitute qubit and qubit-interfaces. Interestingly, the existence of SDE, both in Josephson junctions and in junction-free bulk superconductors, opens another route for the scalability of quantum technology hardware. That is, the existing superconducting circuitry is heavily relying on the conventional Josephson-junctions, with extension to dipole SQUIDs and multi-pole Josephson elements with a complex circuit design. In the conventional Josephson-junctions, $T_c$ is typically lower than the constituent superconductors, and the information processing is heavily based on magnetic bias and fluxes. To move superconducting technologies beyond conventional circuits based on Josephson elements, SDs are important building blocks for exploring junction-free material platforms with intrinsic nonreciprocal superconductivity, scalable circuit design, and optimization of quantum functionalities. 

\textbf{Intervening hybrids and multi-axis sensing:}
In hybrid quantum technologies, where superconducting circuits interact with other quantum systems \cite{xiang13,mustafa2026-arX}, SDs and SD-based devices can be integrated as intervening devices, including SD-based spin-splitters and spin-filters \cite{Strambini22,sun25-spin,giil24b}, SD-based sensors \cite{sinner24} and detectors \cite{heikkila18,geng23-Rev}, SD-based nonreciprocal antenna \cite{Zhang20}, as well as SD-based nonreciprocal devices for signal routing such as topologically protected isolators \cite{wang25-iso} and quantum-limited circulators \cite{chapman17,ranzani19}.

In contrast to supercurrent spin Josephson diodes (SSJDs) \cite{sun25-spin}, the nonreciprocity of the quasi-particle current can be induced by breaking the electron-hole symmetry through spin-splitting and spin-filtering in superconductor–ferromagnet hybrids, allowing superconducting spintronic tunnel diodes (SSTDs) \cite{Strambini22} and corresponding spin-related superconducting logic and memory devices for superconducting spintronics \cite{zutic04,linder15,bergeret18, mel22,cai23-AQT}. This phenomenon also allows superconducting thermoelectric effects \cite{ozaeta14,bergeret18} and superconductor–ferromagnet thermoelectric detectors (SFTEDs) \cite{heikkila18,geng23-Rev}. 
The functionalities of both SSJDs \cite{sun25-spin} and SSTDs \cite{Strambini22}, and thus the corresponding superconducting Josephson diode sensor (SJDS) \cite{sinner24} and self-biased SFTED \cite{geng23-Rev}, are influenced by the combination of time-reversal symmetry breaking and inversion symmetry breaking effects.
Henceforth, the nonreciprocity of spin and thermoelectricity can be made noise-resilient by mapping spin and thermoelectric devices with supercurrent diode effects.

In the symmetry-constrained intrinsic and extrinsic SDs, the directions of the supercurrent, the polar axis, and the magnetization are required to be mutually orthogonal. Under these constraints, symmetry-governed nonreciprocity can only give rise to a single-axis sensor/detector with a restricted dynamic range for magnetic-field sensing/detection, closely resembling sensing schemes based on nitrogen-vacancy centers. However, SDE in various superconducting materials, such as kagome superconductors \cite{wu22-VD}, strained/doped film of the superconductor SrTiO$_3$ \cite{kealhofer23}, and noncentrosymmetric type-II Weyl semimetal T$_d$-MoTe$_2$, can be realized with different orientations of magnetic field. In kagome superconductors, the resulting SDE could be extrinsic, associated with asymmetric vortex dynamics, or intrinsic, associated with electronic chirality. In Weyl semimetals, SDE with different orientations of magnetic field could be associated by different types of Rashba SOI. In SrTiO$_3$ superconducting thin films, on the other hand, the origin of SDE both with magnetic field orientation perpendicular and parallel to the current direction suggests the coexistence of polar and chiral effects. Moreover, in Rashba nanowires \cite{Legg22}, the emergence of an intrinsic SDE necessitates the simultaneous application of magnetic fields oriented both longitudinally (along the wire axis) and transversely (perpendicular to the wire). Nonreciprocal superconductors, in which SDE is realizable through magnetization with a wide dynamic range, can be employed for multi-axis magnetic sensors. However, the underlying quantum mechanisms leading to SDE with multiple orientations of magnetization are not yet fully understood. To steer the engineering of multi-axis superconducting magnetic sensors, a broad search of superconducting materials and a more comprehensive understanding of quantum origin of SDE are required.

In the context of recent advances in SDs and SD-based quantum circuits, this perspective highlights several innovations and technological impacts, including innovative device design that exhibits an optimized and quantum-enhanced SD, a novel concept of TQSD that guaranties quantum-limited noise, and the technological readiness of SDs with a simultaneous quantum-level tunability of dissipationless nonlinearity, large rectification efficiency, and broad operational frequency spectrum. The future research in this area expects to formulate an integrated theoretical framework – based on innovative design, modeling, simulations, and optimization – that translate the underlying quantum functionalities of SDs into meaningful impacts for noise-resilient, high-performance, and scalable c-QED environment. Such a unified framework will expect the development of refined mechanisms for SDs, the establishment of improved circuit design and approaches for SD-integrated c-QED, the generation of new knowledge through the delivery of advanced interdisciplinary techniques spanning quantum condensed matter physics, materials science, materials engineering, and electronic engineering to enable transformative advances in quantum technologies. We hope that this perspective will offer a pathway toward fabricating SD-integrated quantum circuits, in which device-level nonlinearity and nonreciprocity will enable on-chip integration of qubit and qubit-interfaces with perfect anharmonicity, high-fidelity, and large coherence.  


\bibliography{apssamp}

@PREAMBLE{
 "\providecommand{\noopsort}[1]{}" 
 # "\providecommand{\singleletter}[1]{#1}%" 
}

@article{IEA24,
  title={Electricity 2024 analysis and forecast to 2026},
  author={E. Cam and Z. Hungerford and N. Schoch and F. P. Miranda and C. D. Y. de Le´on},
  journal={INTERNATIONAL ENERGY AGENCY},
  year={2024}
}

@article{SRC21,
  title={Decadal Plan for Semiconductors},
  author={SRC},
  journal={Semiconductor Research Corporation, https://www.src.org/about/decadal-plan/},
  year={2021}
}

@article{johansson12,
  title={QuTiP: An open-source Python framework for the dynamics of open quantum systems},
  author={Johansson, J Robert and Nation, Paul D and Nori, Franco},
  journal={Computer physics communications},
  volume={183},
  number={8},
  pages={1760--1772},
  year={2012},
  publisher={Elsevier}
}

@book{tinkham04,
  title={Introduction to Superconductivity},
  author={Tinkham, M.},
  isbn={9780486134727},
  series={Dover Books on Physics Series},
  year={2004},
  publisher={Dover Publications}
}

@book{tafuri19,
  title={Fundamentals and frontiers of the Josephson effect},
  author={Tafuri, Francesco},
  volume={286},
  year={2019},
  publisher={Springer Nature}
}

@article{linder15,
  title={Superconducting spintronics},
  author={Linder, Jacob and Robinson, Jason WA},
  journal={Nature Physics},
  volume={11},
  number={4},
  pages={307--315},
  year={2015},
  publisher={Nature Publishing Group}
}

@article{cai23-AQT,
  title={Superconductor/ferromagnet heterostructures: A platform for superconducting spintronics and quantum computation},
  author={Cai, Ranran and {\v{Z}}uti{\'c}, Igor and Han, Wei},
  journal={Advanced Quantum Technologies},
  volume={6},
  number={1},
  pages={2200080},
  year={2023},
  publisher={Wiley Online Library}
}

@article{braginski19,
  title={Superconductor electronics: status and outlook},
  author={Braginski, Alex I},
  journal={Journal of superconductivity and novel magnetism},
  volume={32},
  number={1},
  pages={23--44},
  year={2019},
  publisher={Springer}
}

@article{mel22,
  title={Superconducting spintronics: state of the art and prospects},
  author={Mel’nikov, AS and Mironov, Sergei Viktorovich and Samokhvalov, Aleksei Vladimirovich and Buzdin, Alexander Ivanovich},
  journal={Uspekhi Fiz. Nauk},
  volume={192},
  pages={1339--1384},
  year={2022}
}

@article{soloviev17,
  title={Beyond Moore’s technologies: operation principles of a superconductor alternative},
  author={Soloviev, Igor I and Klenov, Nikolay V and Bakurskiy, Sergey V and Kupriyanov, Mikhail Yu and Gudkov, Alexander L and Sidorenko, Anatoli S},
  journal={Beilstein journal of nanotechnology},
  volume={8},
  number={1},
  pages={2689--2710},
  year={2017},
  publisher={Beilstein-Institut}
}

@article{holmes13,
  title={Energy-efficient superconducting computing—Power budgets and requirements},
  author={Holmes, D Scott and Ripple, Andrew L and Manheimer, Marc A},
  journal={IEEE Transactions on Applied Superconductivity},
  volume={23},
  number={3},
  pages={1701610--1701610},
  year={2013},
  publisher={IEEE}
}

@article{anders10,
  title={European roadmap on superconductive electronics--status and perspectives},
  author={Anders, S and Blamire, MG and Buchholz, F-Im and Cr{\'e}t{\'e}, D-G and Cristiano, R and Febvre, P and Fritzsch, L and Herr, Anna and Il’Ichev, E and Kohlmann, J and others},
  journal={Physica C: Superconductivity},
  volume={470},
  number={23-24},
  pages={2079--2126},
  year={2010},
  publisher={Elsevier}
}

@article{radebaugh09,
  title={Cryocoolers: the state of the art and recent developments},
  author={Radebaugh, Ray},
  journal={Journal of Physics: Condensed Matter},
  volume={21},
  number={16},
  pages={164219},
  year={2009},
  publisher={IOP Publishing}
}

@article{kirichenko11,
  title={Zero static power dissipation biasing of RSFQ circuits},
  author={Kirichenko, DE and Sarwana, Saad and Kirichenko, AF},
  journal={IEEE Transactions on Applied Superconductivity},
  volume={21},
  number={3},
  pages={776--779},
  year={2011},
  publisher={IEEE}
}

@article{mukhanov11,
  title={Energy-efficient single flux quantum technology},
  author={Mukhanov, Oleg A},
  journal={IEEE Transactions on Applied Superconductivity},
  volume={21},
  number={3},
  pages={760--769},
  year={2011},
  publisher={IEEE}
}

@inproceedings{mukhanov19,
  title={Scalable quantum computing infrastructure based on superconducting electronics},
  author={Mukhanov, O and Kirichenko, A and Howington, C and Walter, J and Hutchings, M and Vernik, I and Yohannes, D and Dodge, K and Ballard, A and Plourde, BLT and others},
  booktitle={2019 IEEE International Electron Devices Meeting (IEDM)},
  pages={31--2},
  year={2019},
  organization={IEEE}
}

@article{mcdermott18,
  title={Quantum--classical interface based on single flux quantum digital logic},
  author={McDermott, Robert and Vavilov, Mikhail Grigorievich and Plourde, Benjamin Louis Thomas and Wilhelm, Frank Karl and Liebermann, Peter James and Mukhanov, Oleg Alexandrovich and Ohki, Tatsuya Akira},
  journal={Quantum science and technology},
  volume={3},
  number={2},
  pages={024004},
  year={2018},
  publisher={IOP Publishing}
}

@article{zutic04,
  title={Spintronics: Fundamentals and applications},
  author={{\v{Z}}uti{\'c}, Igor and Fabian, Jaroslav and Sarma, S Das},
  journal={Reviews of modern physics},
  volume={76},
  number={2},
  pages={323},
  year={2004},
  publisher={APS}
}

@article{blais21,
  title={Circuit quantum electrodynamics},
  author={Blais, Alexandre and Grimsmo, Arne L and Girvin, Steven M and Wallraff, Andreas},
  journal={Reviews of Modern Physics},
  volume={93},
  number={2},
  pages={025005},
  year={2021},
  publisher={APS}
}

@article{wendin17,
  title={Quantum information processing with superconducting circuits: a review},
  author={Wendin, G{\"o}ran},
  journal={Reports on Progress in Physics},
  volume={80},
  number={10},
  pages={106001},
  year={2017},
  publisher={IOP Publishing}
}

@article{liu19,
  title={2D materials for quantum information science},
  author={Liu, Xiaolong and Hersam, Mark C},
  journal={Nature Reviews Materials},
  volume={4},
  number={10},
  pages={669--684},
  year={2019},
  publisher={Nature Publishing Group}
}

@article{dixit21,
  title={Searching for dark matter with a superconducting qubit},
  author={Dixit, Akash V and Chakram, Srivatsan and He, Kevin and Agrawal, Ankur and Naik, Ravi K and Schuster, David I and Chou, Aaron},
  journal={Physical review letters},
  volume={126},
  number={14},
  pages={141302},
  year={2021},
  publisher={APS}
}

@article{farrah19,
  title={Far-infrared instrumentation and technological development for the next decade},
  author={Farrah, Duncan and Smith, Kimberly Ennico and Ardila, David and Bradford, Charles M and DiPirro, Michael J and Ferkinhoff, Carl and Glenn, Jason and Goldsmith, Paul F and Leisawitz, David T and Nikola, Thomas and others},
  journal={Journal of Astronomical Telescopes, Instruments, and Systems},
  volume={5},
  number={2},
  pages={020901},
  year={2019},
  publisher={SPIE}
}

@article{nadeem23,
  title={The superconducting diode effect},
  author={Nadeem, Muhammad and Fuhrer, Michael S and Wang, Xiaolin},
  journal={Nature Reviews Physics},
  volume={5},
  number={10},
  pages = {558-577},
  year={2023},
  publisher={Nature Publishing Group UK London}
}

@article{nadeem23-arXiv,
  title={Superconducting Diode Effect--Fundamental Concepts, Material Aspects, and Device Prospects},
  author={Nadeem, Muhammad and Fuhrer, Michael S and Wang, Xiaolin},
  journal={arXiv preprint arXiv:2301.13564},
  year={2023}
}

@article{ma25-APR,
  title={Superconducting Diode Effects: Mechanisms, Materials and Applications},
  author={Ma, Jiajun and Zhan, Ruiya and Lin, Xiao},
  journal={Advanced Physics Research},
  pages={2400180},
  year={2025},
  publisher={Wiley Online Library}
}

@article{shaffer25-arXiv,
  title={Theories of Superconducting Diode Effects},
  author={Shaffer, Daniel and Levchenko, Alex},
  journal={arXiv preprint arXiv:2510.25864},
  year={2025}
}

@article{davydova24,
  title={Nonreciprocal superconductivity},
  author={Davydova, Margarita and Geier, Max and Fu, Liang},
  journal={Science Advances},
  volume={10},
  number={48},
  pages={eadr4817},
  year={2024},
  publisher={American Association for the Advancement of Science}
}

@article{Daido22,
  title = {Intrinsic Superconducting Diode Effect},
  author = {Daido, Akito and Ikeda, Yuhei and Yanase, Youichi},
  journal = {Phys. Rev. Lett.},
  volume = {128},
  issue = {3},
  pages = {037001},
  numpages = {6},
  year = {2022},
  month = {Jan},
  publisher = {American Physical Society},
  doi = {10.1103/PhysRevLett.128.037001}
}

@article{Noah22,
author = {Noah F. Q. Yuan  and Liang Fu },
title = {Supercurrent diode effect and finite-momentum superconductors},
journal = {Proceedings of the National Academy of Sciences},
volume = {119},
number = {15},
pages = {e2119548119},
year = {2022},
doi = {10.1073/pnas.2119548119}
}

@article{He22,
doi = {10.1088/1367-2630/ac6766},
year = {2022},
month = {may},
publisher = {IOP Publishing},
volume = {24},
number = {5},
pages = {053014},
author = {James Jun He and Yukio Tanaka and Naoto Nagaosa},
title = {A phenomenological theory of superconductor diodes},
journal = {New Journal of Physics}
}

@article{Ilic22,
  title = {Theory of the Supercurrent Diode Effect in Rashba Superconductors with Arbitrary Disorder},
  author = {Ili\ifmmode \acute{c}\else \'{c}\fi{}, S. and Bergeret, F. S.},
  journal = {Phys. Rev. Lett.},
  volume = {128},
  issue = {17},
  pages = {177001},
  numpages = {6},
  year = {2022},
  month = {Apr},
  publisher = {American Physical Society},
  doi = {10.1103/PhysRevLett.128.177001}
}

@article{mayer20gate,
  title={Gate controlled anomalous phase shift in Al/InAs Josephson junctions},
  author={Mayer, William and Dartiailh, Matthieu C and Yuan, Joseph and Wickramasinghe, Kaushini S and Rossi, Enrico and Shabani, Javad},
  journal={Nature communications},
  volume={11},
  number={1},
  pages={1--6},
  year={2020},
  publisher={Nature Publishing Group}
}

@article{Baumgartner22-JPCM,
doi = {10.1088/1361-648X/ac4d5e},
year = {2022},
month = {feb},
publisher = {IOP Publishing},
volume = {34},
number = {15},
pages = {154005},
author = {C Baumgartner and L Fuchs and A Costa and Jordi Picó-Cortés and S Reinhardt and S Gronin and G C Gardner and T Lindemann and M J Manfra and P E Faria Junior and D Kochan and J Fabian and N Paradiso and C Strunk},
title = {Effect of Rashba and Dresselhaus spin–orbit coupling on supercurrent rectification and magnetochiral anisotropy of ballistic Josephson junctions},
journal = {Journal of Physics: Condensed Matter}
}

@article{Ando20,
author = {Ando, Fuyuki and Miyasaka, Yuta and Li, Tian and Ishizuka, Jun and Arakawa, Tomonori and Shiota, Yoichi and Moriyama, Takahiro and Yanase, Youichi and Ono, Teruo},
year = {2020},
month = {08},
pages = {373-376},
title = {Observation of superconducting diode effect},
volume = {584},
journal = {Nature},
doi = {10.1038/s41586-020-2590-4}
}

@article{Lorenz22,
author = {Lorenz, Bauriedl and Christian, Bäuml and Lorenz, Fuchs and Christian, Baumgartner and Nicolas, Paulik and Jonas, M. Bauer and Kai-Qiang, Lin and John, M. Lupton and Takashi, Taniguchi and Kenji, Watanabe and Christoph, Strunk and Nicola Paradiso},
year = {2022},
month = {07},
pages = {4266},
title = {Supercurrent diode effect and magnetochiral anisotropy in few-layer NbSe$_2$},
volume = {13},
journal = {Nature Communications},
}

@article{JOSEPHSON62,
title = {Possible new effects in superconductive tunnelling},
journal = {Physics Letters},
volume = {1},
number = {7},
pages = {251-253},
year = {1962},
issn = {0031-9163},
doi = {https://doi.org/10.1016/0031-9163(62)91369-0},
author = {B.D. Josephson}
}

@article{Davydova22,
author = {Margarita Davydova  and Saranesh Prembabu  and Liang Fu },
title = {Universal Josephson diode effect},
journal = {Science Advances},
volume = {8},
number = {23},
pages = {eabo0309},
year = {2022},
doi = {10.1126/sciadv.abo0309}
}

@article{Lyu21,
author = {Lyu, Yang-Yang and Jiang, Ji and Wang, Yong-Lei and Xiao, Zhi-Li and Dong, Sining and Chen, Qing-Hu and Milošević, Milorad and Wang, Huabing and Divan, Ralu and Pearson, John and Wu, Ph and Peeters, Francois and Kwok, Wai-Kwong},
year = {2021},
month = {05},
pages = {2703},
title = {Superconducting diode effect via conformal-mapped nanoholes},
volume = {12},
journal = {Nature Communications},
}

@article{Baumgartner22,
author = {Baumgartner, Christian and Fuchs, Lorenz and Costa, Andreas and Reinhardt, Simon and Gronin, S.V. and Gardner, Geoffrey and Lindemann, Tyler and Manfra, M.J. and Faria Junior, Paulo and Kochan, Denis and Fabian, Jaroslav and Paradiso, Nicola and Strunk, Christoph},
year = {2022},
month = {01},
pages = {39–44},
title = {Supercurrent rectification and magnetochiral effects in symmetric Josephson junctions},
volume = {17},
journal = {Nature Nanotechnology},
doi = {10.1038/s41565-021-01009-9}
}

@article{yokoyama14,
  title={Anomalous Josephson effect induced by spin-orbit interaction and Zeeman effect in semiconductor nanowires},
  author={Yokoyama, Tomohiro and Eto, Mikio and Nazarov, Yuli V},
  journal={Physical Review B},
  volume={89},
  number={19},
  pages={195407},
  year={2014},
  publisher={APS}
}

@article{dolcini15,
  title={Topological Josephson $\phi$ 0 junctions},
  author={Dolcini, Fabrizio and Houzet, Manuel and Meyer, Julia S},
  journal={Physical Review B},
  volume={92},
  number={3},
  pages={035428},
  year={2015},
  publisher={APS}
}

@article{chen18,
  title={Asymmetric Josephson effect in inversion symmetry breaking topological materials},
  author={Chen, Chui-Zhen and He, James Jun and Ali, Mazhar N and Lee, Gil-Ho and Fong, Kin Chung and Law, Kam Tuen},
  journal={Physical Review B},
  volume={98},
  number={7},
  pages={075430},
  year={2018},
  publisher={APS}
}

@article{Tanaka22d-wave,
  title = {Theory of giant diode effect in $d$-wave superconductor junctions on the surface of a topological insulator},
  author = {Tanaka, Yukio and Lu, Bo and Nagaosa, Naoto},
  journal = {Phys. Rev. B},
  volume = {106},
  issue = {21},
  pages = {214524},
  numpages = {13},
  year = {2022},
  month = {Dec},
  publisher = {American Physical Society},
  doi = {10.1103/PhysRevB.106.214524}
}

@article{Kopasov21,
  title = {Geometry controlled superconducting diode and anomalous Josephson effect triggered by the topological phase transition in curved proximitized nanowires},
  author = {Kopasov, A. A. and Kutlin, A. G. and Mel'nikov, A. S.},
  journal = {Phys. Rev. B},
  volume = {103},
  issue = {14},
  pages = {144520},
  numpages = {13},
  year = {2021},
  month = {Apr},
  publisher = {American Physical Society},
  doi = {10.1103/PhysRevB.103.144520}
}

@article{Strambini22,
author = {Strambini, E. and Spies, M. and Ligato, Nadia and Ilić, S. and Rouco, M. and González Orellana, Carmen and Ilyn, Maxim and Rogero, Celia and Bergeret, F. and Moodera, J. and Virtanen, P. and Heikkila, T. and Giazotto, Francesco},
year = {2022},
month = {05},
pages = {2431},
title = {Superconducting spintronic tunnel diode},
volume = {13},
journal = {Nature Communications},
doi = {10.1038/s41467-022-29990-2}
}

@article{Zhai22,
  title = {Prediction of ferroelectric superconductors with reversible superconducting diode effect},
  author = {Zhai, Baoxing and Li, Bohao and Wen, Yao and Wu, Fengcheng and He, Jun},
  journal = {Phys. Rev. B},
  volume = {106},
  issue = {14},
  pages = {L140505},
  numpages = {7},
  year = {2022},
  month = {Oct},
  publisher = {American Physical Society},
  doi = {10.1103/PhysRevB.106.L140505}
}

@article{yang25,
  title={Field-resilient supercurrent diode in a multiferroic Josephson junction},
  author={Yang, Hung-Yu and Cuozzo, Joseph J and Bokka, Anand Johnson and Qiu, Gang and Eckberg, Christopher and Lyu, Yanfeng and Huyan, Shuyuan and Chu, Ching-Wu and Watanabe, Kenji and Taniguchi, Takashi and others},
  journal={Nature communications},
  volume={16},
  number={1},
  pages={9287},
  year={2025},
  publisher={Nature Publishing Group UK London}
}

@book{plattner12,
  title={In-memory data management: technology and applications},
  author={Plattner, Hasso and Zeier, Alexander},
  year={2012},
  publisher={Springer Science \& Business Media}
}

@article{golod15,
  title={Single Abrikosov vortices as quantized information bits},
  author={Golod, Taras and Iovan, Adrian and Krasnov, Vladimir M},
  journal={Nature communications},
  volume={6},
  number={1},
  pages={8628},
  year={2015},
  publisher={Nature Publishing Group UK London}
}

@article{Golod22,
author = {Taras, Golod and Vladimir, M. Krasnov},
year = {2022},
month = {05},
pages = {3658},
title = {Demonstration of a superconducting diode-with-memory, operational at zero magnetic field with switchable nonreciprocity},
volume = {13},
journal = {Nature Communications},
doi = {10.48550/arXiv.2205.12196}
}

@article{Wu22,
author = {Wu, Heng and Wang, Yaojia and Xu, Yuanfeng and Sivakumar, Pranava and Pasco, Chris and Filippozzi, Ulderico and Parkin, Stuart and Zeng, Yujia and McQueen, Tyrel and Ali, Mazhar},
year = {2022},
month = {04},
pages = {653-656},
title = {The field-free Josephson diode in a van der Waals heterostructure},
volume = {604},
journal = {Nature},
doi = {10.1038/s41586-022-04504-8}
}

@article{khan18,
  title={Novel magnetic burn-in for retention and magnetic tolerance testing of STTRAM},
  author={Khan, Mohammad Nasim Imtiaz and Iyengar, Anirudh S and Ghosh, Swaroop},
  journal={IEEE Transactions on Very Large Scale Integration (VLSI) Systems},
  volume={26},
  number={8},
  pages={1508--1517},
  year={2018},
  publisher={IEEE}
}

@article{gunzler25,
  title={Spin environment of a superconducting qubit in high magnetic fields},
  author={G{\"u}nzler, Simon and Beck, Johannes and Rieger, Dennis and Gosling, Nicolas and Zapata, Nicolas and Field, Mitchell and Geisert, Simon and Bacher, Andreas and Hohmann, Judith K and Spiecker, Martin and others},
  journal={Nature Communications},
  volume={16},
  number={1},
  pages={9564},
  year={2025},
  publisher={Nature Publishing Group UK London}
}

@article{gao24,
  title={An antiferromagnetic diode effect in even-layered MnBi2Te4},
  author={Gao, Anyuan and Chen, Shao-Wen and Ghosh, Barun and Qiu, Jian-Xiang and Liu, Yu-Fei and Onishi, Yugo and Hu, Chaowei and Qian, Tiema and B{\'e}rub{\'e}, Damien and Dinh, Thao and others},
  journal={Nature Electronics},
  pages={1--9},
  year={2024},
  publisher={Nature Publishing Group UK London}
}

@article{Legg22-nano,
author = {Legg, Henry and Rößler, Matthias and Münning, Felix and Fan, Dingxun and Breunig, Oliver and Bliesener, Andrea and Lippertz, Gertjan and Uday, Anjana and Taskin, A. and Loss, Daniel and Klinovaja, Jelena and Ando, Yoichi},
year = {2022},
month = {07},
pages = {696–700},
title = {Giant magnetochiral anisotropy from quantum-confined surface states of topological insulator nanowires},
volume = {17},
journal = {Nature Nanotechnology},
doi = {10.1038/s41565-022-01124-1}
}

@article{Takasan22,
  title = {Supercurrent-induced topological phase transitions},
  author = {Takasan, Kazuaki and Sumita, Shuntaro and Yanase, Youichi},
  journal = {Phys. Rev. B},
  volume = {106},
  issue = {1},
  pages = {014508},
  numpages = {14},
  year = {2022},
  month = {Jul},
  publisher = {American Physical Society},
  doi = {10.1103/PhysRevB.106.014508}
}

@article{Ye22,
author = {Ye, Chen and Xie, Xiangnan and Lv, Wenxing and Huang, Ke and Yang, Allen Jian and Jiang, Sicong and Liu, Xue and Zhu, Dapeng and Qiu, Xuepeng and Tong, Mingyu and Zhou, Tong and Hsu, Chuang-Han and Chang, Guoqing and Lin, Hsin and Li, Peisen and Yang, Kesong and Wang, Zhenyu and Jiang, Tian and Renshaw Wang, Xiao},
title = {Nonreciprocal Transport in a Bilayer of MnBi2Te4 and Pt},
journal = {Nano Letters},
volume = {22},
number = {3},
pages = {1366-1373},
year = {2022},
doi = {10.1021/acs.nanolett.1c04756}
}

@article{Legg22,
  title = {Superconducting diode effect due to magnetochiral anisotropy in topological insulators and Rashba nanowires},
  author = {Legg, Henry F. and Loss, Daniel and Klinovaja, Jelena},
  journal = {Phys. Rev. B},
  volume = {106},
  issue = {10},
  pages = {104501},
  numpages = {8},
  year = {2022},
  month = {Sep},
  publisher = {American Physical Society},
  doi = {10.1103/PhysRevB.106.104501}
}

@article{masuko22,
  title={Nonreciprocal charge transport in topological superconductor candidate Bi2Te3/PdTe2 heterostructure},
  author={Masuko, Makoto and Kawamura, Minoru and Yoshimi, Ryutaro and Hirayama, Motoaki and Ikeda, Yuya and Watanabe, Ryota and He, James Jun and Maryenko, Denis and Tsukazaki, Atsushi and Takahashi, Kei S and others},
  journal={npj Quantum Materials},
  volume={7},
  number={104},
  pages={1--7},
  year={2022},
  publisher={Nature Publishing Group}
}

@article{yasuda16,
  title={Large unidirectional magnetoresistance in a magnetic topological insulator},
  author={Yasuda, Kenji and Tsukazaki, Atsushi and Yoshimi, Ryutaro and Takahashi, KS and Kawasaki, Masashi and Tokura, Yoshinori},
  journal={Physical review letters},
  volume={117},
  number={12},
  pages={127202},
  year={2016},
  publisher={APS}
}

@article{he18,
  title={Bilinear magnetoelectric resistance as a probe of three-dimensional spin texture in topological surface states},
  author={He, Pan and Zhang, Steven S-L and Zhu, Dapeng and Liu, Yang and Wang, Yi and Yu, Jiawei and Vignale, Giovanni and Yang, Hyunsoo},
  journal={Nature Physics},
  volume={14},
  number={5},
  pages={495--499},
  year={2018},
  publisher={Nature Publishing Group UK London}
}

@article{yasuda20,
  title={Large non-reciprocal charge transport mediated by quantum anomalous Hall edge states},
  author={Yasuda, Kenji and Morimoto, Takahiro and Yoshimi, Ryutaro and Mogi, Masataka and Tsukazaki, Atsushi and Kawamura, Minoru and Takahashi, Kei S and Kawasaki, Masashi and Nagaosa, Naoto and Tokura, Yoshinori},
  journal={Nature Nanotechnology},
  volume={15},
  number={10},
  pages={831--835},
  year={2020},
  publisher={Nature Publishing Group UK London}
}

@article{Zhaowei22,
author = {Zhang, Zhaowei and Wang, Naizhou and Cao, Ning and Wang, Aifeng and Zhou, Xiaoyuan and Watanabe, Kenji and Taniguchi, Takashi and Yan, Binghai and Gao, Wei-bo},
year = {2022},
month = {03},
pages = {6191},
title = {Non-reciprocal charge transport in an intrinsic magnetic topological insulator MnBi2Te4},
journal = {Nature Communications}
}

@article{li24-NM,
  title={Observation of giant non-reciprocal charge transport from quantum Hall states in a topological insulator},
  author={Li, Chunfeng and Wang, Rui and Zhang, Shuai and Qin, Yuyuan and Ying, Zhe and Wei, Boyuan and Dai, Zheng and Guo, Fengyi and Chen, Wei and Zhang, Rong and others},
  journal={Nature Materials},
  volume={23},
  number={9},
  pages={1208--1213},
  year={2024},
  publisher={Nature Publishing Group UK London}
}

@book{mei24,
  title={Linear and Nonlinear Responses in Magnetic Topological Materials},
  author={Mei, Ruobing},
  year={2024},
  publisher={The Pennsylvania State University}
}

@article{morimoto16,
  title={Chiral anomaly and giant magnetochiral anisotropy in noncentrosymmetric Weyl semimetals},
  author={Morimoto, Takahiro and Nagaosa, Naoto},
  journal={Physical review letters},
  volume={117},
  number={14},
  pages={146603},
  year={2016},
  publisher={APS}
}

@article{wang22,
  title={Gigantic magnetochiral anisotropy in the topological semimetal ZrTe 5},
  author={Wang, Yongjian and Legg, Henry F and B{\"o}merich, Thomas and Park, Jinhong and Biesenkamp, Sebastian and Taskin, AA and Braden, Markus and Rosch, Achim and Ando, Yoichi},
  journal={Physical Review Letters},
  volume={128},
  number={17},
  pages={176602},
  year={2022},
  publisher={APS}
}

@article{wang24-PRL,
  title={Spontaneous Inversion Symmetry Breaking and Emergence of Berry Curvature and Orbital Magnetization in Topological ZrTe 5 Films},
  author={Wang, Erqing and Zeng, Hui and Duan, Wenhui and Huang, Huaqing},
  journal={Physical Review Letters},
  volume={132},
  number={26},
  pages={266802},
  year={2024},
  publisher={APS}
}

@article{Pal22,
author = {Pal, Banabir and Chakraborty, Anirban and Sivakumar, Pranava and Davydova, Margarita and Gopi, Ajesh and Pandeya, Avanindra and Krieger, Jonas and Zhang, Yang and Date, Mihir and Ju, Sailong and Yuan, Noah and Schröter, Niels and Fu, Liang and Parkin, Stuart},
year = {2022},
month = {08},
pages = {1228–1233},
title = {Josephson diode effect from Cooper pair momentum in a topological semimetal},
volume = {18},
journal = {Nature Physics},
doi = {10.1038/s41567-022-01699-5}
}

@article{Lu21,
  title = {Full electric-field tuning of the nonreciprocal transport effect in massive chiral fermions with trigonal warping},
  author = {Lu, Elisha Cho-Hao and Cheng, Cheng-Tung and Li, Liang and Lee, Wei-Li},
  journal = {Phys. Rev. Research},
  volume = {3},
  issue = {3},
  pages = {033160},
  numpages = {15},
  year = {2021},
  month = {Aug},
  publisher = {American Physical Society},
  doi = {10.1103/PhysRevResearch.3.033160}
}

@article{Wei22,
  title = {Supercurrent rectification effect in graphene-based Josephson junctions},
  author = {Wei, Ya-Jun and Liu, Han-Lin and Wang, J. and Liu, Jun-Feng},
  journal = {Phys. Rev. B},
  volume = {106},
  issue = {16},
  pages = {165419},
  numpages = {7},
  year = {2022},
  month = {Oct},
  publisher = {American Physical Society},
  doi = {10.1103/PhysRevB.106.165419}
}

@article{Lin22,
author = {Lin, Jiang-Xiazi and Siriviboon, Phum and Scammell, Harley and Liu, Song and Rhodes, Daniel and Watanabe, K. and Taniguchi, Takashi and Hone, James and Scheurer, Mathias and Li, J.},
year = {2022},
month = {08},
pages = {1221–1227},
title = {Zero-field superconducting diode effect in small-twist-angle trilayer graphene},
volume = {18},
journal = {Nature Physics},
doi = {10.1038/s41567-022-01700-1}
}

@article{Scammell22,
doi = {10.1088/2053-1583/ac5b16},
year = {2022},
month = {mar},
publisher = {IOP Publishing},
volume = {9},
number = {2},
pages = {025027},
author = {Harley D Scammell and J I A Li and Mathias S Scheurer},
title = {Theory of zero-field superconducting diode effect in twisted trilayer graphene},
journal = {2D Materials}
}

@article{Vries21,
author = {de Vries, Folkert and Portolés, Elías and Zheng, Giulia and Taniguchi, Takashi and Watanabe, Kenji and Ihn, Thomas and Ensslin, Klaus and Rickhaus, Peter},
year = {2021},
month = {07},
pages = {760–763},
title = {Gate-defined Josephson junctions in magic-angle twisted bilayer graphene},
volume = {16},
journal = {Nature Nanotechnology},
doi = {10.1038/s41565-021-00896-2}
}

@article{diez23,
  title={Symmetry-broken Josephson junctions and superconducting diodes in magic-angle twisted bilayer graphene},
  author={D{\'\i}ez-M{\'e}rida, J and D{\'\i}ez-Carl{\'o}n, A and Yang, SY and Xie, Y-M and Gao, X-J and Senior, J and Watanabe, K and Taniguchi, T and Lu, X and Higginbotham, AP and others},
  journal={Nature Communications},
  volume={14},
  number={1},
  pages={2396},
  year={2023},
  publisher={Nature Publishing Group UK London}
}

@article{han25,
  title={Signatures of chiral superconductivity in rhombohedral graphene},
  author={Han, Tonghang and Lu, Zhengguang and Hadjri, Zach and Shi, Lihan and Wu, Zhenghan and Xu, Wei and Yao, Yuxuan and Cotten, Armel A and Sharifi Sedeh, Omid and Weldeyesus, Henok and others},
  journal={Nature},
  volume={643},
  number={8072},
  pages={654--661},
  year={2025},
  publisher={Nature Publishing Group UK London}
}

@article{chen25,
  title={Intrinsic superconducting diode effect and nonreciprocal superconductivity in rhombohedral graphene multilayers},
  author={Chen, Yinqi and Scheurer, Mathias S and Schrade, Constantin},
  journal={Physical Review B},
  volume={112},
  number={6},
  pages={L060505},
  year={2025},
  publisher={APS}
}

@article{wang25,
  title={Theoretical study of superconducting diode effect in planar Td-MoTe2 Josephson junctions},
  author={Wang, Gong-Qi and Miao, Jian-Jian and Chen, Wei-Qiang},
  journal={Physical Review B},
  volume={112},
  number={1},
  pages={014508},
  year={2025},
  publisher={APS}
}

@article{Tokura18,
author = {Yoshinori, Tokura and Naoto, Nagaosa},
year = {2018},
month = {09},
pages = {3740},
title = {Nonreciprocal responses from non-centrosymmetric quantum materials},
volume = {9},
journal = {Nature communications},
doi = {10.1038/s41467-018-05759-4}
}

@article{Toshiya21,
author = {Ideue, Toshiya and Iwasa, Yoshihiro},
title = {Symmetry Breaking and Nonlinear Electric Transport in van der Waals Nanostructures},
journal = {Annual Review of Condensed Matter Physics},
volume = {12},
number = {1},
pages = {201-223},
year = {2021},
doi = {10.1146/annurev-conmatphys-060220-100347}
}

@article{isobe20,
  title={High-frequency rectification via chiral Bloch electrons},
  author={Isobe, Hiroki and Xu, Su-Yang and Fu, Liang},
  journal={Science advances},
  volume={6},
  number={13},
  pages={eaay2497},
  year={2020},
  publisher={American Association for the Advancement of Science}
}

@article{Rikken01,
  title = {Electrical Magnetochiral Anisotropy},
  author = {Rikken, G. L. J. A. and F\"olling, J. and Wyder, P.},
  journal = {Phys. Rev. Lett.},
  volume = {87},
  issue = {23},
  pages = {236602},
  numpages = {4},
  year = {2001},
  month = {Nov},
  publisher = {American Physical Society},
  doi = {10.1103/PhysRevLett.87.236602}
}

@article{Krstic02,
author = {Krstić,V.  and Roth,S.  and Burghard,M.  and Kern,K.  and Rikken,G. L. J. A. },
title = {Magneto-chiral anisotropy in charge transport through single-walled carbon nanotubes},
journal = {The Journal of Chemical Physics},
volume = {117},
number = {24},
pages = {11315-11319},
year = {2002},
doi = {10.1063/1.1523895}
}

@article{Pop14,
author = {Pop, Flavia and Auban-Senzier, Pascale and Canadell, Enric and Rikken, Geert and Avarvari, Narcis},
year = {2014},
month = {05},
pages = {3757},
title = {Electrical magnetochiral anisotropy in a bulk chiral molecular conductor},
volume = {5},
journal = {Nature communications},
doi = {10.1038/ncomms4757}
}

@article{Rikken05,
  title = {Magnetoelectric Anisotropy in Diffusive Transport},
  author = {Rikken, G. L. J. A. and Wyder, P.},
  journal = {Phys. Rev. Lett.},
  volume = {94},
  issue = {1},
  pages = {016601},
  numpages = {4},
  year = {2005},
  month = {Jan},
  publisher = {American Physical Society},
  doi = {10.1103/PhysRevLett.94.016601}
}

@article{Ideue17,
author = {Ideue, Toshiya and Hamamoto, K. and Koshikawa, S. and Ezawa, M. and Shimizu, Sunao and Kaneko, Yoshio and Tokura, Y. and Nagaosa, N. and Iwasa, Y.},
year = {2017},
month = {03},
pages = {578–583},
title = {Bulk rectification effect in a polar semiconductor},
volume = {13},
journal = {Nature Physics},
doi = {10.1038/nphys4056}
}

@article{Kenji19,
title={Nonreciprocal charge transport at topological insulator/superconductor interface},
  author={Yasuda, Kenji and Yasuda, Hironori and Liang, Tian and Yoshimi, Ryutaro and Tsukazaki, Atsushi and Takahashi, Kei S and Nagaosa, Naoto and Kawasaki, Masashi and Tokura, Yoshinori},
  journal={Nature communications},
  volume={10},
  number={1},
  pages={2734},
  year={2019},
  publisher={Nature Publishing Group UK London}
}

@article{Wickramaratne20,
  title = {Ising Superconductivity and Magnetism in ${\mathrm{NbSe}}_{2}$},
  author = {Wickramaratne, Darshana and Khmelevskyi, Sergii and Agterberg, Daniel F. and Mazin, I. I.},
  journal = {Phys. Rev. X},
  volume = {10},
  issue = {4},
  pages = {041003},
  numpages = {13},
  year = {2020},
  month = {Oct},
  publisher = {American Physical Society},
  doi = {10.1103/PhysRevX.10.041003}
}

@article{Zhang20,
author = {Zhang, Enze and Xu, Xian and Zou, Yichao and Ai, Linfeng and Dong, Xiang and Huang, Ce and Leng, Pengliang and Liu, Shanshan and Zhang, Yuda and Jia, Zehao and Peng, Xinyue and Zhao, Minhao and Yang, Yunkun and Li, Zihan and Guo, Hangwen and Haigh, Sarah and Nagaosa, Naoto and Shen, Jian and Xiu, Faxian},
year = {2020},
month = {11},
pages = {5634},
title = {Nonreciprocal superconducting NbSe2 antenna},
volume = {11},
journal = {Nature Communications},
doi = {10.1038/s41467-020-19459-5}
}

@article{Noah21,
  title={Topological metals and finite-momentum superconductors},
  author={Yuan, Noah FQ and Fu, Liang},
  journal={Proceedings of the National Academy of Sciences},
  volume={118},
  number={3},
  pages={e2019063118},
  year={2021},
  publisher={National Acad Sciences}
}

@article{Qin17,
author = {Qin, F. and Shi, W. and Ideue, Toshiya and Yoshida, M. and Zak, A. and Tenne, Reshef and Kikitsu, T. and Inoue, D. and Hashizume, D. and Iwasa, Y.},
year = {2017},
month = {02},
pages = {14465},
title = {Superconductivity in a chiral nanotube},
volume = {8},
journal = {Nature Communications},
doi = {10.1038/ncomms14465}
}

@article{Zinkl22,
  title = {Symmetry conditions for the superconducting diode effect in chiral superconductors},
  author = {Zinkl, Bastian and Hamamoto, Keita and Sigrist, Manfred},
  journal = {Phys. Rev. Research},
  volume = {4},
  issue = {3},
  pages = {033167},
  numpages = {12},
  year = {2022},
  month = {Aug},
  publisher = {American Physical Society},
  doi = {10.1103/PhysRevResearch.4.033167}
}

@article{Kaneyasu10,
author = {Kaneyasu ,Hirono and Hayashi ,Nobuhiko and Gut ,Bruno and Makoshi ,Kenji and Sigrist ,Manfred},
title = {Phase Transition in the 3-Kelvin Phase of Eutectic Sr2RuO4–Ru},
journal = {Journal of the Physical Society of Japan},
volume = {79},
number = {10},
pages = {104705},
year = {2010},
doi = {10.1143/JPSJ.79.104705}
}

@article{Hooper04,
  title = {Anomalous Josephson network in the $\mathrm{Ru}\text{\penalty1000-\hskip0pt}{\mathrm{Sr}}_{2}{\mathrm{Ru}\mathrm{O}}_{4}$ eutectic system},
  author = {Hooper, J. and Mao, Z. Q. and Nelson, K. D. and Liu, Y. and Wada, M. and Maeno, Y.},
  journal = {Phys. Rev. B},
  volume = {70},
  issue = {1},
  pages = {014510},
  numpages = {4},
  year = {2004},
  month = {Jul},
  publisher = {American Physical Society},
  doi = {10.1103/PhysRevB.70.014510}
}

@article{rikken19,
  title={Strong electrical magnetochiral anisotropy in tellurium},
  author={Rikken, GLJA and Avarvari, N},
  journal={Physical Review B},
  volume={99},
  number={24},
  pages={245153},
  year={2019},
  publisher={APS}
}

@article{pal19,
  title={Quantized Josephson phase battery},
  author={Pal, Subhajit and Benjamin, Colin},
  journal={Europhysics Letters},
  volume={126},
  number={5},
  pages={57002},
  year={2019},
  publisher={EDP Sciences, IOP Publishing and Societ{\`a} Italiana di Fisica}
}

@article{strambini20,
  title={A Josephson phase battery},
  author={Strambini, Elia and Iorio, Andrea and Durante, Ofelia and Citro, Roberta and Sanz-Fern{\'a}ndez, Cristina and Guarcello, Claudio and Tokatly, Ilya V and Braggio, Alessandro and Rocci, Mirko and Ligato, Nadia and others},
  journal={Nature Nanotechnology},
  volume={15},
  number={8},
  pages={656--660},
  year={2020},
  publisher={Nature Publishing Group}
}

@article{bergeret15,
  title={Theory of diffusive $\varphi$ 0 Josephson junctions in the presence of spin-orbit coupling},
  author={Bergeret, FS and Tokatly, IV},
  journal={Europhysics Letters},
  volume={110},
  number={5},
  pages={57005},
  year={2015},
  publisher={EDP Sciences, IOP Publishing and Societ{\`a} Italiana di Fisica}
}

@article{campaioli24-RMP,
  title={Colloquium: quantum batteries},
  author={Campaioli, Francesco and Gherardini, Stefano and Quach, James Q and Polini, Marco and Andolina, Gian Marcello},
  journal={Reviews of Modern Physics},
  volume={96},
  number={3},
  pages={031001},
  year={2024},
  publisher={APS}
}

@article{ferraro26-NRP,
  title={Opportunities and challenges of quantum batteries},
  author={Ferraro, Dario and Cavaliere, Fabio and Genoni, Marco G and Benenti, Giuliano and Sassetti, Maura},
  journal={Nature Reviews Physics},
  pages={1--13},
  year={2026},
  publisher={Nature Publishing Group}
}

@article{kurman26,
  title={Powering Quantum Computation with Quantum Batteries},
  author={Kurman, Yaniv and Hymas, Kieran and Fedorov, Arkady and Munro, William J and Quach, James},
  journal={Physical Review X},
  volume={16},
  number={1},
  pages={011016},
  year={2026},
  publisher={APS}
}

@article{quach23-joul,
  title={Quantum batteries: The future of energy storage?},
  author={Quach, James Q and Cerullo, Giulio and Virgili, Tersilla},
  journal={Joule},
  volume={7},
  number={10},
  pages={2195--2200},
  year={2023},
  publisher={Elsevier}
}

@article{camposeo25-AM,
  title={Quantum batteries: A materials science perspective},
  author={Camposeo, Andrea and Virgili, Tersilla and Lombardi, Floriana and Cerullo, Giulio and Pisignano, Dario and Polini, Marco},
  journal={Advanced Materials},
  volume={37},
  number={17},
  pages={2415073},
  year={2025},
  publisher={Wiley Online Library}
}

@article{ferraro18,
  title={High-power collective charging of a solid-state quantum battery},
  author={Ferraro, Dario and Campisi, Michele and Andolina, Gian Marcello and Pellegrini, Vittorio and Polini, Marco},
  journal={Physical review letters},
  volume={120},
  number={11},
  pages={117702},
  year={2018},
  publisher={APS}
}

@article{dicke54,
  title={Coherence in spontaneous radiation processes},
  author={Dicke, Robert H},
  journal={Physical review},
  volume={93},
  number={1},
  pages={99},
  year={1954},
  publisher={APS}
}

@article{andolina19,
  title={Quantum versus classical many-body batteries},
  author={Andolina, Gian Marcello and Keck, Maximilian and Mari, Andrea and Giovannetti, Vittorio and Polini, Marco},
  journal={Physical Review B},
  volume={99},
  number={20},
  pages={205437},
  year={2019},
  publisher={APS}
}

@article{andolina18,
  title={Charger-mediated energy transfer in exactly solvable models for quantum batteries},
  author={Andolina, Gian Marcello and Farina, Donato and Mari, Andrea and Pellegrini, Vittorio and Giovannetti, Vittorio and Polini, Marco},
  journal={Physical Review B},
  volume={98},
  number={20},
  pages={205423},
  year={2018},
  publisher={APS}
}

@article{farina19,
  title={Charger-mediated energy transfer for quantum batteries: An open-system approach},
  author={Farina, Donato and Andolina, Gian Marcello and Mari, Andrea and Polini, Marco and Giovannetti, Vittorio},
  journal={Physical Review B},
  volume={99},
  number={3},
  pages={035421},
  year={2019},
  publisher={APS}
}

@article{rossini20,
  title={Quantum advantage in the charging process of Sachdev-Ye-Kitaev batteries},
  author={Rossini, Davide and Andolina, Gian Marcello and Rosa, Dario and Carrega, Matteo and Polini, Marco},
  journal={Physical Review Letters},
  volume={125},
  number={23},
  pages={236402},
  year={2020},
  publisher={APS}
}

@article{rosenhaus19,
  title={An introduction to the SYK model},
  author={Rosenhaus, Vladimir},
  journal={Journal of Physics A: Mathematical and Theoretical},
  volume={52},
  number={32},
  pages={323001},
  year={2019},
  publisher={IOP Publishing}
}

@article{chew17,
  title={Approximating the sachdev-ye-kitaev model with majorana wires},
  author={Chew, Aaron and Essin, Andrew and Alicea, Jason},
  journal={Physical Review B},
  volume={96},
  number={12},
  pages={121119},
  year={2017},
  publisher={APS}
}

@article{pikulin17,
  title={Black hole on a chip: Proposal for a physical realization of the Sachdev-Ye-Kitaev model in a solid-state system},
  author={Pikulin, DI and Franz, Marcel},
  journal={Physical Review X},
  volume={7},
  number={3},
  pages={031006},
  year={2017},
  publisher={APS}
}

@article{julia20,
  title={Bounds on the capacity and power of quantum batteries},
  author={Juli{\`a}-Farr{\'e}, Sergi and Salamon, Tymoteusz and Riera, Arnau and Bera, Manabendra N and Lewenstein, Maciej},
  journal={Physical Review Research},
  volume={2},
  number={2},
  pages={023113},
  year={2020},
  publisher={APS}
}

@article{mandelstam45,
  title={The uncertainty relation between energy and time in nonrelativistic quantum mechanics},
  author={Mandelstam, Leonid and Tamm, Ig},
  journal={J. Phys. USSR},
  volume={9},
  pages={249},
  year={1945}
}

@article{margolus98,
  title={The maximum speed of dynamical evolution},
  author={Margolus, Norman and Levitin, Lev B},
  journal={Physica D: Nonlinear Phenomena},
  volume={120},
  number={1-2},
  pages={188--195},
  year={1998},
  publisher={Elsevier}
}

@article{giovannetti03,
  title={The role of entanglement in dynamical evolution},
  author={Giovannetti, Vittorio and Lloyd, Seth and Maccone, Lorenzo},
  journal={EPL (Europhysics Letters)},
  volume={62},
  number={5},
  pages={615--621},
  year={2003}
}

@article{deffner17,
  title={Quantum speed limits: from Heisenberg’s uncertainty principle to optimal quantum control},
  author={Deffner, Sebastian and Campbell, Steve},
  journal={Journal of Physics A: Mathematical and Theoretical},
  volume={50},
  number={45},
  pages={453001},
  year={2017},
  publisher={IOP Publishing}
}

@article{andolina25,
  title={Genuine quantum advantage in anharmonic bosonic quantum batteries},
  author={Andolina, Gian Marcello and Stanzione, Vittoria and Giovannetti, Vittorio and Polini, Marco},
  journal={Physical Review Letters},
  volume={134},
  number={24},
  pages={240403},
  year={2025},
  publisher={APS}
}

@article{armour13,
  title={Universal quantum fluctuations of a cavity mode driven by a Josephson junction},
  author={Armour, AD and Blencowe, MP and Brahimi, E and Rimberg, AJ},
  journal={Physical review letters},
  volume={111},
  number={24},
  pages={247001},
  year={2013},
  publisher={APS}
}

@article{downing25,
  title={Energy storage in a continuous-variable quantum battery with nonlinear coupling},
  author={Downing, CA and Ukhtary, MS},
  journal={Physical Review E},
  volume={112},
  number={4},
  pages={044143},
  year={2025},
  publisher={APS}
}

@article{ahmadi24,
  title={Nonreciprocal quantum batteries},
  author={Ahmadi, Borhan and Mazurek, Pawe{\l} and Horodecki, Pawe{\l} and Barzanjeh, Shabir},
  journal={Physical Review Letters},
  volume={132},
  number={21},
  pages={210402},
  year={2024},
  publisher={APS}
}

@article{khan25,
  title={Collective enhancement in nonreciprocal multimode quantum batteries},
  author={Khan, Niaz Ali and Zhang, Xingyu and Huang, Chenlong and Liu, Yue and He, Dahai},
  journal={Physical Review B},
  volume={112},
  number={10},
  pages={104318},
  year={2025},
  publisher={APS}
}

@article{guo25,
  title={Nonreciprocal open quantum battery network in a photonic waveguide array},
  author={Guo, Yaowu and Cao, Lianzhen and Zhao, Jiaqiang},
  journal={Physical Review A},
  volume={111},
  number={6},
  pages={063520},
  year={2025},
  publisher={APS}
}

@article{sun25-NRQB,
  title={Nonreciprocal charging in a quantum battery via a mediator},
  author={Sun, Cheng-Ze and Wang, Zai-Kun and Yan, Wei-Bin and Zhang, Ying-Jie and Man, Zhong-Xiao and Cai, Qing-Yu},
  journal={Physical Review A},
  volume={112},
  number={1},
  pages={012429},
  year={2025},
  publisher={APS}
}

@article{zhao25-NRQB,
  title={Enhanced charging in multibattery systems by nonreciprocity},
  author={Zhao, Hua-Wei and Xie, Yong and Huang, Xinyao and Zhang, Guo-Feng},
  journal={Physical Review A},
  volume={112},
  number={2},
  pages={022214},
  year={2025},
  publisher={APS}
}

@article{zafar26,
  title={Loss-Induced Nonreciprocal Quantum Battery},
  author={Zafar, Muhammad Zaeem and Irfan, Muhammad},
  journal={Advanced Quantum Technologies},
  volume={9},
  number={2},
  pages={e00845},
  year={2026},
  publisher={Wiley Online Library}
}

@article{lin26,
  title={Enhanced charging power in nonreciprocal quantum battery by reservoir engineering},
  author={Lin, Qi-Yin and Ye, Guang-Zheng and Li, Can and Su, Wan-Jun and Wu, Huai-Zhi},
  journal={Physica Scripta},
  volume={101},
  number={2},
  pages={025104},
  year={2026},
  publisher={IOP Publishing}
}

@article{hu22,
  title={Optimal charging of a superconducting quantum battery},
  author={Hu, Chang-Kang and Qiu, Jiawei and Souza, Paulo JP and Yuan, Jiahao and Zhou, Yuxuan and Zhang, Libo and Chu, Ji and Pan, Xianchuang and Hu, Ling and Li, Jian and others},
  journal={Quantum Science \& Technology},
  volume={7},
  number={4},
  pages={045018},
  year={2022},
  publisher={IOP Publishing}
}

@article{gemme22-ibm,
  title={IBM quantum platforms: A quantum battery perspective},
  author={Gemme, Giulia and Grossi, Michele and Ferraro, Dario and Vallecorsa, Sofia and Sassetti, Maura},
  journal={Batteries},
  volume={8},
  number={5},
  pages={43},
  year={2022},
  publisher={MDPI}
}

@article{dou23,
  title={Superconducting transmon qubit-resonator quantum battery},
  author={Dou, Fu-Quan and Yang, Fang-Mei},
  journal={Physical Review A},
  volume={107},
  number={2},
  pages={023725},
  year={2023},
  publisher={APS}
}

@article{elghaayda25,
  title={Performance of a superconducting quantum battery},
  author={Elghaayda, Samira and Ali, Asad and Al-Kuwari, Saif and Czerwinski, Artur and Mansour, Mostafa and Haddadi, Saeed},
  journal={Advanced Quantum Technologies},
  volume={8},
  number={9},
  pages={2400651},
  year={2025},
  publisher={Wiley Online Library}
}

@article{hu26,
  title={Quantum charging advantage in superconducting solid-state batteries},
  author={Hu, Chang-Kang and Liu, Chilong and Zhao, Jingchao and Zhong, Liuzhu and Zhou, Yuxuan and Liu, Mingze and Yuan, Haolan and Lin, Yongchang and Xu, Yue and Hu, Guantian and others},
  journal={Physical Review Letters},
  volume={136},
  number={6},
  pages={060401},
  year={2026},
  publisher={APS}
}

@article{yi22crossover,
  title={Crossover from Ising-to Rashba-type superconductivity in epitaxial Bi2Se3/monolayer NbSe2 heterostructures},
  author={Yi, Hemian and Hu, Lun-Hui and Wang, Yuanxi and Xiao, Run and Cai, Jiaqi and Hickey, Danielle Reifsnyder and Dong, Chengye and Zhao, Yi-Fan and Zhou, Ling-Jie and Zhang, Ruoxi and others},
  journal={Nature Materials},
  volume={21},
  number={12},
  pages={1366--1372},
  year={2022},
  publisher={Nature Publishing Group UK London}
}

@article{haldane88,
  title={Model for a quantum Hall effect without Landau levels: Condensed-matter realization of the" parity anomaly"},
  author={Haldane, F Duncan M},
  journal={Physical review letters},
  volume={61},
  number={18},
  pages={2015},
  year={1988},
  publisher={APS}
}

@article{weng15,
  title={Quantum anomalous Hall effect and related topological electronic states},
  author={Weng, Hongming and Yu, Rui and Hu, Xiao and Dai, Xi and Fang, Zhong},
  journal={Advances in Physics},
  volume={64},
  number={3},
  pages={227--282},
  year={2015},
  publisher={Taylor \& Francis}
}

@article{liu16,
  title={The quantum anomalous Hall effect: theory and experiment},
  author={Liu, Chao-Xing and Zhang, Shou-Cheng and Qi, Xiao-Liang},
  journal={Annual Review of Condensed Matter Physics},
  volume={7},
  pages={301--321},
  year={2016},
  publisher={Annual Reviews}
}

@article{bernevig22,
  title={Progress and prospects in magnetic topological materials},
  author={Bernevig, B Andrei and Felser, Claudia and Beidenkopf, Haim},
  journal={Nature},
  volume={603},
  number={7899},
  pages={41--51},
  year={2022},
  publisher={Nature Publishing Group UK London}
}

@article{chang23,
  title={Colloquium: Quantum anomalous hall effect},
  author={Chang, Cui-Zu and Liu, Chao-Xing and MacDonald, Allan H},
  journal={Reviews of Modern Physics},
  volume={95},
  number={1},
  pages={011002},
  year={2023},
  publisher={APS}
}

@article{shabbir18,
  title={Long range intrinsic ferromagnetism in two dimensional materials and dissipationless future technologies},
  author={Shabbir, Babar and Nadeem, Muhammad and Dai, Zhigao and Fuhrer, Michael S and Xue, Qi-Kun and Wang, Xiaolin and Bao, Qiaoliang},
  journal={Applied Physics Reviews},
  volume={5},
  number={4},
  year={2018},
  publisher={AIP Publishing}
}

@article{nadeem20,
  title={Quantum Anomalous Hall Effect in Magnetic Doped Topological Insulators and Ferromagnetic Spin-Gapless Semiconductors—A Perspective Review},
  author={Nadeem, Muhammad and Hamilton, Alex R and Fuhrer, Michael S and Wang, Xiaolin},
  journal={Small},
  volume={16},
  number={42},
  pages={1904322},
  year={2020},
  publisher={Wiley Online Library}
}

@article{nadeem24,
  title={Spin gapless quantum materials and devices},
  author={Nadeem, Muhammad and Wang, Xiaolin},
  journal={Advanced Materials},
  volume={36},
  number={33},
  pages={2402503},
  year={2024},
  publisher={Wiley Online Library}
}

@article{nadeem21,
  title={Overcoming Boltzmann’s tyranny in a transistor via the topological quantum field effect},
  author={Nadeem, Muhammad and Di Bernardo, Iolanda and Wang, Xiaolin and Fuhrer, Michael S and Culcer, Dimitrie},
  journal={Nano Letters},
  volume={21},
  number={7},
  pages={3155--3161},
  year={2021},
  publisher={ACS Publications}
}

@inproceedings{fuhrer21,
  title={Proposal for a negative capacitance topological quantum field-effect transistor},
  author={Fuhrer, Michael S and Edmonds, Mark T and Culcer, Dimitrie and Nadeem, Muhammad and Wang, Xiaolin and Medhekar, Nikhil and Yin, Yuefeng and Cole, Jared H},
  booktitle={2021 IEEE International Electron Devices Meeting (IEDM)},
  pages={38--2},
  year={2021},
  organization={IEEE}
}

@article{nadeem22,
  title={Optimizing topological switching in confined 2D-Xene nanoribbons via finite-size effects},
  author={Nadeem, Muhammad and Zhang, Chao and Culcer, Dimitrie and Hamilton, Alex R and Fuhrer, Michael S and Wang, Xiaolin},
  journal={Applied Physics Reviews},
  volume={9},
  number={1},
  pages={011411},
  year={2022},
  publisher={AIP Publishing LLC}
}

@article{weber24,
  title={2024 roadmap on 2D topological insulators},
  author={Weber, Bent and Fuhrer, Michael S and Sheng, Xian-Lei and Yang, Shengyuan A and Thomale, Ronny and Shamim, Saquib and Molenkamp, Laurens W and Cobden, David and Pesin, Dmytro and Zandvliet, Harold JW and others},
  journal={Journal of Physics: Materials},
  volume={7},
  number={2},
  pages={022501},
  year={2024},
  publisher={IOP Publishing}
}

@article{kayyalha20gate,
  title={Highly skewed current--phase relation in superconductor--topological insulator--superconductor Josephson junctions},
  author={Kayyalha, Morteza and Kazakov, Aleksandr and Miotkowski, Ireneusz and Khlebnikov, Sergei and Rokhinson, Leonid P and Chen, Yong P},
  journal={npj Quantum Materials},
  volume={5},
  number={1},
  pages={1--7},
  year={2020},
  publisher={Nature Publishing Group}
}

@article{gupta22,
  title={Gate-tunable superconducting diode effect in a three-terminal Josephson device},
  author={Gupta, Mohit and Graziano, Gino V and Pendharkar, Mihir and Dong, Jason T and Dempsey, Connor P and Palmstr{\o}m, Chris and Pribiag, Vlad S},
  journal={Nature communications},
  volume={14},
  number={1},
  pages={3078},
  year={2023},
  publisher={Nature Publishing Group UK London}
}

@article{haenel22,
  title={Superconducting diode from flux biased Josephson junction arrays},
  author={Haenel, Rafael and Can, Oguzhan},
  journal={arXiv preprint arXiv:2212.02657},
  year={2022}
}

@article{amundsen24-RMP,
  title={Colloquium: Spin-orbit effects in superconducting hybrid structures},
  author={Amundsen, Morten and Linder, Jacob and Robinson, Jason WA and {\v{Z}}uti{\'c}, Igor and Banerjee, Niladri},
  journal={Reviews of Modern Physics},
  volume={96},
  number={2},
  pages={021003},
  year={2024},
  publisher={APS}
}

@article{dartiailh21,
  title={Phase signature of topological transition in Josephson junctions},
  author={Dartiailh, Matthieu C and Mayer, William and Yuan, Joseph and Wickramasinghe, Kaushini S and Matos-Abiague, Alex and {\v{Z}}uti{\'c}, Igor and Shabani, Javad},
  journal={Physical Review Letters},
  volume={126},
  number={3},
  pages={036802},
  year={2021},
  publisher={APS}
}

@article{karabassov22,
  title={Hybrid helical state and superconducting diode effect in superconductor/ferromagnet/topological insulator heterostructures},
  author={Karabassov, T and Bobkova, IV and Golubov, AA and Vasenko, AS},
  journal={Physical Review B},
  volume={106},
  number={22},
  pages={224509},
  year={2022},
  publisher={APS}
}

@article{legg23,
  title={Parity-protected superconducting diode effect in topological Josephson junctions},
  author={Legg, Henry F and Laubscher, Katharina and Loss, Daniel and Klinovaja, Jelena},
  journal={Physical Review B},
  volume={108},
  number={21},
  pages={214520},
  year={2023},
  publisher={APS}
}

@article{paolucci23,
  title={A gate-and flux-controlled supercurrent diode effect},
  author={Paolucci, F and De Simoni, G and Giazotto, F},
  journal={Applied Physics Letters},
  volume={122},
  number={4},
  pages={042601},
  year={2023},
  publisher={AIP Publishing LLC}
}

@article{alvarado23,
  title={Intrinsic nonmagnetic $\phi$ 0 Josephson junctions in twisted bilayer graphene},
  author={Alvarado, Miguel and Burset, Pablo and Yeyati, A Levy},
  journal={Physical Review Research},
  volume={5},
  number={3},
  pages={L032033},
  year={2023},
  publisher={APS}
}

@article{rothstein26,
  title={Gate-tunable Josephson diodes in magic-angle twisted bilayer graphene},
  author={Rothstein, Alexander and Dolleman, Robin Joey and Klebl, Lennart and Achtermann, Anthony and Volmer, Frank and Watanabe, Kenji and Taniguchi, Takashi and Hassler, Fabian and Banszerus, Luca and Beschoten, Bernd and others},
  journal={Nano Letters},
  volume={26},
  number={6},
  pages={2119--2128},
  year={2026},
  publisher={ACS Publications}
}

@article{antola25,
  title={Streamline-controlled rectification of supercurrent in thin-film asymmetric weak links},
  author={Antola, Filippo and Battisti, Sebastiano and Braggio, Alessandro and Giazotto, Francesco and De Simoni, Giorgio},
  journal={Physical Review Applied},
  volume={24},
  number={6},
  pages={064003},
  year={2025},
  publisher={APS}
}

@article{likharev79-RMP,
  title={Superconducting weak links},
  author={Likharev, KK},
  journal={Reviews of Modern Physics},
  volume={51},
  number={1},
  pages={101},
  year={1979},
  publisher={APS}
}

@article{nunchot24,
  title={Chiral superconducting diode effect by Dzyaloshinsky-Moriya interaction},
  author={Nunchot, Naratip and Yanase, Youichi},
  journal={Physical Review B},
  volume={109},
  number={5},
  pages={054508},
  year={2024},
  publisher={APS}
}

@article{sakamoto24,
  title={Antiferromagnetic spin-torque diode effect in a kagome Weyl semimetal},
  author={Sakamoto, Shoya and Nomoto, Takuya and Higo, Tomoya and Hibino, Yuki and Yamamoto, Tatsuya and Tamaru, Shingo and Kotani, Yoshinori and Kosaki, Hidetoshi and Shiga, Masanobu and Nishio-Hamane, Daisuke and others},
  journal={Nature Nanotechnology},
  pages={1--6},
  year={2024},
  publisher={Nature Publishing Group}
}

@article{geng23-Rev,
  title={Superconductor-ferromagnet hybrids for non-reciprocal electronics and detectors},
  author={Geng, Zhuoran and Hijano, Alberto and Ili{\'c}, Stefan and Ilyn, Maxim and Maasilta, Ilari and Monfardini, Alessandro and Spies, Maria and Strambini, Elia and Virtanen, Pauli and Calvo, Martino and others},
  journal={Superconductor Science and Technology},
  volume={36},
  number={12},
  pages={123001},
  year={2023},
  publisher={IOP Publishing}
}

@article{heikkila18,
  title={Thermoelectric radiation detector based on superconductor-ferromagnet systems},
  author={Heikkil{\"a}, TT and Ojaj{\"a}rvi, Risto and Maasilta, IJ and Strambini, Elia and Giazotto, Francesco and Bergeret, FS},
  journal={Physical Review Applied},
  volume={10},
  number={3},
  pages={034053},
  year={2018},
  publisher={APS}
}

@article{ozaeta14,
  title={Predicted very large thermoelectric effect in ferromagnet-superconductor junctions in the presence of a spin-splitting magnetic field},
  author={Ozaeta, A and Virtanen, Pauli and Bergeret, F Sebastian and Heikkil{\"a}, TT},
  journal={Physical review letters},
  volume={112},
  number={5},
  pages={057001},
  year={2014},
  publisher={APS}
}

@article{bergeret18,
  title={Colloquium: Nonequilibrium effects in superconductors with a spin-splitting field},
  author={Bergeret, F Sebastian and Silaev, Mikhail and Virtanen, Pauli and Heikkil{\"a}, Tero T},
  journal={Reviews of Modern Physics},
  volume={90},
  number={4},
  pages={041001},
  year={2018},
  publisher={APS}
}

@article{wu22-VD,
  title={Nonreciprocal charge transport in topological kagome superconductor CsV3Sb5},
  author={Wu, Yueshen and Wang, Qi and Zhou, Xiang and Wang, Jinghui and Dong, Peng and He, Jiadian and Ding, Yifan and Teng, Bolun and Zhang, Yiwen and Li, Yifei and others},
  journal={npj Quantum Materials},
  volume={7},
  number={1},
  pages={105},
  year={2022},
  publisher={Nature Publishing Group UK London}
}

@article{kealhofer23,
  title={Anomalous superconducting diode effect in a polar superconductor},
  author={Kealhofer, Robert and Jeong, Hanbyeol and Rashidi, Arman and Balents, Leon and Stemmer, Susanne},
  journal={Physical Review B},
  volume={107},
  number={10},
  pages={L100504},
  year={2023},
  publisher={APS}
}

@article{hou23,
  title={Ubiquitous superconducting diode effect in superconductor thin films},
  author={Hou, Yasen and Nichele, Fabrizio and Chi, Hang and Lodesani, Alessandro and Wu, Yingying and Ritter, Markus F and Haxell, Daniel Z and Davydova, Margarita and Ili{\'c}, Stefan and Glezakou-Elbert, Ourania and others},
  journal={Physical Review Letters},
  volume={131},
  number={2},
  pages={027001},
  year={2023},
  publisher={APS}
}

@article{vodolazov05-FM/SC,
  title={Considerable enhancement of the critical current in a superconducting film by a magnetized magnetic strip},
  author={Vodolazov, D Yu and Gribkov, BA and Gusev, SA and Klimov, A Yu and Nozdrin, Yu N and Rogov, VV and Vdovichev, SN},
  journal={Physical Review B},
  volume={72},
  number={6},
  pages={064509},
  year={2005},
  publisher={APS}
}

@article{hu23,
  title={Josephson Diode Effect Induced by Valley Polarization in Twisted Bilayer Graphene},
  author={Hu, Jin-Xin and Sun, Zi-Ting and Xie, Ying-Ming and Law, KT},
  journal={Physical Review Letters},
  volume={130},
  number={26},
  pages={266003},
  year={2023},
  publisher={APS}
}

@article{costa23-NN,
  title={Sign reversal of the Josephson inductance magnetochiral anisotropy and 0--$\pi$-like transitions in supercurrent diodes},
  author={Costa, A and Baumgartner, C and Reinhardt, S and Berger, J and Gronin, S and Gardner, GC and Lindemann, T and Manfra, MJ and Fabian, J and Kochan, D and others},
  journal={Nature Nanotechnology},
  pages={1--7},
  year={2023},
  publisher={Nature Publishing Group UK London}
}

@article{he23,
  title={The supercurrent diode effect and nonreciprocal paraconductivity due to the chiral structure of nanotubes},
  author={He, James Jun and Tanaka, Yukio and Nagaosa, Naoto},
  journal={Nature Communications},
  volume={14},
  number={1},
  pages={3330},
  year={2023},
  publisher={Nature Publishing Group UK London}
}

@article{li25,
  title={Microscopic study of supercurrent diode effect in chiral nanotubes},
  author={Li, Chuang and He, James Jun},
  journal={Physical Review B},
  volume={111},
  number={22},
  pages={224504},
  year={2025},
  publisher={APS}
}

@article{lu23,
  title={Tunable Josephson Diode Effect on the Surface of Topological Insulators},
  author={Lu, Bo and Ikegaya, Satoshi and Burset, Pablo and Tanaka, Yukio and Nagaosa, Naoto},
  journal={Physical Review Letters},
  volume={131},
  number={9},
  pages={096001},
  year={2023},
  publisher={APS}
}

@article{wei23,
  title={Josephson diode effect in a line-centered honeycomb lattice based superconductor junction},
  author={Wei, Ya-Jun and Wang, Juan-Juan and Wang, J},
  journal={Physical Review B},
  volume={108},
  number={5},
  pages={054521},
  year={2023},
  publisher={APS}
}

@article{mondal25-jde,
  title={Josephson diode effect with Andreev and Majorana bound states},
  author={Mondal, Sayan and Fu, Pei-Hao and Cayao, Jorge},
  journal={Physical Review B},
  volume={112},
  number={14},
  pages={144506},
  year={2025},
  publisher={APS}
}

@article{nikodem25,
  title={Tunable superconducting diode effect in a topological nano-SQUID},
  author={Nikodem, Ella and Schluck, Jakob and Geier, Max and Papaj, Micha{\l} and Legg, Henry F and Feng, Junya and Bagchi, Mahasweta and Fu, Liang and Ando, Yoichi},
  journal={Science Advances},
  volume={11},
  number={38},
  pages={eadw4898},
  year={2025},
  publisher={American Association for the Advancement of Science}
}

@article{liu24,
  title={Josephson diode effect in topological superconductors},
  author={Liu, Zhaochen and Huang, Linghao and Wang, Jing},
  journal={Physical Review B},
  volume={110},
  number={1},
  pages={014519},
  year={2024},
  publisher={APS}
}

@article{hosur23,
  title={Proximity-induced equilibrium supercurrent and perfect superconducting diode effect due to band asymmetry},
  author={Hosur, Pavan and Palacios, Daniel},
  journal={Physical Review B},
  volume={108},
  number={9},
  pages={094513},
  year={2023},
  publisher={APS}
}

@article{wang25-iso,
  title={Topological microwave isolator with> 100-dB isolation},
  author={Wang, Gang and Lu, Ling},
  journal={Nature Photonics},
  volume={19},
  number={10},
  pages={1064--1069},
  year={2025},
  publisher={Nature Publishing Group UK London}
}

@article{chapman17,
  title={Widely tunable on-chip microwave circulator for superconducting quantum circuits},
  author={Chapman, Benjamin J and Rosenthal, Eric I and Kerckhoff, Joseph and Moores, Bradley A and Vale, Leila R and Mates, JAB and Hilton, Gene C and Lalumiere, Kevin and Blais, Alexandre and Lehnert, KW},
  journal={Physical Review X},
  volume={7},
  number={4},
  pages={041043},
  year={2017},
  publisher={APS}
}

@article{ranzani19,
  title={Circulators at the quantum limit: Recent realizations of quantum-limited superconducting circulators and related approaches},
  author={Ranzani, Leonardo and Aumentado, Jose},
  journal={IEEE Microwave Magazine},
  volume={20},
  number={4},
  pages={112--122},
  year={2019},
  publisher={IEEE}
}

@book{pozar24,
  title={Microwave Engineering},
  author={Pozar, David M},
  year={2024},
  publisher={John wiley \& sons}
}

@article{vodolazov05,
  title={Superconducting rectifier based on the asymmetric surface barrier effect},
  author={Vodolazov, D Yu and Peeters, FM},
  journal={Physical Review B},
  volume={72},
  number={17},
  pages={172508},
  year={2005},
  publisher={APS}
}

@article{cerbu13,
  title={Vortex ratchet induced by controlled edge roughness},
  author={Cerbu, Dorin and Gladilin, VN and Cuppens, Johan and Fritzsche, Joachim and Tempere, Jacques and Devreese, JT and Moshchalkov, VV and Silhanek, AV and Van de Vondel, Joris},
  journal={New Journal of Physics},
  volume={15},
  number={6},
  pages={063022},
  year={2013},
  publisher={IOP Publishing}
}

@article{sivakov18,
  title={Spatial characterization of the edge barrier in wide superconducting films},
  author={Sivakov, AG and Turutanov, OG and Kolinko, AE and Pokhila, AS},
  journal={Low Temperature Physics},
  volume={44},
  number={3},
  pages={226--232},
  year={2018},
  publisher={AIP Publishing}
}

@article{suri22,
  title={Non-reciprocity of vortex-limited critical current in conventional superconducting micro-bridges},
  author={Suri, Dhavala and Kamra, Akashdeep and Meier, Thomas NG and Kronseder, Matthias and Belzig, Wolfgang and Back, Christian H and Strunk, Christoph},
  journal={Applied Physics Letters},
  volume={121},
  number={10},
  year={2022},
  publisher={AIP Publishing}
}

@article{ohkuma23,
  title={Nonreciprocal supercurrent in thin film of type II superconducting Sn},
  author={Ohkuma, Masahiro and Matsumoto, Ryo and Takano, Yoshihiko},
  journal={Applied Physics Express},
  volume={16},
  number={4},
  pages={043004},
  year={2023},
  publisher={IOP Publishing}
}

@Article{Zhang24-NbSe,
AUTHOR = {Yiwen Zhang and Jiliang Cai and Peng Dong and Jiadian He and Yifan Ding and Jinghui Wang and Xiang Zhou and Kecheng Cao and Yueshen Wu and Jun Li},
TITLE = {Intrinsic supercurrent diode effect in NbSe<sub>2</sub> nanobridge},
JOURNAL = {Microstructures},
VOLUME = {4},
YEAR = {2024},
NUMBER = {2},
ARTICLE-NUMBER = {2024018}
}

@article{bhowmik25,
  title={Topological Majorana zero modes and the superconducting diode effect driven by Fulde-Ferrell-Larkin-Ovchinnikov pairing in a helical Shiba chain},
  author={Bhowmik, Sayak and Saha, Arijit},
  journal={Physical Review B},
  volume={111},
  number={16},
  pages={L161402},
  year={2025},
  publisher={APS}
}

@article{cayao24,
  title={Enhancing the Josephson diode effect with Majorana bound states},
  author={Cayao, Jorge and Nagaosa, Naoto and Tanaka, Yukio},
  journal={Physical Review B},
  volume={109},
  number={8},
  pages={L081405},
  year={2024},
  publisher={APS}
}

@article{anh24,
  title={Large superconducting diode effect in ion-beam patterned Sn-based superconductor nanowire/topological Dirac semimetal planar heterostructures},
  author={Anh, Le Duc and Ishihara, Keita and Hotta, Tomoki and Inagaki, Kohdai and Maki, Hideki and Saeki, Takahiro and Kobayashi, Masaki and Tanaka, Masaaki},
  journal={Nature Communications},
  volume={15},
  number={1},
  pages={8014},
  year={2024},
  publisher={Nature Publishing Group UK London}
}

@article{nagahama25,
  title={Two-dimensional superconducting diode effect in topological insulator/superconductor heterostructure},
  author={Nagahama, Soma and Sato, Yuki and Kawamura, Minoru and Belopolski, Ilya and Yoshimi, Ryutaro and Tsukazaki, Atsushi and Kanazawa, Naoya and Takahashi, Kei S and Kawasaki, Masashi and Tokura, Yoshinori},
  journal={Physical Review Letters},
  volume={135},
  number={24},
  pages={246003},
  year={2025},
  publisher={APS}
}

@article{chenPingbo23,
  title={Edelstein Effect Induced Superconducting Diode Effect in Inversion Symmetry Breaking MoTe2 Josephson Junctions},
  author={Chen, Pingbo and Wang, Gongqi and Ye, Bicong and Wang, Jinhua and Zhou, Liang and Tang, Zhenzhong and Wang, Le and Wang, Jiannong and Zhang, Wenqing and Mei, Jiawei and others},
  journal={Advanced Functional Materials},
  pages={2311229},
  year={2023},
  publisher={Wiley Online Library}
}

@article{chenPingbo24,
  title={Asymmetric edge supercurrents in MoTe 2 Josephson junctions},
  author={Chen, Pingbo and Wang, Jinhua and Wang, Gongqi and Ye, Bicong and Zhou, Liang and Wang, Le and Wang, Jiannong and Zhang, Wenqing and Chen, Weiqiang and Mei, Jiawei and others},
  journal={Nanoscale Advances},
  volume={6},
  number={2},
  pages={690--696},
  year={2024},
  publisher={Royal Society of Chemistry}
}

@article{banerjee24,
  title={Enhanced superconducting diode effect due to coexisting phases},
  author={Banerjee, Sayan and Scheurer, Mathias S},
  journal={Physical Review Letters},
  volume={132},
  number={4},
  pages={046003},
  year={2024},
  publisher={APS}
}

@article{cai23,
  title={Nonreciprocal transport of superconductivity in a Bi/Ni bilayer},
  author={Cai, Ranran and Yue, Di and Qiao, Weiliang and Guo, Liangliang and Chen, Zidan and Xie, XC and Jin, Xiaofeng and Han, Wei},
  journal={Physical Review B},
  volume={108},
  number={6},
  pages={064501},
  year={2023},
  publisher={APS}
}

@article{daido25,
  title={Unidirectional superconductivity and superconducting diode effect induced by dissipation},
  author={Daido, Akito and Yanase, Youichi},
  journal={Physical Review B},
  volume={111},
  number={2},
  pages={L020508},
  year={2025},
  publisher={APS}
}

@article{chiles23,
  title={Nonreciprocal supercurrents in a field-free graphene Josephson triode},
  author={Chiles, John and Arnault, Ethan G and Chen, Chun-Chia and Larson, Trevyn FQ and Zhao, Lingfei and Watanabe, Kenji and Taniguchi, Takashi and Amet, Fran{\c{c}}ois and Finkelstein, Gleb},
  journal={Nano Letters},
  volume={23},
  number={11},
  pages={5257--5263},
  year={2023},
  publisher={ACS Publications}
}

@article{xie23,
  title={Orbital Fulde-Ferrell pairing state in moir{\'e} Ising superconductors},
  author={Xie, Ying-Ming and Law, KT},
  journal={Physical Review Letters},
  volume={131},
  number={1},
  pages={016001},
  year={2023},
  publisher={APS}
}

@article{nakamura24,
  title={Orbital effect on the intrinsic superconducting diode effect},
  author={Nakamura, Kyohei and Daido, Akito and Yanase, Youichi},
  journal={Physical Review B},
  volume={109},
  number={9},
  pages={094501},
  year={2024},
  publisher={APS}
}

@article{zhao23,
  title={Evidence of finite-momentum pairing in a centrosymmetric bilayer},
  author={Zhao, Dong and Debbeler, Lukas and K{\"u}hne, Matthias and Fecher, Sven and Gross, Nils and Smet, Jurgen},
  journal={Nature Physics},
  pages={1--6},
  year={2023},
  publisher={Nature Publishing Group UK London}
}

@article{asaba24,
  title={Evidence for a finite-momentum Cooper pair in tricolor d-wave superconducting superlattices},
  author={Asaba, T and Naritsuka, M and Asaeda, H and Kosuge, Y and Ikemori, S and Suetsugu, S and Kasahara, Y and Kohsaka, Y and Terashima, T and Daido, A and others},
  journal={Nature Communications},
  volume={15},
  number={1},
  pages={3861},
  year={2024},
  publisher={Nature Publishing Group UK London}
}

@article{buck56,
  title={The cryotron-a superconductive computer component},
  author={Buck, Dudley A},
  journal={Proceedings of the IRE},
  volume={44},
  number={4},
  pages={482--493},
  year={1956},
  publisher={IEEE}
}

@article{matisoo67,
  title={The tunneling cryotron—A superconductive logic element based on electron tunneling},
  author={Matisoo, J},
  journal={Proceedings of the IEEE},
  volume={55},
  number={2},
  pages={172--180},
  year={1967},
  publisher={IEEE}
}

@article{alam23,
  title={Cryogenic memory technologies},
  author={Alam, Shamiul and Hossain, Md Shafayat and Srinivasa, Srivatsa Rangachar and Aziz, Ahmedullah},
  journal={Nature Electronics},
  volume={6},
  number={3},
  pages={185--198},
  year={2023},
  publisher={Nature Publishing Group UK London}
}

@inproceedings{tannu17,
  title={Cryogenic-DRAM based memory system for scalable quantum computers: A feasibility study},
  author={Tannu, Swamit S and Carmean, Douglas M and Qureshi, Moinuddin K},
  booktitle={Proceedings of the International Symposium on Memory Systems},
  pages={189--195},
  year={2017}
}

@article{hornibrook15,
  title={Cryogenic control architecture for large-scale quantum computing},
  author={Hornibrook, JM and Colless, JI and Conway Lamb, ID and Pauka, SJ and Lu, H and Gossard, AC and Watson, JD and Gardner, GC and Fallahi, S and Manfra, MJ and others},
  journal={Physical Review Applied},
  volume={3},
  number={2},
  pages={024010},
  year={2015},
  publisher={APS}
}

@article{wu26-arXiv,
  title={Discovery of an electrically-controllable superconducting memory effect},
  author={Wu, Zheyu and Chen, Hanyi and Long, Mengmeng and Shaffer, Daniel and Chichinadze, Dmitry V and Cabala, Andrej and Weinberger, Theodore I and Hickey, Alexander J and Pu, Jinxu and Graf, Dave and others},
  journal={arXiv preprint arXiv:2603.02450},
  year={2026}
}

@article{he23-ADI,
  title={Spin-Related Superconducting Devices for Logic and Memory Applications},
  author={He, Yu and Li, Jiaxu and Wang, Qiusha and Matsuki, Hisakazu and Yang, Guang},
  journal={Advanced Devices \& Instrumentation},
  volume={4},
  pages={0035},
  year={2023},
  publisher={AAAS}
}

@article{semenov19,
  title={Very large scale integration of Josephson-junction-based superconductor random access memories},
  author={Semenov, Vasili K and Polyakov, Yuri A and Tolpygo, Sergey K},
  journal={IEEE Transactions on Applied Superconductivity},
  volume={29},
  number={5},
  pages={1--9},
  year={2019},
  publisher={IEEE}
}

@article{ligato21,
  title={Preliminary demonstration of a persistent Josephson phase-slip memory cell with topological protection},
  author={Ligato, Nadia and Strambini, Elia and Paolucci, Federico and Giazotto, Francesco},
  journal={Nature communications},
  volume={12},
  number={1},
  pages={5200},
  year={2021},
  publisher={Nature Publishing Group UK London}
}

@article{sinner24,
  title={Superconducting diode sensor},
  author={Sinner, A and Wang, X-G and Parkin, SSP and Ernst, A and Dugaev, V and Chotorlishvili, L},
  journal={Physical Review B},
  volume={109},
  number={21},
  pages={214510},
  year={2024},
  publisher={APS}
}

@article{sinner26,
  title={Skyrmion Footprint in the Thermodynamics of the Josephson Superconducting Diode},
  author={Sinner, Andreas and Wang, Xi-Guang and Chotorlishvili, Levan},
  journal={Annalen der Physik},
  volume={538},
  number={1},
  pages={e00488},
  year={2026},
  publisher={Wiley Online Library}
}

@article{hess23,
  title={Josephson transistor from the superconducting diode effect in domain wall and skyrmion magnetic racetracks},
  author={Hess, Richard and Legg, Henry F and Loss, Daniel and Klinovaja, Jelena},
  journal={Physical Review B},
  volume={108},
  number={17},
  pages={174516},
  year={2023},
  publisher={APS}
}

@article{xiong24,
  title={Electrical switching of Ising-superconducting nonreciprocity for quantum neuronal transistor},
  author={Xiong, Junlin and Xie, Jiao and Cheng, Bin and Dai, Yudi and Cui, Xinyu and Wang, Lizheng and Liu, Zenglin and Zhou, Ji and Wang, Naizhou and Xu, Xianghan and others},
  journal={Nature Communications},
  volume={15},
  number={1},
  pages={4953},
  year={2024},
  publisher={Nature Publishing Group UK London}
}

@article{wang25-wte2,
  title={Nonvolatile electrical switching of nonreciprocal transport in ferroelectric polar metal WTe2},
  author={Wang, Ruihan and Chen, Haoyun and Wang, Hui and Liu, Bingyan and Chen, Xin and Cao, Jin and Han, Ziyi and Yang, Zherui and Renshaw Wang, X and Zhao, Xiaoxu and others},
  journal={Nature communications},
  year={2025},
  publisher={Nature Publishing Group UK London}
}

@article{sun25-spin,
  title={Voltage-tunable spin supercurrent nonreciprocity reaching 100\% efficiency},
  author={Sun, Chi and Tjernshaugen, Johanne Bratland and Linder, Jacob},
  journal={Physical Review B},
  volume={112},
  number={6},
  pages={064504},
  year={2025},
  publisher={APS}
}

@article{chen24,
  title={All-electrical skyrmionic magnetic tunnel junction},
  author={Chen, Shaohai and Lourembam, James and Ho, Pin and Toh, Alexander KJ and Huang, Jifei and Chen, Xiaoye and Tan, Hang Khume and Yap, Sherry LK and Lim, Royston JJ and Tan, Hui Ru and others},
  journal={Nature},
  volume={627},
  number={8004},
  pages={522--527},
  year={2024},
  publisher={Nature Publishing Group UK London}
}

@article{zhou25-AM,
  title={Topological spin textures: Basic physics and devices},
  author={Zhou, Yuqing and Li, Shuang and Liang, Xue and Zhou, Yan},
  journal={Advanced Materials},
  volume={37},
  number={2},
  pages={2312935},
  year={2025},
  publisher={Wiley Online Library}
}

@article{leblanc24,
  title={From nonreciprocal to charge-4e supercurrent in Ge-based Josephson devices with tunable harmonic content},
  author={Leblanc, Axel and Tangchingchai, Chotivut and Momtaz, Zahra Sadre and Kiyooka, Elyjah and Hartmann, Jean-Michel and Fernandez-Bada, Gonzalo Troncoso and Scher{\"u}bl, Zolt{\'a}n and Brun, Boris and Schmitt, Vivien and Zihlmann, Simon and others},
  journal={Physical Review Research},
  volume={6},
  number={3},
  pages={033281},
  year={2024},
  publisher={APS}
}

@article{araujo24,
  title={Superconducting spintronic heat engine},
  author={Araujo, Clodoaldo Irineu Levartoski de and Virtanen, Pauli and Spies, Maria and Gonz{\'a}lez-Orellana, Carmen and Kerschbaumer, Samuel and Ilyn, Maxim and Rogero, Celia and Heikkil{\"a}, Tero Tapio and Giazotto, Francesco and Strambini, Elia},
  journal={Nature Communications},
  volume={15},
  number={1},
  pages={4823},
  year={2024},
  publisher={Nature Publishing Group UK London}
}

@article{chahid23,
  title={High-frequency diode effect in superconducting Nb 3 Sn microbridges},
  author={Chahid, Sara and Teknowijoyo, Serafim and Mowgood, Iris and Gulian, Armen},
  journal={Physical Review B},
  volume={107},
  number={5},
  pages={054506},
  year={2023},
  publisher={APS}
}

@article{kochan25,
  title={Low-loss electronics with superconducting diodes},
  author={Kochan, Denis and Strunk, Christoph},
  journal={Nature Electronics},
  pages={1--2},
  year={2025},
  publisher={Nature Publishing Group}
}

@article{castellani25,
  title={A superconducting full-wave bridge rectifier},
  author={Castellani, Matteo and Medeiros, Owen and Buzzi, Alessandro and Foster, Reed A and Colangelo, Marco and Berggren, Karl K},
  journal={Nature Electronics},
  volume={8},
  pages={417–425},
  year={2025},
  publisher={Nature Publishing Group}
}

@article{ingla25,
  title={Efficient superconducting diodes and rectifiers for quantum circuitry},
  author={Ingla-Ayn{\'e}s, Josep and Hou, Yasen and Wang, Sarah and Chu, En-De and Mukhanov, Oleg A and Wei, Peng and Moodera, Jagadeesh S},
  journal={Nature Electronics},
  volume={8},
  pages={411–416},
  year={2025},
  publisher={Nature Publishing Group}
}

@article{guarcello25,
  title={Effect of a second-harmonic current--phase relation on the behavior of a Josephson traveling-wave parametric amplifier},
  author={Guarcello, Claudio and Barone, Carlo and Carapella, Giovanni and Filatrella, Giovanni and Giachero, Andrea and Pagano, Sergio},
  journal={Applied Physics Letters},
  volume={126},
  number={16},
  year={2025},
  publisher={AIP Publishing}
}

@article{banerjee24-AM,
  title={Altermagnetic superconducting diode effect},
  author={Banerjee, Sayan and Scheurer, Mathias S},
  journal={Physical Review B},
  volume={110},
  number={2},
  pages={024503},
  year={2024},
  publisher={APS}
}

@article{giil24b,
  title={Quasiclassical theory of superconducting spin-splitter effects and spin-filtering via altermagnets},
  author={Giil, Hans Gl{\o}ckner and Brekke, Bj{\o}rnulf and Linder, Jacob and Brataas, Arne},
  journal={Physical Review B},
  volume={110},
  number={14},
  pages={L140506},
  year={2024},
  publisher={APS}
}

@article{chakraborty25,
  title={Perfect superconducting diode effect in altermagnets},
  author={Chakraborty, Debmalya and Black-Schaffer, Annica M},
  journal={Physical Review Letters},
  volume={135},
  number={2},
  pages={026001},
  year={2025},
  publisher={APS}
}

@article{shaffer24,
  title={Superconducting diode effect in multiphase superconductors},
  author={Shaffer, Daniel and Chichinadze, Dmitry V and Levchenko, Alex},
  journal={Physical Review B},
  volume={110},
  number={18},
  pages={184509},
  year={2024},
  publisher={APS}
}

@article{hosur25-arXiv,
  title={Criticality as a Universal Thermodynamic Requirement for Perfect Intrinsic Superconducting Diodes},
  author={Hosur, Pavan},
  journal={arXiv preprint arXiv:2512.21384},
  year={2025}
}

@article{roig24,
  title={Superconducting diodes from magnetization gradients},
  author={Roig, Merc{\`e} and Kotetes, Panagiotis and Andersen, Brian M},
  journal={Physical Review B},
  volume={109},
  number={14},
  pages={144503},
  year={2024},
  publisher={APS}
}

@article{cadorim24,
  title={Harnessing the superconducting diode effect through inhomogeneous magnetic fields},
  author={Cadorim, Leonardo Rodrigues and Sardella, Edson and Silva, Cl{\'e}cio C de Souza},
  journal={Physical Review Applied},
  volume={21},
  number={5},
  pages={054040},
  year={2024},
  publisher={APS}
}

@article{zhao23-Sci,
title={Time-reversal symmetry breaking superconductivity between twisted cuprate superconductors},
  author={Zhao, SY Frank and Cui, Xiaomeng and Volkov, Pavel A and Yoo, Hyobin and Lee, Sangmin and Gardener, Jules A and Akey, Austin J and Engelke, Rebecca and Ronen, Yuval and Zhong, Ruidan and others},
  journal={Science},
  volume={382},
  number={6677},
  pages={1422--1427},
  year={2023},
  publisher={American Association for the Advancement of Science}
}

@article{ghosh24,
  title={High-temperature Josephson diode},
  author={Ghosh, Sanat and Patil, Vilas and Basu, Amit and Kuldeep and Dutta, Achintya and Jangade, Digambar A and Kulkarni, Ruta and Thamizhavel, A and Steiner, Jacob F and von Oppen, Felix and others},
  journal={Nature Materials},
  volume={23},
  number={5},
  pages={612--618},
  year={2024},
  publisher={Nature Publishing Group UK London}
}

@article{qi25,
  title={High-temperature field-free superconducting diode effect in high-T c cuprates},
  author={Qi, Shichao and Ge, Jun and Ji, Chengcheng and Ai, Yiwen and Ma, Gaoxing and Wang, Ziqiao and Cui, Zihan and Liu, Yi and Wang, Ziqiang and Wang, Jian},
  journal={Nature Communications},
  volume={16},
  number={1},
  pages={531},
  year={2025},
  publisher={Nature Publishing Group UK London}
}

@article{oh24,
  title={Nonreciprocal transport in U (1) gauge theory of high-T c cuprates},
  author={Oh, Taekoo and Nagaosa, Naoto},
  journal={Physical Review B},
  volume={110},
  number={13},
  pages={134507},
  year={2024},
  publisher={APS}
}

@article{lee21,
  title={Twisted van der Waals Josephson junction based on a high-T c superconductor},
  author={Lee, Jongyun and Lee, Wonjun and Kim, Gi-Yeop and Choi, Yong-Bin and Park, Jinho and Jang, Seong and Gu, Genda and Choi, Si-Young and Cho, Gil Young and Lee, Gil-Ho and others},
  journal={Nano Letters},
  volume={21},
  number={24},
  pages={10469--10477},
  year={2021},
  publisher={ACS Publications}
}

@book{barone82,
  title={Physics and Applications of the Josephson Effect},
  author={Barone, A. and Paterno, G.},
  isbn={9780471014690},
  lccn={lc81007554},
  series={A Wiley-interscience publication},
  year={1982},
  publisher={Wiley}
}

@article{souto22,
  title={Josephson diode effect in supercurrent interferometers},
  author={Souto, Rub{\'e}n Seoane and Leijnse, Martin and Schrade, Constantin},
  journal={Physical Review Letters},
  volume={129},
  number={26},
  pages={267702},
  year={2022},
  publisher={APS}
}

@article{reinhardt24,
  title={Link between supercurrent diode and anomalous Josephson effect revealed by gate-controlled interferometry},
  author={Reinhardt, Simon and Ascherl, Tim and Costa, Andreas and Berger, Johanna and Gronin, Sergei and Gardner, Geoffrey C and Lindemann, Tyler and Manfra, Michael J and Fabian, Jaroslav and Kochan, Denis and others},
  journal={Nature Communications},
  volume={15},
  number={1},
  pages={4413},
  year={2024},
  publisher={Nature Publishing Group UK London}
}

@article{karabassov24,
  title={Anisotropic Josephson diode effect in the topological hybrid junctions with the hexagonal warping},
  author={Karabassov, Tairzhan},
  journal={JETP Letters},
  volume={119},
  number={4},
  pages={316--323},
  year={2024},
  publisher={Springer}
}

@article{zhong25,
  title={Twisted superconducting quantum diodes: Towards anharmonicity and high fidelity},
  author={Zhong, Han and Kochan, Denis and Zutic, Igor and Wu, Yingying},
  journal={arXiv preprint arXiv:2510.19627},
  year={2025}
}

@article{dirnegger25,
  title={Nonreciprocal quantum information processing with superconducting diodes in circuit quantum electrodynamics},
  author={Dirnegger, Nicolas and Narang, Prineha and Arora, Arpit},
  journal={arXiv preprint arXiv:2511.20758},
  year={2025}
}

@article{kafri17,
  title={Tunable inductive coupling of superconducting qubits in the strongly nonlinear regime},
  author={Kafri, Dvir and Quintana, Chris and Chen, Yu and Shabani, Alireza and Martinis, John M and Neven, Hartmut},
  journal={Physical Review A},
  volume={95},
  number={5},
  pages={052333},
  year={2017},
  publisher={APS}
}

@article{solgun19,
  title={Simple impedance response formulas for the dispersive interaction rates in the effective Hamiltonians of low anharmonicity superconducting qubits},
  author={Solgun, Firat and DiVincenzo, David P and Gambetta, Jay M},
  journal={IEEE transactions on microwave theory and techniques},
  volume={67},
  number={3},
  pages={928--948},
  year={2019},
  publisher={IEEE}
}

@article{labarca24,
  title={Toolbox for nonreciprocal dispersive models in circuit quantum electrodynamics},
  author={Labarca, Lautaro and Benhayoune-Khadraoui, Othmane and Blais, Alexandre and Parra-Rodriguez, Adrian},
  journal={Physical Review Applied},
  volume={22},
  number={3},
  pages={034038},
  year={2024},
  publisher={APS}
}

@article{ren25,
  title={Nonreciprocal interaction and entanglement between two superconducting qubits},
  author={Ren, Yu-Meng and Pan, Xue-Feng and Yao, Xiao-Yu and Huo, Xiao-Wen and Zheng, Jun-Cong and Hei, Xin-Lei and Qiao, Yi-Fan and Li, Peng-Bo},
  journal={Physical Review Research},
  volume={7},
  number={2},
  pages={023287},
  year={2025},
  publisher={APS}
}

@article{bozkurt23,
  title={Double-Fourier engineering of Josephson energy-phase relationships applied to diodes},
  author={Bozkurt, A Mert and Brookman, Jasper and Fatemi, Valla and Akhmerov, Anton R},
  journal={SciPost Physics},
  volume={15},
  number={5},
  pages={204},
  year={2023}
}

@article{schrade24,
  title={Dissipationless nonlinearity in quantum material Josephson diodes},
  author={Schrade, Constantin and Fatemi, Valla},
  journal={Physical Review Applied},
  volume={21},
  number={6},
  pages={064029},
  year={2024},
  publisher={APS}
}

@article{beenakker91,
  title={Universal limit of critical-current fluctuations in mesoscopic Josephson junctions},
  author={Beenakker, CWJ},
  journal={Physical review letters},
  volume={67},
  number={27},
  pages={3836},
  year={1991},
  publisher={APS}
}

@article{castellanos08,
  title={Amplification and squeezing of quantum noise with a tunable Josephson metamaterial},
  author={Castellanos-Beltran, Manuel A and Irwin, KD and Hilton, GC and Vale, LR and Lehnert, KW},
  journal={Nature Physics},
  volume={4},
  number={12},
  pages={929--931},
  year={2008},
  publisher={Nature Publishing Group UK London}
}

@article{frattini17,
  title={3-wave mixing Josephson dipole element},
  author={Frattini, NE and Vool, U and Shankar, S and Narla, A and Sliwa, KM and Devoret, MH},
  journal={Applied Physics Letters},
  volume={110},
  number={22},
  year={2017},
  publisher={AIP Publishing}
}

@article{abdo13,
  title={Nondegenerate three-wave mixing with the Josephson ring modulator},
  author={Abdo, Baleegh and Kamal, Archana and Devoret, Michel},
  journal={Physical Review B},
  volume={87},
  number={1},
  pages={014508},
  year={2013},
  publisher={APS}
}

@article{frattini18,
  title={Optimizing the nonlinearity and dissipation of a SNAIL parametric amplifier for dynamic range},
  author={Frattini, NE and Sivak, VV and Lingenfelter, A and Shankar, S and Devoret, MH},
  journal={Physical Review Applied},
  volume={10},
  number={5},
  pages={054020},
  year={2018},
  publisher={APS}
}

@article{sivak19,
  title={Kerr-free three-wave mixing in superconducting quantum circuits},
  author={Sivak, VV and Frattini, NE and Joshi, VR and Lingenfelter, A and Shankar, S and Devoret, MH},
  journal={Physical Review Applied},
  volume={11},
  number={5},
  pages={054060},
  year={2019},
  publisher={APS}
}

@article{miano22,
  title={Frequency-tunable Kerr-free three-wave mixing with a gradiometric SNAIL},
  author={Miano, A and Liu, G and Sivak, VV and Frattini, NE and Joshi, VR and Dai, W and Frunzio, L and Devoret, MH},
  journal={Applied Physics Letters},
  volume={120},
  number={18},
  year={2022},
  publisher={AIP Publishing}
}

@article{zorin16,
  title={Josephson traveling-wave parametric amplifier with three-wave mixing},
  author={Zorin, AB},
  journal={Physical Review Applied},
  volume={6},
  number={3},
  pages={034006},
  year={2016},
  publisher={APS}
}

@article{hu25-adma,
  title={Polarity-Reversible Zero-Field Diode Effect in van der Waals Ferromagnetic Josephson Junction for Logic Operation},
  author={Hu, Guojing and Han, Yechao and Guo, Hui and Lv, Senhao and Gao, Tianqi and Wang, Yunhao and Zhao, Zhen and Zhu, Ke and Qi, Qi and Xian, Guoyu and others},
  journal={Advanced Materials},
  pages={e13434},
  year={2025},
  publisher={Wiley Online Library}
}

@article{cuozzo24,
  title={Microwave-tunable diode effect in asymmetric SQUIDs with topological Josephson junctions},
  author={Cuozzo, Joseph J and Pan, Wei and Shabani, Javad and Rossi, Enrico},
  journal={Physical Review Research},
  volume={6},
  number={2},
  pages={023011},
  year={2024},
  publisher={APS}
}

@article{shaffer25,
  title={Josephson diode effect from nonequilibrium current in a superconducting interferometer},
  author={Shaffer, Daniel and Li, Songci and Hasan, Jaglul and Titov, Mikhail and Levchenko, Alex},
  journal={Physical Review B},
  volume={112},
  number={9},
  pages={094509},
  year={2025},
  publisher={APS}
}

@article{wang25-NP,
  title={Quantum superconducting diode effect with perfect efficiency above liquid-nitrogen temperature},
  author={Wang, Heng and Zhu, Yuying and Bai, Zhonghua and Lyu, Zhaozheng and Yang, Jiangang and Zhao, Lin and Zhou, XJ and Xue, Qi-Kun and Zhang, Ding},
  journal={Nature Physics},
  pages={1--7},
  year={2025},
  publisher={Nature Publishing Group UK London}
}

@article{valentini24,
  title={Parity-conserving Cooper-pair transport and ideal superconducting diode in planar germanium},
  author={Valentini, Marco and Sagi, Oliver and Baghumyan, Levon and de Gijsel, Thijs and Jung, Jason and Calcaterra, Stefano and Ballabio, Andrea and Aguilera Servin, Juan and Aggarwal, Kushagra and Janik, Marian and others},
  journal={Nature Communications},
  volume={15},
  number={1},
  pages={169},
  year={2024},
  publisher={Nature Publishing Group UK London}
}

@article{seoane24,
  title={Tuning the Josephson diode response with an ac current},
  author={Seoane Souto, Rub{\'e}n and Leijnse, Martin and Schrade, Constantin and Valentini, Marco and Katsaros, Georgios and Danon, Jeroen},
  journal={Physical Review Research},
  volume={6},
  number={2},
  pages={L022002},
  year={2024},
  publisher={APS}
}

@article{zhao23-npj,
  title={Engineering quantum diode in one-dimensional time-varying superconducting circuits},
  author={Zhao, Xuedong and Xing, Yan and Cao, Ji and Liu, Shutian and Cui, Wen-Xue and Wang, Hong-Fu},
  journal={npj Quantum Information},
  volume={9},
  number={1},
  pages={59},
  year={2023},
  publisher={Nature Publishing Group UK London}
}

@article{monroe24,
  title={Phase jumps in Josephson junctions with time-dependent spin--orbit coupling},
  author={Monroe, David and Shen, Chenghao and Tringali, Dario and Alidoust, Mohammad and Zhou, Tong and {\v{Z}}uti{\'c}, Igor},
  journal={Applied Physics Letters},
  volume={125},
  number={1},
  year={2024},
  publisher={AIP Publishing}
}

@article{yang23-NRM,
  title={Terahertz control of many-body dynamics in quantum materials},
  author={Yang, Chia-Jung and Li, Jingwen and Fiebig, Manfred and Pal, Shovon},
  journal={Nature Reviews Materials},
  volume={8},
  number={8},
  pages={518--532},
  year={2023},
  publisher={Nature Publishing Group UK London}
}

@incollection{martinis04,
  title={Superconducting qubits and the physics of Josephson junctions},
  author={Martinis, John M},
  booktitle={Les Houches},
  volume={79},
  pages={487--520},
  year={2004},
  publisher={Elsevier}
}

@article{golubov04,
  title={The current-phase relation in Josephson junctions},
  author={Golubov, Alexandre Avraamovitch and Kupriyanov, M Yu and Il’Ichev, E},
  journal={Reviews of modern physics},
  volume={76},
  number={2},
  pages={411},
  year={2004},
  publisher={APS}
}

@article{krantz19,
  title={A quantum engineer's guide to superconducting qubits},
  author={Krantz, Philip and Kjaergaard, Morten and Yan, Fei and Orlando, Terry P and Gustavsson, Simon and Oliver, William D},
  journal={Applied physics reviews},
  volume={6},
  number={2},
  year={2019},
  publisher={AIP Publishing}
}

@article{aumentado20,
  title={Superconducting parametric amplifiers: The state of the art in Josephson parametric amplifiers},
  author={Aumentado, Jose},
  journal={IEEE Microwave magazine},
  volume={21},
  number={8},
  pages={45--59},
  year={2020},
  publisher={IEEE}
}

@article{esposito21,
  title={Perspective on traveling wave microwave parametric amplifiers},
  author={Esposito, Martina and Ranadive, Arpit and Planat, Luca and Roch, Nicolas},
  journal={Applied Physics Letters},
  volume={119},
  number={12},
  year={2021},
  publisher={AIP Publishing}
}

@article{leroux22-arXiv,
  title={Nonreciprocal devices based on voltage-tunable junctions},
  author={Leroux, C and Parra-Rodriguez, A and Shillito, R and Di Paolo, A and Oliver, WD and Marcus, CM and Kjaergaard, M and Gyenis, A and Blais, A},
  year={2022},
  journal={arXiv preprint arXiv:2209.06194}
}

@article{fadavi23,
  title={Three-wave mixing traveling-wave parametric amplifier with periodic variation of the circuit parameters},
  author={Fadavi Roudsari, Anita and Shiri, Daryoush and Renberg Nilsson, Hampus and Tancredi, Giovanna and Osman, Amr and Svensson, Ida-Maria and Kudra, Marina and Rommel, Marcus and Bylander, Jonas and Shumeiko, Vitaly and others},
  journal={Applied Physics Letters},
  volume={122},
  number={5},
  year={2023},
  publisher={AIP Publishing}
}

@article{ranadive25,
  title={A travelling-wave parametric amplifier isolator},
  author={Ranadive, Arpit and Fazliji, Bekim and Le Gal, Gwenael and Cappelli, Giulio and Butseraen, Guilliam and Bonet, Edgar and Eyraud, Eric and B{\"o}hling, Sina and Planat, Luca and Metelmann, A and others},
  journal={Nature Electronics},
  pages={1--10},
  year={2025},
  publisher={Nature Publishing Group UK London}
}

@article{kippenberg08,
  title={Cavity optomechanics: back-action at the mesoscale},
  author={Kippenberg, Tobias J and Vahala, Kerry J},
  journal={science},
  volume={321},
  number={5893},
  pages={1172--1176},
  year={2008},
  publisher={American Association for the Advancement of Science}
}

@article{teufel11,
  title={Sideband cooling of micromechanical motion to the quantum ground state},
  author={Teufel, John D and Donner, Tobias and Li, Dale and Harlow, Jennifer W and Allman, MS and Cicak, Katarina and Sirois, Adam J and Whittaker, Jed D and Lehnert, Konrad W and Simmonds, Raymond W},
  journal={Nature},
  volume={475},
  number={7356},
  pages={359--363},
  year={2011},
  publisher={Nature Publishing Group UK London}
}

@article{bienfait16,
  title={Reaching the quantum limit of sensitivity in electron spin resonance},
  author={Bienfait, A and Pla, JJ and Kubo, Y and Stern, M and Zhou, X and Lo, CC and Weis, CD and Schenkel, T and Thewalt, MLW and Vion, D and others},
  journal={Nature nanotechnology},
  volume={11},
  number={3},
  pages={253--257},
  year={2016},
  publisher={Nature Publishing Group UK London}
}

@article{xiao04,
  title={Electrical detection of the spin resonance of a single electron in a silicon field-effect transistor},
  author={Xiao, M and Martin, I and Yablonovitch, E and Jiang, HW},
  journal={Nature},
  volume={430},
  number={6998},
  pages={435--439},
  year={2004},
  publisher={Nature Publishing Group UK London}
}

@inproceedings{bryerton13,
  title={Ultra low noise cryogenic amplifiers for radio astronomy},
  author={Bryerton, Eric W and Morgan, Matthew and Pospieszalski, Marian W},
  booktitle={2013 IEEE Radio and Wireless Symposium},
  pages={358--360},
  year={2013},
  organization={IEEE}
}

@article{smith13,
  title={Low noise amplifier for radio astronomy},
  author={Smith, David MP and Bakker, Laurens and Witvers, Roel H and Woestenburg, Bert EM and Palmer, Keith D},
  journal={International Journal of Microwave and Wireless Technologies},
  volume={5},
  number={4},
  pages={453--461},
  year={2013},
  publisher={Cambridge University Press}
}

@inproceedings{pospieszalski18,
  title={Extremely low-noise cryogenic amplifiers for radio astronomy: Past, present and future},
  author={Pospieszalski, Marian W},
  booktitle={2018 22nd International Microwave and Radar Conference (MIKON)},
  pages={1--6},
  year={2018},
  organization={IEEE}
}

@article{backes21,
  title={A quantum enhanced search for dark matter axions},
  author={Backes, Kelly M and Palken, Daniel A and Kenany, S Al and Brubaker, Benjamin M and Cahn, SB and Droster, A and Hilton, Gene C and Ghosh, Sumita and Jackson, H and Lamoreaux, Steve K and others},
  journal={Nature},
  volume={590},
  number={7845},
  pages={238--242},
  year={2021},
  publisher={Nature Publishing Group UK London}
}

@article{caldwell17,
  title={Dielectric haloscopes: A new way to detect axion dark matter},
  author={Caldwell, Allen and Dvali, Gia and Majorovits, B{\'e}la and Millar, Alexander and Raffelt, Georg and Redondo, Javier and Reimann, Olaf and Simon, Frank and Steffen, Frank and (MADMAX Working Group)},
  journal={Physical review letters},
  volume={118},
  number={9},
  pages={091801},
  year={2017},
  publisher={APS}
}

@article{jeong20,
  title={Search for invisible axion dark matter with a multiple-cell haloscope},
  author={Jeong, Junu and Youn, SungWoo and Bae, Sungjae and Kim, Jihngeun and Seong, Taehyeon and Kim, Jihn E and Semertzidis, Yannis K},
  journal={Physical Review Letters},
  volume={125},
  number={22},
  pages={221302},
  year={2020},
  publisher={APS}
}

@article{rybka14,
  title={Direct detection searches for axion dark matter},
  author={Rybka, Gray and ADMX Collaboration and others},
  journal={Physics of the Dark Universe},
  volume={4},
  pages={14--16},
  year={2014},
  publisher={Elsevier}
}

@article{caves82,
  title={Quantum limits on noise in linear amplifiers},
  author={Caves, Carlton M},
  journal={Physical Review D},
  volume={26},
  number={8},
  pages={1817},
  year={1982},
  publisher={APS}
}

@article{clerk10,
  title={Introduction to quantum noise, measurement, and amplification},
  author={Clerk, Aashish A and Devoret, Michel H and Girvin, Steven M and Marquardt, Florian and Schoelkopf, Robert J},
  journal={Reviews of Modern Physics},
  volume={82},
  number={2},
  pages={1155--1208},
  year={2010},
  publisher={APS}
}

@article{bergeal10,
  title={Phase-preserving amplification near the quantum limit with a Josephson ring modulator},
  author={Bergeal, N and Schackert, F and Metcalfe, M and Vijay, R and Manucharyan, VE and Frunzio, L and Prober, DE and Schoelkopf, RJ and Girvin, SM and Devoret, MH},
  journal={Nature},
  volume={465},
  number={7294},
  pages={64--68},
  year={2010},
  publisher={Nature Publishing Group UK London}
}

@article{bergeal10-NP,
  title={Analog information processing at the quantum limit with a Josephson ring modulator},
  author={Bergeal, N and Vijay, R and Manucharyan, VE and Siddiqi, I and Schoelkopf, RJ and Girvin, SM and Devoret, MH},
  journal={Nature Physics},
  volume={6},
  number={4},
  pages={296--302},
  year={2010},
  publisher={Nature Publishing Group UK London}
}

@article{hatridge11,
  title={Dispersive magnetometry with a quantum limited SQUID parametric amplifier},
  author={Hatridge, M and Vijay, R and Slichter, DH and Clarke, John and Siddiqi, I},
  journal={Physical Review B—Condensed Matter and Materials Physics},
  volume={83},
  number={13},
  pages={134501},
  year={2011},
  publisher={APS}
}

@article{roch12,
  title={Widely Tunable, Nondegenerate Three-Wave Mixing Microwave Device Operating<? format?> near the Quantum Limit},
  author={Roch, Nicolas and Flurin, Emmanuel and Nguyen, Francois and Morfin, Pascal and Campagne-Ibarcq, Philippe and Devoret, Michel H and Huard, Benjamin},
  journal={Physical review letters},
  volume={108},
  number={14},
  pages={147701},
  year={2012},
  publisher={APS}
}

@article{macklin15,
  title={A near--quantum-limited Josephson traveling-wave parametric amplifier},
  author={Macklin, Chris and O’brien, K and Hover, D and Schwartz, ME and Bolkhovsky, V and Zhang, X and Oliver, WD and Siddiqi, I},
  journal={Science},
  volume={350},
  number={6258},
  pages={307--310},
  year={2015},
  publisher={American Association for the Advancement of Science}
}

@article{heinsoo18,
  title={Rapid high-fidelity multiplexed readout of superconducting qubits},
  author={Heinsoo, Johannes and Andersen, Christian Kraglund and Remm, Ants and Krinner, Sebastian and Walter, Theodore and Salath{\'e}, Yves and Gasparinetti, Simone and Besse, Jean-Claude and Poto{\v{c}}nik, Anton and Wallraff, Andreas and others},
  journal={Physical Review Applied},
  volume={10},
  number={3},
  pages={034040},
  year={2018},
  publisher={APS}
}

@article{zmuidzinas12,
  title={Superconducting microresonators: Physics and applications},
  author={Zmuidzinas, Jonas},
  journal={Annu. Rev. Condens. Matter Phys.},
  volume={3},
  number={1},
  pages={169--214},
  year={2012},
  publisher={Annual Reviews}
}

@article{mcrae20,
  title={Materials loss measurements using superconducting microwave resonators},
  author={McRae, Corey Rae Harrington and Wang, Haozhi and Gao, Jiansong and Vissers, Michael R and Brecht, Teresa and Dunsworth, Andrew and Pappas, David P and Mutus, Josh},
  journal={Review of Scientific Instruments},
  volume={91},
  number={9},
  year={2020},
  publisher={AIP Publishing}
}

@article{gurevich23,
  title={Tuning microwave losses in superconducting resonators},
  author={Gurevich, Alex},
  journal={Superconductor Science and Technology},
  volume={36},
  number={6},
  pages={063002},
  year={2023},
  publisher={IOP Publishing}
}

@article{xiang13,
  title={Hybrid quantum circuits: Superconducting circuits interacting<? format?> with other quantum systems},
  author={Xiang, Ze-Liang and Ashhab, Sahel and You, JQ and Nori, Franco},
  journal={Reviews of Modern Physics},
  volume={85},
  number={2},
  pages={623--653},
  year={2013},
  publisher={APS}
}

@article{mustafa2026-arX,
  title={Interfacing Superconductor and Semiconductor Digital Electronics},
  author={Mustafa, Yerzhan and K{\"o}se, Sel{\c{c}}uk},
  journal={arXiv preprint arXiv:2601.09969},
  year={2026}
}

@article{wallraff04,
  title={Strong coupling of a single photon to a superconducting qubit using circuit quantum electrodynamics},
  author={Wallraff, Andreas and Schuster, David I and Blais, Alexandre and Frunzio, Luigi and Huang, R-S and Majer, Johannes and Kumar, Sameer and Girvin, Steven M and Schoelkopf, Robert J},
  journal={Nature},
  volume={431},
  number={7005},
  pages={162--167},
  year={2004},
  publisher={Nature Publishing Group UK London}
}
\end{document}